\documentclass[structabstract]{aa}

\usepackage{graphicx}
\usepackage{txfonts}
\usepackage{natbib}
\usepackage{longtable}
\usepackage{lscape}
\usepackage{rotating}
\bibpunct{(}{)}{;}{a}{}{,}

\begin{document}

\newcommand{\kms}{\,km\,s$^{-1}$}
\newcommand{\Wmq}{\,Wm$^{-2}$}
\newcommand{\ergps}{\,erg\,s$^{-1}$}
\newcommand{\mum}{\,$\mu$m}
\newcommand{\degree}{$^{\circ}$}
\newcommand{\Msun}{\,M$_{\odot}$}
\newcommand{\Lsun}{\,L$_{\odot}$}
\newcommand{\Msunyr}{\,M$_{\odot}$\,yr$^{-1}$}

\newcommand{\OI}{[O\,I]}
\newcommand{\SII}{[S\,II]}
\newcommand{\NII}{[N\,II]}
\newcommand{\CaII}{Ca\,II}
\newcommand{\HeI}{He\,I}
\newcommand{\LiI}{Li\,I}

\newcommand{\Teff}{$T_{\rm{eff}}$}

\newcommand{\nodata}{...}
\newcommand{\accunit}{$\odot$\,yr$^{-1}$}
\newcommand{\Mdot}{$\dot M$}
\newcommand{\Mdotacc}{$\dot M_{\rm{acc}}$}
\newcommand{\AV}{$A_{\rm{V}}$}

\newcommand{\fig}{Fig.\,}
\newcommand{\figs}{Figs.\,}
\newcommand{\sect}{Sect.\,}
\newcommand{\tab}{Table\,}
\newcommand{\tabs}{Tables\,}
\newcommand{\alfasixtwentyfour}{$\alpha_{\rm{5.8-24}}$}
\newcommand{\alfathreeeight}{$\alpha_{\rm{3.6-8}}$}
\newcommand{\alfathreetwentyfour}{$\alpha_{\rm{3.6-24}}$}

\newcommand{\rev}{ }
\newcommand{\newrev}{ }
\newcommand{\lrev}{ }

\title{Star and protoplanetary disk properties in Orion's suburbs\thanks{Based on observations performed at ESO's La Silla-Paranal observatory under programme 078.C-0696, and on observations collected at the German-Spanish Astronomical Center, Calar Alto, jointly operated by the Max-Planck-Institut f\"ur Astronomie Heidelberg and the  Instituto de Astrofísica de Andalucía (CSIC)}}

\author{M.~Fang\inst{1,2} \and R.~van~Boekel\inst{1} \and W.~Wang\inst{1} \and A.~Carmona\inst{3,4,1} \and A.~Sicilia-Aguilar\inst{1} \and Th.~Henning\inst{1} }

\institute{Max-Planck Institute for Astronomy, K\"onigstuhl 17,
           D-69117 Heidelberg, Germany
           \and
           Purple Mountain Observatory, 2 West Beijing Road,
           210008 Nanjing, China
           \and
           ISDC Data Centre for Astrophysics, University of Geneva,
           chemin d'Ecogia 16, CH-1290 Versoix, Switzerland
           \and
           Observatoire de Genève, University of Geneva,
           chemin des Maillettes 51, CH-1290 Sauverny, Switzerland
}

   \date{Received 11 May 2009; accepted 26 June 2009}
  \abstract
{Knowledge of the evolution of circumstellar accretion disks is pivotal to our
  understanding of star and planet formation; and yet despite intensive theoretical and observational studies, the disk dissipation process is not well understood. Infrared observations of large numbers of young stars, as performed by the \emph{Spitzer Space Telescope}, may advance our knowledge of this inherently complex process. While infrared data reveal the evolutionary status of the disk, they hold little information on the properties of the central star and the accretion characteristics.}
{Existing 2MASS and Spitzer archive data of the Lynds~1630N and 1641 clouds in the Orion GMC provide disk properties of a large number of young stars. We wish to complement these data with optical data that provide the physical stellar parameters and accretion characteristics.}
{We performed a large optical spectroscopic and photometric survey of the aforementioned clouds. Spectral types, as well as accretion and outflow characteristics, are derived from our VLT/VIMOS spectra. Optical SDSS and CAHA/LAICA imaging was combined with  2MASS, Spitzer IRAC, and MIPS imaging to obtain spectral energy distributions from 0.4 to 24\,\mum. Reddened model atmospheres were fitted to the optical/NIR photometric data, keeping $T{_{\rm eff}}$ fixed at the spectroscopic value. Mass and age estimates of individual objects were made through placement in the HR diagram and comparison to several sets of pre-main sequence evolutionary tracks.}
{We provide a catalog of 132 confirmed young stars in L1630N and 267 such objects in L1641. We identify 28 transition disk systems, 20 of which were previously unknown, as well as 42 new transition disk candidates for which we have broad-band photometry but no optical spectroscopy. We give mass and age estimates for the individual stars, as well as equivalent widths of optical emission lines, the extinction, and   measures of the evolutionary state of the circumstellar dusty disk. We estimate mass accretion rates \Mdotacc \ from the equivalent widths of the H$\alpha$, H$\beta$, and \HeI\,5876\AA \ emission lines, and find a dependence of \Mdotacc\,$\propto$\,$M_*^{\alpha}$, with $\alpha\sim$3.1 in the subsolar mass range that we probe. An investigation of a large literature sample of mass accretion rate estimates yields a similar slope of $\alpha\sim$2.8 in the subsolar regime, but a shallower slope of $\alpha\sim$2.0 if the whole mass range of 0.04\Msun$\leq$$M_*$$\leq$5\Msun \ is included. The fraction of stars with transition disks that show significant accretion activity is relatively low compared to stars with still optically thick disks (26$\pm$11\% vs. 57$\pm$6\%, respectively). However, those transition disks that \emph{do} show significant accretion have the same median accretion rate as normal optically thick disks of 3-4$\times$10$^{-9}$\Msunyr. Analyzing the age distribution of various populations, we find that the ages of the CTTSs and the WTTSs with disks are statistically indistinguishable, the WTTSs without disks are significantly older than the CTTSs, and the ages of the transition disks and the WTTSs without disks are statistically indistinguishable. These results argue against disk-binary interaction or gravitational instability as mechanisms causing a transition disk appearance. Our observations indicate that disk lifetimes in the clustered population are shorter than in the distributed population. In addition to the spectroscopic sample analyzed in this paper, we provide a photometric catalog of sources detected in the optical and infrared, but without spectroscopic observations. As judged by their infrared colors, many of these are YSO candidates. In our survey we identify 2 new aggregates in L1641. We find 4 apparently subluminous objects with extremely high equivalent widths of H$\alpha$ and other emission lines, and 1 previously unknown FU~Orionis object. We find that the low-density molecular cloud emission that surrounds the star-forming cores has significant substructure on scales of $\lesssim$0.2\,pc in L1641 but not in L1630. We propose refined H$\alpha$ equivalent width criteria to distinguish WTTSs from CTTSs in which the boundary EW is lowered significantly for late M spectral types.\thanks{Tables 7, 8, and 11-14 are only available in electronic form at the CDS via anonymous ftp to cdsarc.u-strasbg.fr (130.79.128.5) or via http://cdsweb.u-strasbg.fr/cgi-bin/qcat?J/A+A/}}
{}

 \maketitle

\section{Introduction}
\label{sec:Introduction}
Circumstellar disks play a key role in the formation of new stars and planetary systems, and form as a result of angular momentum conservation during the protostellar core collapse \citep{1977ApJ...214..488S,2005ASPC..337....3H}. Through the disks, a significant fraction of the stellar mass is accreted, while the excess angular momentum is transported outward.

After the main accretion phase has ended, gas-rich circumstellar disks can survive at low accretion rates until they are eroded by stellar winds, photo-evaporation, or interaction with giant gas planets or stellar companions {\rev \citep{1994ApJ...421..651A,2001MNRAS.328..485C,2006MNRAS.369..229A,2003MNRAS.342...79R,2004ApJ...612L.137Q,2008PhST..130a4024H}}. It is during this phase that planets are believed to form inside the disks. A direct observational link between planet formation and disk evolution was recently proposed by \cite{2008Natur.451...38S}, who found evidence for a massive planet in close orbit around TW~Hya, whose disk is currently being dissipated and is thought to be in a transition state between a gas-rich disk typical of classical T-Tauri stars and a gas-poor debris disk. This finding was, however, questioned by \citet{2008A&A...489L...9H}, {\rev who instead attribute the observed radial velocities to a large stellar spot that rotationally modulates the signal. However, the brightness variations that are inherent to the spot model were not detected in 2008 by \cite{2008MNRAS.391.1913R} while the radial velocity variations first seen by \cite{2008Natur.451...38S} remain present (Setiawan et al. in prep), seriously challenging the spot model of \citet{2008A&A...489L...9H}.} At any rate, knowledge of disk evolution is clearly pivotal to understanding planet formation \citep{2008PhST..130a4019H}.

The disk dissipation process has been constrained observationally by investigating the fraction of young stars that have strong excess emission from disk material at near-infrared wavelengths, in clusters exhibiting a range of ages. Near-infrared imaging surveys have suggested that the inner disk frequency is $>$50\% at ages of 1-2~Myr, and dramatically decreases to $\sim$10\% at ages of 5-10~Myr, indicating that the lifetime of inner disks is a few~Myr \citep{1989AJ.....97.1451S,2001ApJ...553L.153H, 2002astro.ph.10520H}. Observations at longer mid-infrared to millimeter wavelengths, tracing cooler dust at larger radii within the disks, have suggested that the outer parts of the disks survive somewhat longer than the regions close to the central star  \citep{1995ApJ...439..288O,2000LNP...548..341M,2004ApJ...612..496M,2005ApJ...631.1134A,2005AJ....129.1049C}.

The rate at which disk material accretes onto the central star can be estimated from the infrared excess, veiling in optical spectra, and emission lines that are thought to be due to magnetospheric accretion \citep{1992ApJ...397..613H,1995ApJ...452..736H,1998ApJ...492..323G,1998AJ....116.2965M,2000prpl.conf..377C,2001ApJ...550..944M}. The accretion rate has been found to scale approximately with the square of the stellar mass from the brown dwarf to the HAeBe star regime \citep{2003ApJ...592..266M,2006A&A...452..245N,2004AJ....128.1294C,2005ApJ...625..906M,2008ApJ...681..594H}. Additionally, a clear trend of decreasing accretion rate with increasing age has been identified \citep{1998ApJ...495..385H,2000prpl.conf..377C,2004AJ....128..805S,2005ASPC..341..131H}.

\begin{table*}
\caption{Observing log for the LAICA imaging observations of L1641.}\label{laica_log}
\centering
\begin{tabular}{lccccc}
\hline\hline
   &  Pointing  center  & Observation date & Airmass & Exposure times & \\
Fields  & (J2000) & (UT)  &  (g'/r'/i'/z')       & (second) & Calibartion fields   \\
\hline
L1641-1 & 05 39 17.29  $-$07 12 41.7&2007-12-15 21:52:15--22:37:36&1.6/1.6/1.6/1.7&1, 45, 300&SA98\\
L1641-2 & 05 40 18.76  $-$07 28 03.5&2007-12-15 22:40:01--23:24:01&1.4/1.5/1.5/1.5&1, 45, 300&SA98\\
L1641-3 & 05 41 20.10  $-$07 43 26.3&2007-12-15 23:27:25--00:12:01&1.4/1.4/1.4/1.4&1, 45, 300&SA98\\
L1641-4 & 05 42 21.46  $-$07 58 48.9&2007-12-16 00:13:57--00:57:30&1.4/1.4/1.4/1.4&1, 45, 300&SA98\\
L1641-5 & 05 43 22.91  $-$08 14 10.5&2007-12-16 00:59:17--01:43:11&1.5/1.5/1.5/1.5&1, 45, 300&SA98\\
L1641-6 & 05 44 24.25  $-$08 29 33.3&2007-12-16 01:45:51--02:29:56&1.7/1.7/1.6/1.6&1, 45, 300&SA98\\
L1641-7 & 05 40 18.20  $-$07 43 35.7&2008-03-11 20:07:04--20:54:23&1.6/1.7/1.7/1.8&1, 45, 300&SA98\\
L1641-8 & 05 41 19.55  $-$07 58 58.6&2008-03-11 20:58:16--21:41:39&2.0/2.0/2.2/1.8&1, 45, 300&SA98\\
L1641-9 & 05 42 20.89  $-$08 14 20.5&2008-03-12 20:19:18--20:34:37&1.6/1.6/1.6/1.7&1, 45, 300&SA95,SA98\\
L1641-10& 05 43 22.35  $-$08 29 43.5&2008-03-12 20:47:01--21:26:38&1.8/1.9/2.0/2.1&1, 45, 300&SA95,SA98\\
\hline
\end{tabular}
\end{table*}

\begin{table*}
\caption{Observing log for the VIMOS optical spectroscopy. }\label{vimos_log}
\centering
\begin{tabular}{rcccc}
\hline\hline
  &  Pointing  center  & Data/time of observations & Integration time &  \\
Fields & (J2000) & (UT)&    (second)  & Airmass  \\
\hline
L1630-1 BLUE& 05 46 57.08  $+$00 23 34.4&2007-02-11 01:14:15/01:40:37/02:07:10&3$\times$1526&1.1/1.1/1.1\\
        RED& 05 46 57.08  $+$00 23 34.2&2007-02-09 02:13:46/02:40:08/03:06:40&3$\times$1526&1.1/1.2/1.3\\
L1630-2 BLUE& 05 46 11.01  $+$00 08 33.3&2007-02-19 01:07:22/01:33:44/02:00:16&3$\times$1526&1.1/1.1/1.2\\
        RED& 05 46 11.00  $+$00 08 33.0&2007-02-18 01:19:07/01:45:29/02:12:00&3$\times$1526&1.1/1.2/1.2\\
L1630-3 BLUE& 05 46 04.81  $-$00 06 54.8&2007-02-20 00:54:26/01:20:48/01:47:19&3$\times$1526&1.1/1.1/1.2\\
        RED& 05 46 04.80  $-$00 06 54.8&2007-02-21 00:52:21/01:18:42/01:45:13&3$\times$1526&1.1/1.1/1.2\\
L1641-1 BLUE& 05 35 50.41  $-$06 20 43.3&2007-01-28 01:35:00/02:01:22/02:27:54&3$\times$1526&1.0/1.1/1.1\\
        RED& 05 35 50.41  $-$06 20 43.3&2007-01-28 03:08:50/03:35:12/04:01:46&3$\times$1526&1.1/1.2/1.2\\
L1641-2 BLUE& 05 36 19.38  $-$06 46 15.1&2006-12-22 04:53:29/05:16:31/05:39:44&3$\times$1326&1.1/1.1/1.1\\
        RED& 05 36 19.38  $-$06 46 15.2&2006-12-21 04:33:45/04:56:47/05:20:01&3$\times$1327&1.1/1.1/1.1\\
L1641-3 BLUE&05 38 25.92  $-$06 59 43.2&2007-02-17 00:58:59/01:25:21/01:51:51&3$\times$1526&1.1/1.1/1.1\\
        RED& 05 38 25.92  $-$06 59 43.2&2007-02-15 01:11:09/01:37:31/02:04:01&3$\times$1526&1.1/1.1/1.1\\
L1641-4 BLUE& 05 42 01.92  $-$08 04 49.0&2007-02-22 01:30:08/01:56:30/02:23:01&3$\times$1526&1.1/1.1/1.2\\
        RED& 05 42 01.94  $-$08 04 49.4&2007-03-12 00:56:23/01:22:45/01:49:16&3$\times$1526&1.2/1.2/1.3\\
L1641-5 BLUE& 05 40 51.60  $-$07 51 01.4&2007-01-25 01:54:00/02:20:21/02:46:55&3$\times$1526&1.0/1.0/1.1\\
        RED& 05 40 51.60  $-$07 51 02.0&2007-01-27 01:16:57/01:43:19/02:09:49&3$\times$1526&1.1/1.0/1.0\\
\hline
\end{tabular}
\end{table*}

With the advent of the Spitzer Space Telescope our ability to study disk evolution has dramatically increased, both by largely increasing of the number of stars and clusters studied, as well as extending the wavelength range at which large samples of objects can be accurately measured further into the infrared. Many young stars in clusters with ages from 1~Myr to tens of Myrs have been studied \citep[e.g.][]{2004ApJS..154..374G,2004ApJS..154..428Y,2005ApJ...629..881H,2005ApJ...634L.113M,2006AJ....131.1574L,2006ApJ...638..897S,2008ApJ...687.1145S,2006ApJ...652..472H,2007AJ....133.2072D,2007ApJ...662.1067H,2008AJ....135..966F}. The inner disk fraction changes from $>$54$\pm$15\% in the core of the 1-Myr-old cluster NGC~7129 \citep{2004ApJS..154..374G}, to $\sim$44$\pm$7\% in the cluster IC~348 with an age of 2-3~Myr \citep{2006AJ....131.1574L}, $\sim$33.9$\pm$3.1\% in the 3-Myr-old cluster $\sigma$~Orionis \citep{2007ApJ...662.1067H}, $\sim$20\% in the 5-Myr-old cluster NGC~2362\citep{2007AJ....133.2072D}, to 4\% in the cluster NGC~7160 with an age of $\sim$10~Myr \citep{2006ApJ...638..897S}. Generally, the disk fraction decreases with increasing cluster age, confirming previous results on the dissipation timescale based mainly on ground-based near-infrared data.

Some studies show that the disk fraction peaks for stars in the T~Tauri mass range \citep{2007ApJ...662.1067H}, especially for those of K6-M2 types with masses around one solar mass, suggesting that planet formation is favored around solar-mass young stars \citep{2006AJ....131.1574L,2007ApJ...671.1784H}. Research on weak-line T Tauri stars (WTTS) with ages of 1-2~Myr shows that up to 50\% of WTTSs no longer possess inner disks, indicating that some young stars can dissipate their disks at very early ages \citep{2007ApJ...667..308C}. Observations at mid-infrared or longer wavelengths identify many stars with prominent excesses at these wavelengths, in some cases also in objects showing no near-infrared excess, whose inner disk regions appear to already have dissipated \citep{2004ApJS..154..379M,2004ApJS..154..428Y,2005ApJ...621..461D,2005ApJ...630L.185C,2008arXiv0804.3113G}.

In this paper, we combine optical spectroscopy with optical, near-, and mid-infrared photometry to characterize the stellar- and disk properties of a large set of young stars located in the Lynds~1630 and Lynds~1641 clouds, located in the Orion molecular cloud complex. The distance to this complex is estimated to be between 400 and 500~pc  \citep[e.g.][]{1982AJ.....87.1213A,2007PASJ...59..897H}, and the region probably has a ``depth'' of at least several tens of parcecs. Throughout this work we will assume a distance of 450~pc for both the L1630 and L1641 clouds. L1630 lies in the northern part of the Orion complex (Orion~B), and L1641 is located in the southern part (Orion~A). Toward L1630, near-infrared surveys show that most young stars are found in four clusters (NGC~2023, NGC~2024, NGC~2068, and NGC~2071) rather than being uniformly distributed \citep{1991ApJ...371..171L,1997ApJ...488..277L}. In NGC~2068/2071, which are located in the northern part of L1630 and named L1630N hereafter, a previous study shows that 53 out of  the 67 identified members have infrared excesses, and all stars with infrared excess also display active accretion \citep{2008AJ....135..966F}. Here,  we extend the number of identified and characterized young stars in this region. In contrast to the L1630 region, the L1641 cloud harbors a large population of young stars existing in relative isolation, in addition to a population of stars in a number of clusters or aggregates \citep{1993ApJ...412..233S,1995PhDT..........A}. Thus, a comparative study of disks around young stars in L1630 and L1641 offers the opportunity to study the effect of a clustered or isolated environment on the disk evolution.

We arrange the paper as follows: in $\S$2 we describe the observations and data reduction, in $\S$3 we delineate our data analysis, we present our result in $\S$4, followed by a discussion in $\S$5.

\section{Observations and data reduction}
\label{sec:observations_and_data_reduction}

The data employed in this work consist of photometry in the 0.4 to 24\,\mum \ range, and optical spectroscopy from $\sim$4000 to $\sim$9000\,$\AA$. The optical photometry of {\rev The south-east half of} L1641 is new, and the rest of the photometry are publicly available archive data. We performed optical spectroscopy with the multi-object spectrograph VIMOS at the ESO-VLT.

\subsection{Optical photometry}
\label{sec:osbservations:optical_photometry}

Optical photometry was taken from the Sloan Digital Sky Survey \citep[SDSS,][]{2000AJ....120.1579Y} in the $u'g'r'i'z'$ bands centered on 0.35, 0.48, 0.62, 0.76 and 0.91~$\mu$m, respectively. The L1630N cloud was covered entirely and scanned multiple times ($\sim$3.3 times on average), resulting in average 10$\sigma$ limiting magnitudes of 21.9, 23.0, 22.4, 21.8, and 20.2, respectively. The L1630N data are publicly available in the SDSS ``Low Galactic Latitude Fields'' data release \citep{2004AJ....128.2577F}. Only about half of the L1641 cloud was covered by SDSS and the covered parts were scanned an average of $\sim$1.3 times, with resulting 10$\sigma$ limiting magnitudes of about 20.5, 21.7, 21.4, 21.1 and 19.7, respectively. The south-east half of the L1641 cloud has not been observed by SDSS.

We complemented the SDSS photometry of L1641 with CCD imaging in the SDSS $g'r'i'z'$ bands performed at the Calar Alto 3.5m telescope, using the Large Area Imager for Calar Alto (LAICA). LAICA is a wide field optical imager employing four 4k$\times$4k CCDs. Part of the south-east half of L1641 was observed on December 15, 2007. Conditions were photometric, but the seeing was poor (2-3\arcsec), somewhat limiting the sensitivity for faint point sources ($\sim$19.6, 19.6, 19.9, and 18.9 mag at $g'r'i'z'$ band, respectively). In order to increase the dynamic range, three exposures were taken at each position and in each filter, with integration times of 1/45/300 seconds (see table~\ref{laica_log}). Standard data reduction for optical CCD imaging consisted of bias subtraction, flat-fielding using sky flats, and removal of fringing due to night sky airglow in the $i'$ and $z'$ bands.

The astrometric solution was determined for each individual CCD frame by correlating stellar positions with the USNO~A-2 catalog, typically using a few dozen stars but never fewer than 5. The resulting positional uncertainties of detected stars are less than 0\farcs5 over the whole field, sufficient for unambiguous cross-identification with sources detected in the infrared data.  Photometric calibration was performed by observing Landolt standard field  SA98, and SA95. These fields have been observed by SDSS. We {\lrev used} the SDSS photometry of all the isolated stars to calibrate our LAICA observation. The photometry for each star {\lrev was} chosen from the longest unsaturated exposure. In total, almost the entire L1630N and L1641 Spitzer fields are covered by our optical imaging.

\subsection{Infrared photometry}
\label{sec:observations:infrared_photometry}
Near-infrared photometry in the $J$, $H$, and $K_S$ bands was taken from the Two-Micron All Sky Survey \citep[2MASS, ][]{2006AJ....131.1163S}, with 10$\sigma$ limiting magnitudes of 16.2, 15.3, and 14.6 magnitudes, respectively. The L1630N and L1641 clouds were imaged using the Spitzer Space Telescope IRAC \citep{2004ApJS..154...10F} and MIPS \citep{2004ApJS..154...25R} cameras.

\subsubsection{IRAC photometry}
\label{sec:observations:irac_photometry}
IRAC images at 3.6, 4.5, 5.8, and 8.0~$\mu$m were made in High-Dynamic Range mode with integration times of 0.4 and 10.4 seconds (Spitzer program ID 43). Four mosaics were made of each cloud, with a fair amount of overlap between the individual exposures in each mosaic. The IRAC data of the L1630 and L1641 clouds were published previously by \cite{2005IAUS..227..383M}.

The Spitzer archive provides pipeline reduced (version S14.0.0) image mosaics as well as the corresponding pixel to pixel flux uncertainty maps. For each field the individual mosaics were combined into a final image by making a weighted average (after cosmic ray rejection by a sigma-clipping procedure, and using the uncertainty maps as weights). Separate images were created from the 0.4 and 10.4 second exposures.

A custom made IDL program was used to search for point sources in the IRAC images. At its heart is the \texttt{find.pro} procedure from the \texttt{astrolib} library. However, rather than searching for point sources in the images themselves, we let \texttt{find} search in an image \emph{minus} a smoothed version of that image, thus effectively removing the sometimes bright and inhomogenous nebular background. Moreover, we {\lrev used} a variable detection threshold that is low in ``clean'' regions, but higher in regions with a high background. After determining proper extraction parameters and extensive testing of the procedure, checking the results by eye, we found that our procedure is very robust and effectively finds all but the very faintest sources in the whole image, without yielding false detections in regions with high and variable background.

The source finder was run in each of the 4 IRAC bands, on the short and long exposures separately. Source detections in the various bands were correlated by spatial coincidence within 1 pixel (1\farcs2). Only sources that were seen in both the 3.6 and 4.5 micron bands were kept in the analysis (source counts amount to $\sim$25000 and $\sim$40000 in the L1630N and L1641 fields, respectively).

PSF photometry was performed on the detected sources. For each band, the point spread function was determined from bright, isolated, non-saturated stars. The psf-fitting program is based on the StarFinder code  \citep{2000A&AS..147..335D}. The psf-fitting extraction box is 15.6\arcsec$\times$15.6\arcsec for each IRAC band. The zero-point magnitudes are  17.30, 16.82, 16.33, and 15.69 in the 3.6, 4.5, 5.8, and 8.0~\mum \ bands, respectively. We {\lrev compared} the photometry at short and long exposures to determine the saturation level for the longer exposures. If the stars are not saturated on the long-exposure images, we select their photometry from these long-exposure images. Otherwise, we select them  from the short-exposure images. \citet{2008AJ....135..966F} present photometry of 69 stars in L1630N. A comparison between their photometric magnitudes and ours shows rms differences of $\sim$0.06-0.1\,mag for the four IRAC bands. Adopting these values as typical uncertainties in our photometry, we conclude that the photometric accuracy of the pipeline processed data as we have used them, without any custom post-processing, is perfectly adequate for our purposes\footnote{The [5.8] magnitude for YSO \#105 in L1630N, as well as the [8.0] magnitudes for the YSOs \#24, 54, 76, 84,105, 116, 125, 128 in L1630N, were adopted from \cite{2008AJ....135..966F}.}.

\subsubsection{MIPS photometry}
\label{sec:observations:mips_photometry}
The L1630N and L1641 clouds were mapped with the MIPS instrument (Spitzer program ID 47). The effective integration time was 80, 40, and 8 seconds at 24, 70, and 160\,$\mu$m, respectively. Following \citet{2008AJ....135..966F}, who previously presented the L1630N data, we {\lrev included} only the 24\,$\mu$m data due to lack of detected Class~II sources at the longer wavelengths\footnote{The MIPS [24] magnitudes for YSOs \#24, 81, 125, 126 in L1630N were adopted from \citet{2008AJ....135..966F}.}. The same searching and psf-fitting programs as used for the IRAC data {\lrev were} used on the 24\,$\mu$m image, with a psf-fitting extraction box of 56.35\arcsec$\times$56.35\arcsec and a zero-point magnitude of 11.76.

\begin{figure}
\centering
\includegraphics[width=\columnwidth]{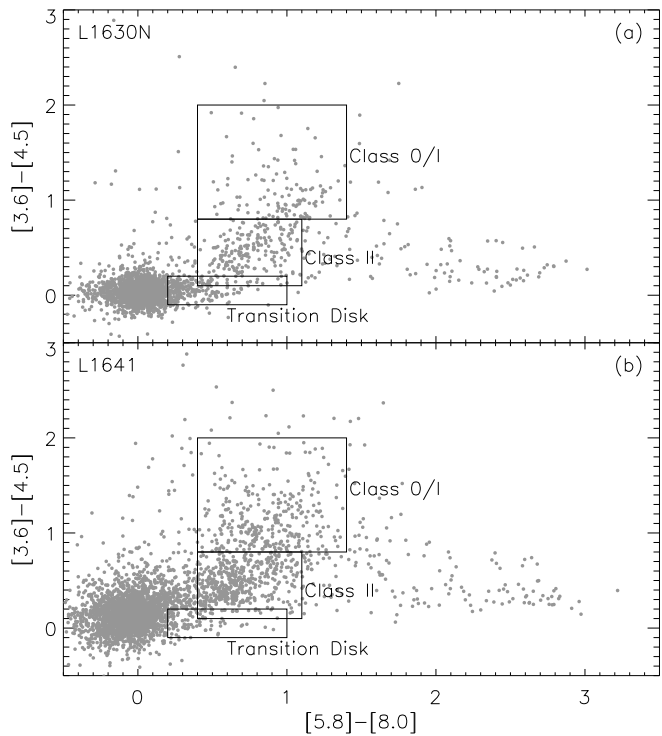}
\caption{\label{fig:allccmap} Spitzer [5.8]-[8.0] vs. [3.6]-[4.5] color-color diagrams for all objects detected at all four IRAC bands in L1630N and L1641. Three boxes enclose the boundaries of candidate selection regions for three types of YSOs, i.e. class\,0/I YSOs, class\,II YSOs, and YSOs with transition disks.}
\end{figure}

\begin{figure*}
\centering
\includegraphics[width=18.7cm]{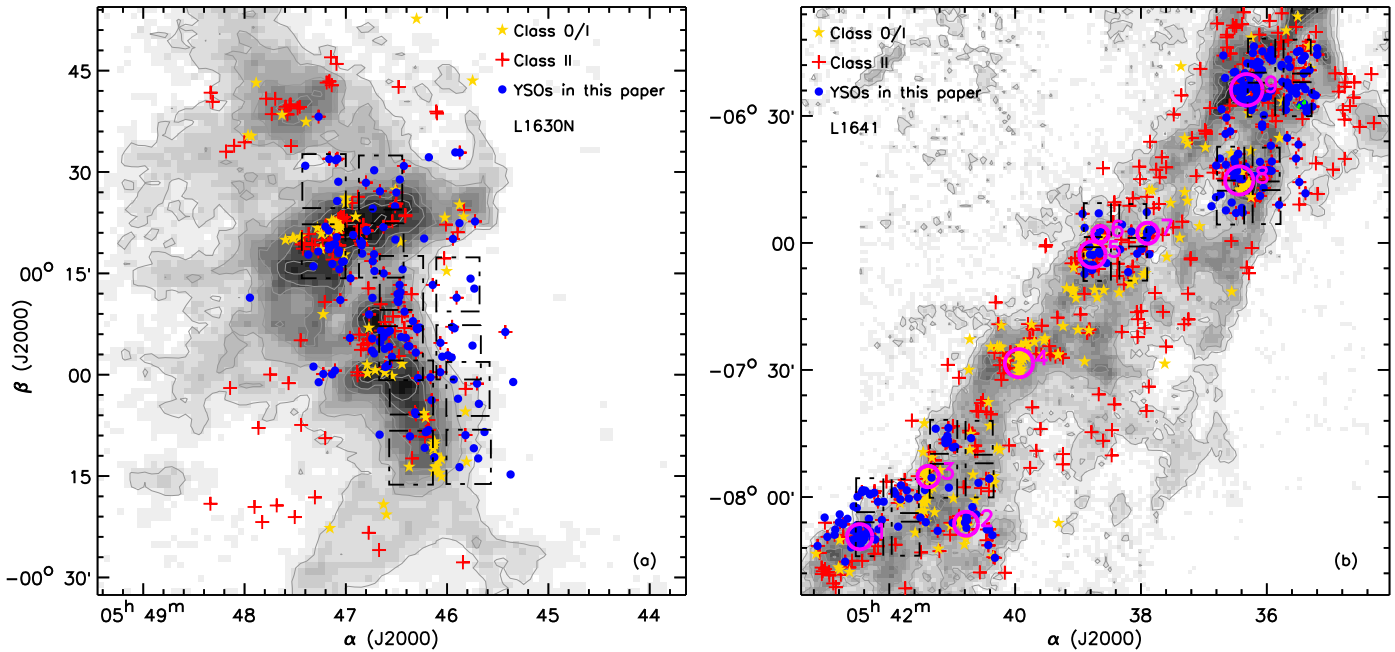}
\caption{\label{fig:YSODIS} The distribution of YSOs in L1630N and L1641 overplotted on a  $^{13}$CO integrated intensity image \citep{1987ApJ...312L..45B,1994ApJ...429..645M}. The filled circles mark the YSOs discussed in this paper, which have confirmed spectral types. The asterisks and pluses represent the class 0/I and class II YSOs, respectively. The boxes show our fields of view of our VIMOS observations, each set of four quadrants representing one VIMOS pointing.}
\end{figure*}

\subsection{Optical spectroscopy}
\label{sec:observations:optical_spectrosocpy}
The main new data set presented in this paper is VLT/VIMOS optical spectroscopy of over 700 targets in the direction of the L1630\,N and L1641 star-forming clouds. Accurate stellar parameters and extinction estimates for individual objects cannot be derived from photometry alone, but instead require the combination of spectroscopy and photometry.

\subsubsection{Target Selection}
\label{sec:observations:target_selection}
The L1630 and L1641 clouds subtend large solid angles on the sky, and to cover them completely with VIMOS observations would require very large amounts of telescope time. Instead, we concentrated on a number of subfields, 3 VIMOS pointings in L1630\,N and 5 in L1641. The pointings were chosen on the basis of the Spitzer IRAC data and we selected those fields with many potential YSO candidates, i.e. IR excess sources, for optical follow-up. In the [5.8]-[8.0] vs. [3.6]-[4.5] color-color diagrams, we selected YSO candidates using the following criteria (see \fig\ref{fig:allccmap}): (1) 0.4$\leq$[5.8]-[8.0] $\leq$1.1 and 0.1$\leq$[3.6]-[4.5] $\leq$0.8 (classical T Tauri candidates), (2) 0.2$\leq$[5.8]-[8.0] $\leq$1.0 and -0.1$\leq$[3.6]-[4.5] $\leq$0.2 (transition disk candidates), or (3) 0.4$\leq$[5.8]-[8.0] $\leq$1.4 and 0.8$\leq$[3.6]-[4.5] $\leq$2.0 (class 0/I candidates), see also \cite{2004ApJS..154..363A}.

Pre-imaging in R-band of the selected fields was performed with VIMOS. Based on these data, targets that are sufficiently bright were selected for spectroscopic follow-up. Here, preference was given to sources with  IRAC colors {\rev that suggest they are young stars possessing disks}, but many sources that do not posses colors within the boxes outlined above were selected as well, mostly sources without obvious IR excess emission. This approach has the advantage that a large fraction of available resources are spent on young stars instead of unrelated background objects, but it also introduces a bias toward IR excess sources in the sample of cluster members (see \sect\ref{sec:observations:biases}).

\subsubsection{VLT/VIMOS observations}
\label{sec:observations:vimos_observations}
We used \emph{VIMOS}, the Visible Multi-Object Spectrograph \citep[][]{2003SPIE.4841.1670L} mounted on the ESO Very Large Telescope to obtain optical spectra. The observations were performed  during the period from December 2006 to March 2007 with two intermediate resolution grisms, the HR blue ($\lambda/\Delta\lambda$=2050 for a slit width of 1\arcsec) and HR red ($\lambda/\Delta\lambda$=2500 for a slit width of 1\arcsec) grisms, covering the wavelength range from 4100$\AA$ to 8750$\AA$ (see table~\ref{vimos_log}). We observed 322 targets in L1630N and 393 in L1641, distributed over 3 and 5 VIMOS pointings, respectively (see \fig\ref{fig:YSODIS}). For each pointing we performed three 25-minute exposures. The VIMOS spectra were reduced using the VIMOS pipeline provided by ESO (Carlo Izzo, private communication). Two methods can be applied to subtract telluric and nebular emission lines from the stellar spectra: the ``skylocal'' and ``skymedian'' options. In the ``skylocal'' method, the sky spectrum is subtracted from CCD science data before the data are corrected for optical distortions. In the ``skymedian'' method, the sky subtraction is performed on the distortion-corrected science data, which have been resampled. We {\lrev prefered} the skylocal method, and only very occasionally {\rev used} spectra from the skymedian method, in cases where the the reductions using the skylocal method contain artifacts.

\subsubsection{Biases introduced in target selection}
\label{sec:observations:biases}
We observed a large number of cloud members with and without infrared excess emission, as well as unrelated field objects, mostly background stars. However, since sources with IR excess emission were more likely (see \sect\ref{sec:observations:target_selection}) to be included in the optical spectrocopic observations, our sample will be biased towards these. Therefore, quantities like the \emph{absolute} disk fraction will be affected, and in general will be over-estimated. However, \emph{relative} trends, such as the disk fraction as a function of stellar mass, are in principle unaffected by our target selection.

\subsection{Matching of optical and IR data.}
 We matched the different photometric data sets and spectra based on spatial coincidence using a 2\arcsec \ tolerance. Since the investigated fields are not very crowded this approach is adequate.

\section{Analysis}
\label{sec:analysis}

In this section we will describe the methods that we applied to extract the physically interesting information from our data. The main goals are the identification of young stars, the determination of their stellar parameters, characterization of their disk geometry and estimation of the rate at which disk material is accreted onto the central star.

\subsection{YSO selection criteria}
\label{sec:analysis:YSO_selection_cirteria}
A star in our sample observed with VIMOS is classified as a young star if it obeys any of the following criteria:

\begin{enumerate}
\item \label{crit:IR_excess} IR excess
\item \label{crit:LiI} Li\,I absorption
\item \label{crit:Halpha} H$\alpha$ emission
\end{enumerate}

\noindent
We find that sources with IR excess emission always show H$\alpha$ emission, though the opposite is not necessarily true, i.e. some sources show H$\alpha$ emission but no IR excess. A large fraction of the sources with H$\alpha$ emission show Li\,I absorption in their spectra, and all sources showing Li\,I also show H$\alpha$ emission. We note that there may be a small contamination of our sample with dMe stars, which are old, M-type stars that show H$\alpha$ emission due to chromospheric activity.

\subsection{Spectral classification}
\label{sec:analysis:spectral_classification}
We determined the spectral types of the stars for which we have VIMOS spectroscopy using the classification scheme developed by \cite{2004AJ....127.1682H}. The method uses empirical relations between the equivalent widths of selected atomic and molecular absorption lines and the effective temperature. In general, each individual line or molecular band is a sensitive measure of \Teff \ over a limited range in spectral types only, but the combination of a number of absorption features yields a unique determination of  the spectral type in the range of early B to late M. The classification scheme consists of 3~subregimes, each  spanning a range in \Teff: the "HAeBe"~type, "G"~type, and "late"~type scheme. For a detailed description of the method, we refer to \citet{2004AJ....127.1682H}.

\begin{figure*}
\centering
\includegraphics[width=14cm]{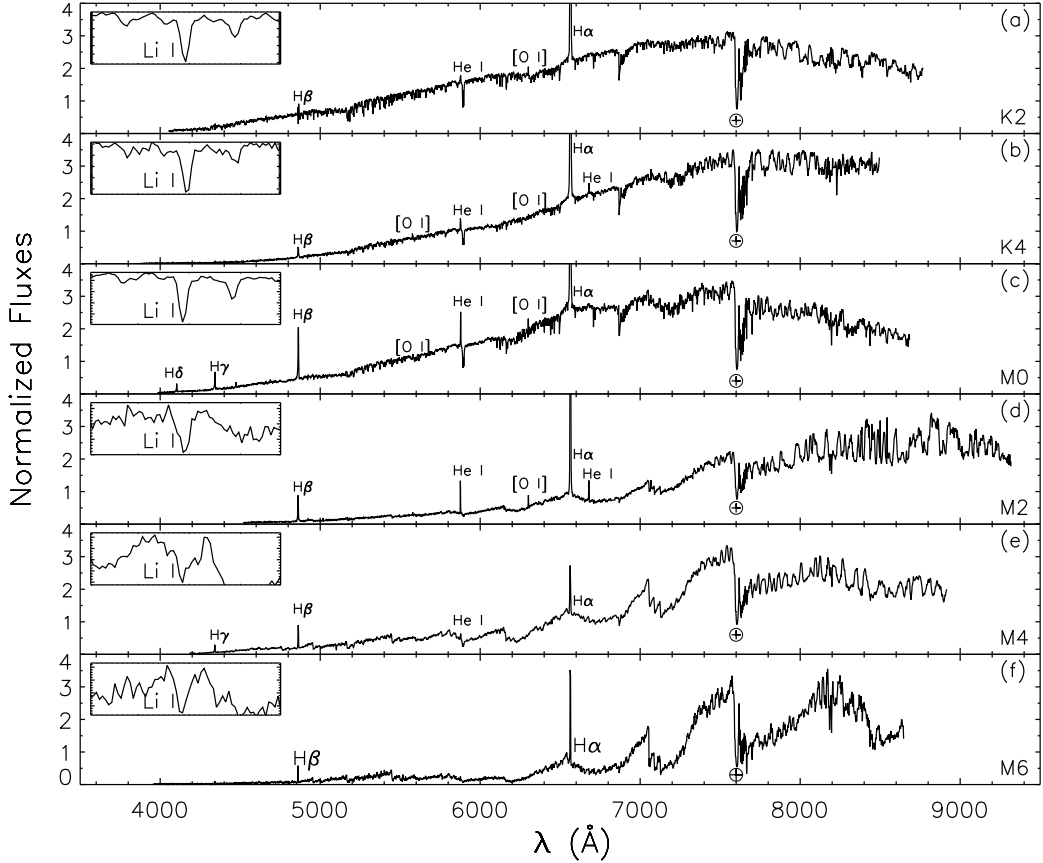}
\caption{\label{fig:spectral_sequence} Example spectra of our VIMOS observations covering the range of spectral types of the young stars discussed in this work. The main emission lines have been marked, and the insets show the Li\,I\,$\lambda6707$ absorption line.}
\end{figure*}

In \fig\ref{fig:spectral_sequence}, we show several examples of VIMOS spectra of our target stars, covering the early~K to mid~M spectral type range, representative of the vast majority of the young stars in our sample. The changes in spectral shape over this range are well visible, in particular TiO absorption bands become prominent in spectral types later than M2. Since VIMOS does not use fibers but rather employs masks with multiple slits that are directly imaged onto the detectors, the spectral range covered depends on the position of a star within the field of view. This is illustrated by, e.g. spectra (b) and (d) in \fig\ref{fig:spectral_sequence}.

\subsubsection{Reliability and accuracy of the classification code}
\label{sec:analysis:accuracy_spectral_classification}

In order to test the accuracy and reliability of our spectral classification code, we ran it on 1273 spectral templates from the Indo-US library \citep{2004ApJS..152..251V} and compared the spectral types derived using our code to those listed in the spectral library. We {\lrev found} that we accurately {\lrev recovered} the spectral types of the templates over the whole spectral region from early B to late M, with a standard deviation of $\sim$1 sub~type. Since the spectral templates in the library have their own intrinsic uncertainty that is not much better than 1~sub~type, we may safely assume 1~sub~type as the intrinsic accuracy of our method.

In order to assess the effect of limited SNR in our spectra, we ran the classification code again after adding normally distributed noise to the template spectra, such that their SNR was reduced to 30 and 10, respectively. From this exercise, we {\lrev found} that for a SNR of 30, we {\lrev recovered} the spectral types nearly as well as for the original templates. For a SNR of 10, we reliably {\lrev recovered} spectral types later than $\sim$F5, with a  accuracy of $\sim$1 sub~type. For spectral types earlier than F5, the method can yield results that are off by as much as 10 sub types for spectra with a SNR of 10, and no robust estimates can be made.

\begin{figure}
\centering
\includegraphics[width=\columnwidth]{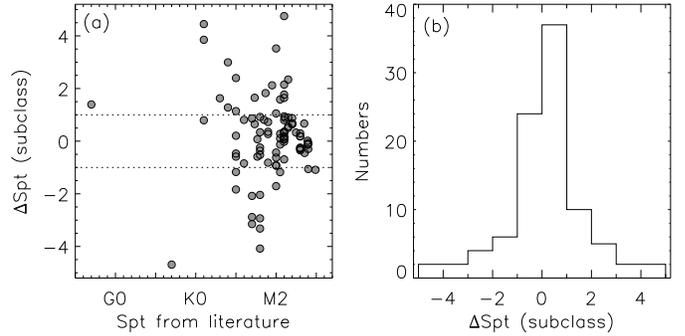}
\caption{\label{fig:spectral_type_dif} Left panel: the differences between the spectral types in this paper and in the literature vs their spectral types in literature. Right panel: the distribution of spectral-type differences in left panel. Among the total of 95 stars, there are 65\% with differences less than one subclass, and $\sim$ 82\% with less than two subclasses.}
\end{figure}

Of the 540 stars for which we could reliably determine the spectral types from our VIMOS data, 94 have been classified previously by other authors \citep[][]{1995PhDT..........A,2008AJ....135..966F,2008A&A...489.1409G}, allowing an independent check on our classification. \fig\ref{fig:spectral_type_dif} shows the direct comparison between the spectral types we derived and those quoted in the literature. 65\% of this sample agrees within 1~subclass, 82\% within 2~subclasses.

\subsubsection{Complementary literature spectroscopy}
\label{sec:analysis:literature_spectroscopy}

A substantial number of young stellar objects with confirmed spectral types can be found in the literature, which we used to complement our sample. \citet{1995PhDT..........A} presents a list of 337 YSO candidates. Among these, there are 78 stars that we did not observe with VIMOS, and which have published equivalent widths of the H$\alpha$ emission line or exhibit IR excess emission in our Spitzer data. In \cite{2008A&A...489.1409G} we found 39 YSOs with H$\alpha$ emission and confirmed spectral types, of which 10 were not covered by our VIMOS observations and were added to our YSO sample in L1641. \citet{2008AJ....135..966F} identify 67 YSOs in L1630N, of which we observed 23 with VIMOS. In total, the literature data add 44 and 88 YSOs to our sample in the L1630N and L1641, respectively. We used only the spectral types and H$\alpha$ equivalent widths given in the quoted papers, and complemented these data with our own optical and infrared photometry. Subsequently, we derived stellar and disk properties in the same way as for the VIMOS sample.

\subsection{Determining the stellar properties}
\label{sec:analysis:determining_stellar_properties}

Our optical spectroscopy and photometry provide the observational basis for determining the stellar parameters of our targets. First, the effective temperatures of the target stars are inferred from their measured spectral types. Reddened model atmospheres are then fitted to the optical photometry to determine the angular diameter and the extinction towards each star. Together with the assumed distance, the angular diameter and effective temperature yield the bolometric luminosity. Subsequent placement in the HR diagram finally yields estimates of the stellar mass and age by comparison to theoretical pre main-sequence evolutionary tracks.

\subsubsection{Effective temperatures, model atmospheres, and extinction}
\label{sec:analysis:Teff_modelatmospheres_extinction}

We transformed the spectral types to effective temperatures using the relation given by \citet{1995ApJS..101..117K} for stars with spectral types earlier than M0, and those by \citet{2003ApJ...593.1093L} for stars of type later than M0. We {\lrev used} model atmospheres to represent the optical SED of our target stars. At temperatures above 4500\,K we {\lrev used} the models by \cite{1979ApJS...40....1K,1994KurCD..19.....K}, at lower temperatures we {\lrev adopted} MARCS models \citep{2008A&A...486..951G}. The surface gravity {\lrev was} assumed to be those of ZAMS stars and we adopted solar metallicity.

The optical spectra of our stars may suffer from strong extinction by intervening dust. We {\lrev used} the extinction law of \cite{1989ApJ...345..245C} to model this, and {\lrev adopted} a total to selective extinction value typical of ISM dust ($R_{\rm{V}}$=3.1).

\subsubsection{Optical SED fitting}
\label{sec:analysis:optical_sed_fitting}
We {\lrev fitted} the optical photometry of each star with a (reddened) model atmosphere, keeping the effective temperature fixed at the spectroscopically determined value. Our SED fit thus has only 2 free parameters: the angular diameter $\theta$ and the extinction in the V~band $A_{\rm{V}}$. In general we {\lrev used} the photometric $g', r', i', z'$, and J bands in our SED fit, but for stars without near-infrared excess emission we also {\lrev included} the H and K$_{s}$ bands. We {\lrev calculated} model fluxes by integrating the intensity of the (reddened) model atmospheres over the spectral response curve of the system for each filter. The synthetic photometry thus obtained {\lrev was} compared to the observations and the resulting $\chi^{2}$ {\lrev was} minimized in an automated iterative procedure, in which the optimum values for $\theta$ and $A_{\rm{V}}$ are found.

\subsubsection{HR diagrams}
\label{sec:analysis:HR_diagrams}

The bolometric luminosity of our stars $L_{\rm{bol}}$ is easily calculated from the effective temperatures and angular diameters, if the distance is known:

\begin{equation}
\label{eq:Lbol}
$$
\ \ \ \ L_{\rm{bol}} = \pi \theta^2 d^2 \sigma T_{\rm{eff}}^4
$$
\end{equation}

\noindent where $\theta$ is the angular diameter, $d$ is the distance, $\sigma$ is the Stefan-Boltzmann constant, and \Teff \ is the effective temperature. We adopt a distance of 450\,pc for both L1630N and L1641 \citep{1981ApJ...244..884G,1982AJ.....87.1213A,1986ApJ...303..375M,2005A&A...430..523W}.  Note that we have implicitly corrected for extinction by calculating the bolometric luminosities in this way.

The thus determined effective temperatures and bolometric luminosities allow placement of our stars in the HR diagram. Stellar masses and ages can then be estimated by comparison to theoretical pre-main sequence (PMS) evolutionary tracks. However, several sets of such tracks exist, by various authors, which yield significantly different results, in particular for the age \cite[see e.g.][for a discussion]{2008ASPC..384..200H}. We estimated masses and ages using four different sets of publicly available PMS evolutionary tracks: those by \cite{1997MmSAI..68..807D} (DM97), \cite{1998A&A...337..403B} (B98), \cite{2000A&A...358..593S} (S00), and \cite{2008ApJS..178...89D} (D08). In the remainder of the discussion, we will adopt the values obtained by employing the PMS evolution tracks from \citet{2008ApJS..178...89D}, as these have the best resolution in both mass and age. We stress, however, that there are substantial systematic differences between the different sets of tracks \citep[e.g.][]{2008ASPC..384..200H}, and our motives for choosing those by \citet{2008ApJS..178...89D} are merely pragmatic.

The uncertainties in mass and age of the individual stars {\lrev were} estimated using a simple Monte-Carlo method, in which we created a large number of synthetic [\Teff,$L_*$] points for each star, assuming the errors in both quantities to be normally distributed. These points {lrev were} then interpolated in the HR diagram, and the standard deviations in the resulting mass and age distribution {\lrev were} adopted as the uncertainties in these quantities. This procedure accounts well for the observational errors, but we stress that systematic uncertainties remain. As noted above, the choice of theoretical PMS tracks can influence the resulting mass and age strongly, by factors upto $\sim$2--3. Also, for highly extincted sources, the shape of the adopted extinction law matters, with higher values for the total to selective extinction $R$ leading to higher stellar luminosities and correspondingly younger ages and, for the earlier spectral types, higher masses.

\subsection{Determining the disk properties}
\label{sec:analysis:determining_disk_properties}
The disk properties that we can determine from our data are twofold: the spatial structure of the disk as inferred by the infrared SED, and the accretion rate as deduced from optical emission lines. Optical emission lines also trace outflow activity, a process intimately related to accretion.

\subsubsection{Disk structure from the infrared SED}
\label{sec:analysis:disk_structure_IR_SED}

Excess emission above the stellar photospheric level is the most employed indicator for dusty circumstellar material. The infrared emission from the disks around T\,Tauri stars is essentially fully comprised of reprocessed stellar radiation, with generally minor contributions from the release of gravitational energy close to the central star. Therefore, they are usually referred to as "passive" disks\footnote{Even in objects where accretion is the dominant source of luminosity (e.g. FUOR and EXOR systems), irradiation by the hot, innermost disk regions is the dominant energy source in most of the disk, except close to the central star. Such disks thus have an ``active'' character within a certain radius and are ``passive'' further out.}. The infrared SED is directly related to the spatial structure of the circumstellar material, even though models fitted solely to SED data may suffer from degeneracies in the derived material distribution. In general, models of circumstellar disks adequately reproduce the SEDs of young stars, and observations that spatially resolve these objects confirm that interpretation \citep[e.g.][]{2003ApJ...588..360E,2004A&A...423..537L,2005ApJ...624..832M,2005ApJ...635.1173A,2005ApJ...622..440A,2006A&A...451..951I,2006ApJ...647..444M,2007ApJ...657..347E,2008A&A...485..209A,2008ApJ...676..490K}, though there is continued discussion on the relative imporance of an \emph{additional} ``spherical halo'' component \citep[e.g.][]{2003MNRAS.346.1151V,2006ApJ...636..348V}.

The near-infrared excess ($\sim$1.5\mum\,$\lesssim$\,$\lambda$\,$\lesssim$\,5\mum) traces hot dust in the inner disk ($\lesssim$\,1\,AU), with a possible contribution from small amounts of material in an optically thin envelope. In the longer part of the observed wavelength range ($\sim$6\mum\,$\lesssim$\,$\lambda$\,$\lesssim$\,24\mum) the slope of the SED is related to the shape of the disk on scales\footnote{This is approximately the spatial range probed in this wavelength regime for YSOs of $\sim$\,1\Lsun, characteristic for our sample. The range scales with $\sqrt{L_*}$, and is correspondingly larger for e.g. HAe stars and smaller for young brown dwarfs.} of $\sim$\,1$-$10\,AU: ``red'' slopes indicate disks with ``flared'' geometries, whereas ``blue'' slopes in this wavelength range point at ``flat'' disks \citep[see e.g.][]{2001A&A...365..476M,2004A&A...417..159D,2004A&A...423..537L}. It has been suggested that disks with flared geometries evolve into disks with flat geometries, and that dust settling and grain growth have proceeded in the latter \cite[e.g.][]{2004A&A...422..621A}. We use \alfasixtwentyfour, the slope between the Spitzer IRAC 5.8\mum \ and MIPS 24\mum \ bands as a measure for the disk flaring.

A distinct population of young stars shows essentially photospheric emission or a small and very blue excess at short wavelengths ($\lambda \lesssim 6$\mum), while simultaneously exhibiting moderate to strong excess emission at longer wavelengths. In these objects, the inner disks appears to have largely or completely dissipated, whereas the disk at larger radii is still relatively intact. They are thought to be in a transition stage between optically thick disks and debris disks. Recently, \cite{Muzerolle_Spitzer} discussed these so-called ``Transition Disks'', and distinguished three types: (1) objects with weak or zero IRAC excess and strong MIPS-24 excess (``canonical transition disks''); (2) objects with moderate IRAC excess and strong MIPS-24 excess (``pre-transition disks''); and (3) objects with weak or zero IRAC excess and weak MIPS-24 excess (“weak” or “evolved” disks). We have visually inspected the SEDs of all stars in our sample and selected all objects that match one of the aforementioned descriptions. These objects all occupy the lower right part of a 2MASS/Spitzer Ks-[5.8] vs. [8.0]-[24] color-color diagram (see \sect\ref{sec:results:disk_geometry}).

\subsubsection{Accretion rates}

\label{sec:analysis:accretion_rates}

Several diagnostics can be used to estimate \Mdotacc, the rate at which disk material is accreted onto the central star. UV excess emission above the photospheric level is generally attributed to hot spots on the star, shock-heated by accreting material hitting the stellar surface at high velocities. If the amount of excess emission can be well determined, it provides a robust estimate of the accretion rate \citep[e.g.][]{1998ApJ...492..323G}. Similarly, veiling of optical stellar spectra in which an additional, featureless continuum from accreting material reduces the contrast of photospheric absorption features can be used \citep[e.g.][]{1991ApJ...382..617H}. Optical and near-infrared emission lines, in particular of hydrogen, are widely used as accretion measures. Their fluxes are relatively easy to measure and they are sensitive to accretion even at very low levels. However, geometrical and optical depth effects can strongly influence the appearance of the emission lines, and while the average line strength correlates strongly with the mass accretion rate, individual objects can scatter up to two orders of magnitude around the average relation \citep[e.g.][]{2006A&A...452..245N}. More detailed studies of emission line profiles, in particular measuring the full width of the line close to the base, provide more robust \Mdotacc \ estimates for individual objects, but are feasible for very broad profiles only in medium resolution spectra like ours.

While many of our target stars are detected in SDSS $u'$ band, these measurements cannot be used to estimate the mass accretion rate, for the following reason. In objects with very red optical SEDs, the observed U-band flux can be severely contaminated by photons that ``leak in'' from longer wavelengths, since here the throughput of the $u'$ filter is close to, but not perfectly, zero. The typical YSO in our sample has K or M spectral type and an optical extinction of several magnitudes, leading to very red optical SEDs and a strong "red leak". We note that the U-band filters in other photometric systems suffer from the same effect. We do \emph{not} employ the $u'$-band photometry to determine accretion rates, neither do we use it in the fits to the optical SEDs (\sect\ref{sec:analysis:optical_sed_fitting}).

Rather, we {\lrev used} the H$\alpha$, H$\beta$, and \HeI\,$\lambda=5876$ emission lines, whose luminosity ($L_{\rm{H}\alpha}$, $L_{\rm{H}\beta}$, and $L_{\rm HeI}$) correlates with the accretion luminosity ($L_{{\rm acc}}$) for YSOs with masses ranging from vastly subsolar to several solar masses \citep{2008AJ....136..521D,2008ApJ...681..594H}. In Appendix~\ref{relation-acc-line} we derive the proportionalities between the line strengths and accretion luminosities. We {\lrev applied} these relations to derive the accretion luminosities for our target stars in L1630N and L1641. These {\lrev were} converted into mass accretion rates via the following relation:
\begin{equation}
\dot{M}_{\rm acc}=\frac{L_{{\rm acc}}R_{\star}}{GM_{\star} (1-\frac{R_{\star}}{R_{\rm in}})},
\end{equation}
\noindent where $R_{\rm in}$ denotes the truncation radius of the disk, and is taken to be 5\,R$_{\star}$ \citep{1998ApJ...492..323G}. The stellar radii ($R_{\star}$) {\lrev were} obtained using the SED fitting procedure described in \sect\ref{sec:analysis:optical_sed_fitting}, and the stellar masses {\lrev were} derived from the location of our YSOs in the HR diagram.

\subsection{Spatial distribution of YSOs}
\label{sec:analysis:spatial_distribution}

The young stars in L1630N are mostly confined to two clusters  \citep[NGC\,2068 and NGC\,2071, see e.g.][]{1991ApJ...371..171L}. In L1641, there are a large number of clusters or aggregates, as well as a more distributed population \citep{1993ApJ...412..233S}. Following \cite{2008ApJ...672..861R}, we {\lrev applied} the nearest neighbor method (NNM) towards the YSO population in L1641 to determine which YSOs are in isolation and which are in aggregates or clusters, and to outline the extension of the clusters. For each star we {\lrev located} the 10$^{\rm{th}}$ nearest neighbor and {\lrev calculated} the surface density of YSOs within the corresponding radius. If this density is higher than the average surface density over the whole cloud, the star is taken to be a member of a cluster/aggregate.

\section{Results}
\label{sec:results}

From our multi-wavelength data of the L1630N and L1641 clouds, we {\lrev derived} the stellar properties of the observed objects as well as properties of their disks. To some extent, we also {\lrev gained} information about the molecular clouds in the form of accurate ``pencil-beam'' extinction measurements towards our targets. In L1641, we frequently discriminate between the ``clustered'' (L1641C) and ``distributed'' (L1641D) populations.

\subsection{Survey products}
\label{sec:results:survey_products}

The main product of our survey is a list of identified YSOs. For each object we give the spectral type, bolometric luminosity, H$\alpha$ and \LiI \ equivalent widths, CTTS/WTTS classification, line of sight extinction, the precence in a clustered or distributed environment, IRAC infrared spectral indice,  mass and age estimates using the PMS evolutionary tracks of \cite{2008ApJS..178...89D}, an SED-based classification of the disk, and estimates of the mass accretion rate using H$\alpha$, H$\beta$ and \HeI \ in \tabs\ref{tab:YSO_parameters_L1630N} and \ref{tab:YSO_parameters_L1641}, for stars in L1630N and L1641, respectively. We list the photometric magnitudes in a number of optical and infrared filters in \tabs\ref{tab:YSO_magnitudes_L1630N} and \ref{tab:YSO_magnitudes_L1641}. Mass and age estimates using the PMS evolutionary tracks of \cite{1997MmSAI..68..807D}, \cite{1998A&A...337..403B}, \cite{2000A&A...358..593S}, and \cite{2008ApJS..178...89D}  are listed in \tabs\ref{tab:L1630_mass_age} and \ref{tab:L1641_mass_age}. The equivalent widths of a number of optical emission lines are listed in \tabs\ref{tab:emission_lines_L1630N} and \ref{tab:emission_lines_L1641}.

We further supply an ensemble of stars that were identified as foreground or background objects by the presence of an H$\alpha$ absorption line and the absence of Li\,I\,$\lambda$6707 absorption, and list their photometric magnitudes, spectral types and optical extinction estimates in \tabs\ref{tab:background_stars_L1630N} and \ref{tab:background_stars_L1641} for the fields of L1630N and L1641, respectively.

Additionally, we provide optical and infrared magnitudes of a total of 21694 sources detected in our optical imaging data with matching sources in the 2MASS catalog. These consist of public SDSS data for L1630N \citep{2004AJ....128.2577F}, previously unpublished SDSS data and new LAICA data for L1641, publicly available NIR photometry from 2MASS \citep{2006AJ....131.1163S}, as well as publicly accessible and previously published \citep{2005IAUS..227..383M} Spitzer data. Without measured spectral types, reliable estimation of the luminosity and effective temperature is hampered by a degeneracy between the stellar temperature and reddening. Therefore, these object are not included in our analysis, and we instead provide these data as a photometric catalog only. This list contains many young stars as well as numerous background objects. We list the optical and infrared magnitudes of these sources in \tabs\ref{tab:all_photometry_L1630N} and \ref{tab:all_photometry_L1641}.

\subsection{Stellar properties}
\label{sec:results:stellar_properties}
We determined the spectral type for each star in our sample as described \sect\ref{sec:analysis:spectral_classification}, and estimated masses and ages using the methods described in \sect\ref{sec:analysis:determining_stellar_properties}. We could obtain reliable spectral types for 71\%  of the observed sample in L1630N and 78\% in L1641, the remainder of the sample usually having too low SNR for a robust analysis.

\subsubsection{Spectral types}
\label{sec:results:spectral types}

\begin{figure}
\centering
\includegraphics[width=\columnwidth]{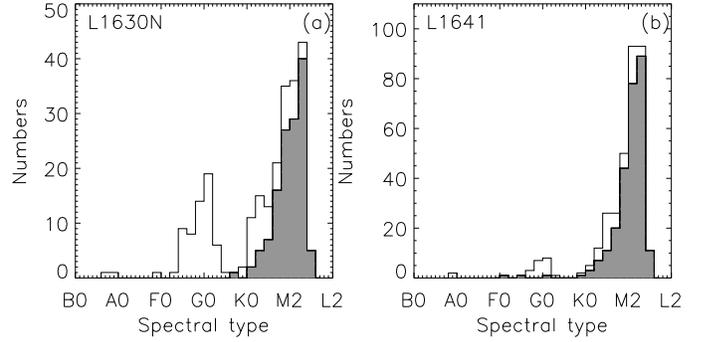}
\caption{\label{fig:spt_dis} The spectral-type distributions of our sample observed with VIMOS in L1630N and L1641. The open histograms show the distribution of all the stars with reliable spectral types. The filled histograms display the YSO distribution}.

\end{figure}

In \fig\ref{fig:spt_dis} we show a histogram of spectral types within our sample. The distribution is bi-modal, with peaks around spectral type G0 and mid~M. In this figure, we distinguish between YSO candidates, selected according to the criteria outlined in \sect\ref{sec:analysis:YSO_selection_cirteria} and shaded gray in the histogram, and probable "field" objects. The young stars in our sample are all of K or M type, with the exception of one F and one G star in L1641. The field stars comprise essentially the entire peak around spectral type G0 and placement of these objects in the HR diagram shows that 11\% (7/61) in L1630 and 25\% (5/20) are foreground dwarfs, the majority of the rest are main sequence stars behind the clouds. The second peak in the field objects distribution, around K and M spectral types, consists mostly of foreground dwarfs and background giants.

\subsubsection{Stellar masses and ages}
\label{sec:results:masses_ages}

\begin{figure*}
\centering
\includegraphics[width=18.5cm]{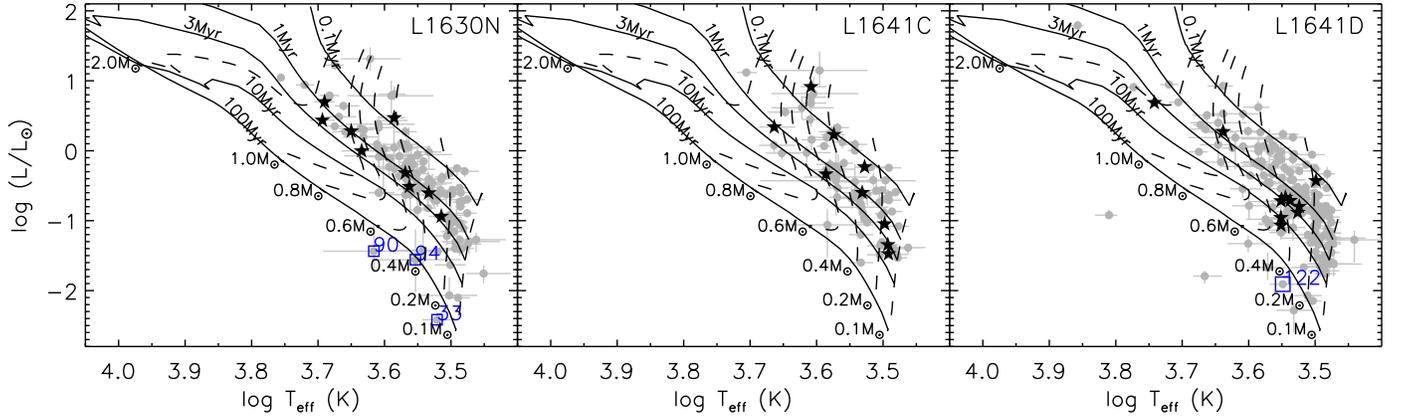}
\caption{\label{fig:HR_diagrams} HR diagrams of YSOs in L1630N (left panel), L1641C (middle panel) and L1641D (right panel). {\newrev Gray circles: YSOs. The error bars for the effective temperatures reflect the uncertainties of the deduced spectral types provided by the classification code. The errors bars for the lumonisity come from the optical SED fitting and include the uncertainties in the photometry as well as the effective temperatures.} {\newrev Asterisks}: YSOs  with transition disks. Open squares: the exotic objects, see \sect\ref{sec:discussion:high_Halpha_EW}.}
\end{figure*}

In \fig\ref{fig:HR_diagrams} we show the HR~diagrams of the L1630N and L1641 clouds, from which we derive mass and age estimates as outlined in \sect\ref{sec:analysis:HR_diagrams}. We show separate diagrams for the ``distributed'' and ``clustered'' population in L1641, as defined using the NNM (\sect\ref{sec:analysis:spatial_distribution}). The vast majority of objects lie between the 0.1 and 3\,Myr isochrones. We use distinct symbols ({\newrev filled asterisks}) for objects with SEDs typical for transition disks (see also \sect\ref{sec:results:disk_geometry}), as well as four ``exotic'' objects that are apparently under-luminous and have very high emission line equivalent widths (blue-boxed points, see Sect.\ref{sec:discussion:high_Halpha_EW}).

\begin{figure}
\centering
\includegraphics[width=\columnwidth]{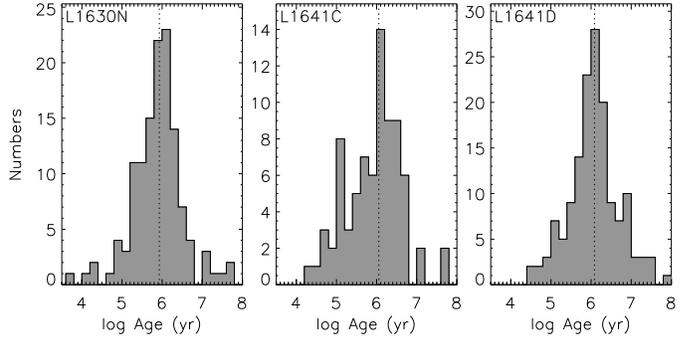}
\caption{\label{fig:age_dis} The age distribution of  YSOs in L1630N, L1641C, and L1641D. The dashed lines show the median masses of YSOs, i.e. $\sim$0.9\,Myr in L1630N, 1.1\,Myr in L1641C ,and 1.3\,Myr  in L1641D.}
\end{figure}

In \fig\ref{fig:age_dis}, we show the age distribution of L1630N, L1641C, and L1641D.   The median ages  are 0.9\,Myr in L1630N, 1.1\,Myr in L1641C, and 1.3\,Myr in L1641D, if a distance of 450\,pc is assumed for both clouds. In this case, a Kolmogorov-Smirnov (KS) test yields a very low probability ($P$\,$\sim$0.03) that the YSOs in {\rev L1641 and L1630N} are drawn randomly from the same age distribution. However, if one assumes a distance of 400\,pc for L1630N, which is within current uncertainties \citep[][]{2008AJ....135..966F,1982AJ.....87.1213A}, the discrepancy between the age distributions of both clouds disappears completely. 

\begin{figure}
\centering
\includegraphics[width=\columnwidth]{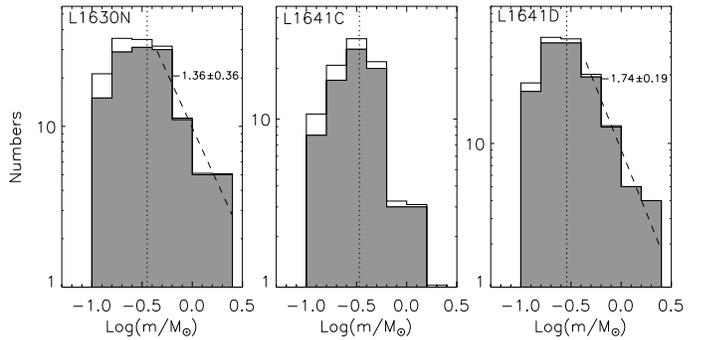}
\caption{\label{fig:mass_dis} The mass distribution of YSOs in L1630N and L1641. The dotted lines show the median masses of YSOs, i.e. $\sim$0.34\,\Msun \ in L1630N, 0.33\Msun \ in L1641C, and 0.28\,\Msun \ in L1641D. {\newrev The shaded histograms are the mass distributions without incompleteness correction, while the open histograms are the mass distribution corrected for incompleteness (see Sect.\ref{sec:results:masses_ages}).} A linear regression toward the {\newrev latter} mass spectra in the range of -0.4\,$\leq$\,log($M_{\star}$/M$_{\odot}$)\,$\leq$\,0.4 gives slopes of -1.36$\pm$0.36 and -1.74$\pm$0.19 for the population in L1630N and L1641D (dashed lines), respectively.}
\end{figure}

In \fig\ref{fig:mass_dis}, we show the mass distributions (filled histograms) derived for our sample. The median masses are $\sim$0.34~M$_{\odot}$ in L1630N, 0.33\,M$_{\odot}$ in L1641C, and  0.28~M$_{\odot}$ in L1641D, respectively.  A KS test toward the mass distributions in L1630N and L1641D reveals a relatively low  probability ($P$\,$\sim$0.08) for stars in both clouds to be drawn randomly from the same  distribution. {\rev In \fig\ref{fig:mass_dis} we see that there are fewer objects in the lowest mass bins in L1641C compared to L1641D and L1630N. This may, however, be caused by higher extinction in the former population,} causing the lowest mass members to be excluded from our sample as they are too faint for optical spectroscopy. The mass distributions for stars above 0.5\,M$_{\odot}$ in L1630N and L1641D are indistinguishable, with a probability of $P$\,$\sim$0.9  for both samples being drawn at random from the same parent distribution, in a KS test.

We {\lrev used} a Monte-Carlo method to estimate the incompleteness of our sample. According to an assumed mass function, we {\lrev constructed} 1000 synthetic stellar populations, each containing 2$\times$10$^{5}$ stars.  Assuming ages of 0.9\,Myr for L1630N, 1.1\,Myr for L1641C and 1.3\,Myr for L1641D, we {\lrev obtained} synthetic photometry for each star using the PMS evolutionary tracks from \cite{2008ApJS..178...89D}. We {\lrev reddened} the photometry using $A_{\rm v}$ values randomly sampled from the observed extinction distributions in both regions, respectively.  We {\lrev used} $r'$=21.5\,mag as a flux limit, and {\lrev assumed} that stars brighter than this limit have spectra of sufficient signal-to-noise for reliable spectral classification. Examination of the VIMOS spectra of our sample shows that this limit is realistic. The incompleteness correction factors for each mass bin can be obtained by comparing the input mass function and the ``observed'' mass function, including all stars brighter than 21.5\,mag at $r'$. The open histograms in \fig\ref{fig:mass_dis} show the mass distribution corrected for incompleteness. A linear regression toward this mass function within the mass range -0.4\,$\leq$\,log($M_{\star}$/M$_{\odot}$)\,$\leq$\,0.4 yields slopes of -1.36$\pm$0.36 and -1.74$\pm$0.19 for the population in L1630N and L1641D, respectively.

\subsection{Disk properties}
\label{sec:results:disk_properties}
From our infrared SED data we {\lrev derived} approximate disk geometries for the observed objects whereas optical emission lines yield accretion and outflow signatures. See \sect\ref{sec:analysis:determining_disk_properties} for a description of the diagnostics used.

\subsubsection{Disk geometry}
\label{sec:results:disk_geometry}

\begin{figure}
\centering
\includegraphics[width=\columnwidth]{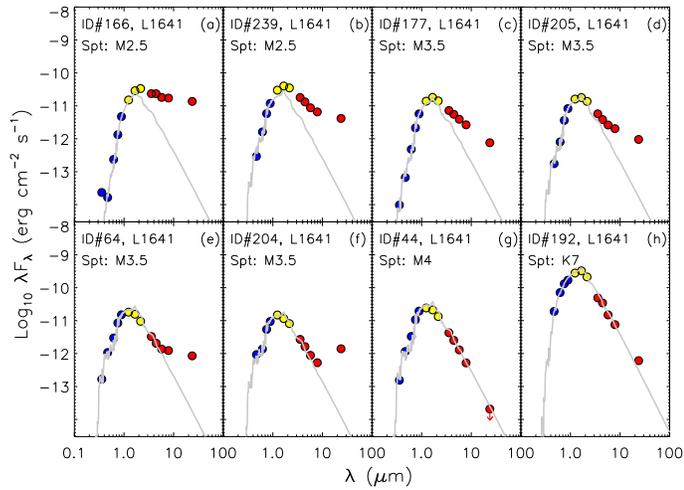}
\caption{\label{fig:SED-sample} Example SEDs of YSOs whose disks are in different evolutionary stages. The photospheric emission level is indiacted with a grey curve in each panel. From (a) to (e-f), the disks evolve from a {\newrev ``strongly flared'' optially thick disk to a ``transition disk''}. (g): a star showing no excess emission in any of the Spitzer bands. (h): a debris-disk candidate with a weak excess at 24\,\mum \ only.}
\end{figure}

\noindent
\textbf{(a) Qualitative description of SEDs \& physical interpretation} \\
\vspace{-0.3cm}

\noindent
In \fig\ref{fig:SED-sample} we show a set of representative SEDs of YSOs displaying a range of spectral shapes that are thought to indicate different stages of disk evolution. In \tab\ref{tab:excess_flux} we list the infrared excess from 3.6\mum \ to 24\mum \ of these same stars in magnitudes.

The object displayed in \fig\ref{fig:SED-sample}(a) shows a strong IR excess with an approximately flat spectrum (in $\lambda F_{\lambda}$). The SED of the system in \fig\ref{fig:SED-sample}(b) shows excesses that are reduced by a similar factor across the whole wavelength range compared to the object in \fig\ref{fig:SED-sample}(a). The excess fluxes of the object shown \fig\ref{fig:SED-sample}(c) are substantially smaller still, in particular at 24\mum. The SEDs shown in \figs\ref{fig:SED-sample}(a) to (c) can be understood in terms of dust growth and settling, which decrease the vertical scale height of the disk and thus the amount of stellar radiation that is absorbed by the disk and re-emitted in the infrared. The corresponding disk geometries could be referred to as "strongly flaring" (a), "mildly flaring" (b), and "flat" (c). Note that the designation "flat" does not imply a geometrically thin disk, but rather indicates that the ratio of the disk vertical scale height and the distance ($H/R$) doesn't increase significantly with radius. For flared disks, this ratio does increase with radius, the classical solution for an optically thick disk being $H/R \propto R^{2/7}$ \citep{1987ApJ...323..714K}.

From \fig\ref{fig:SED-sample}(c) to \fig\ref{fig:SED-sample}(e) the excess emission from 3.6\mum \ to 8.0\mum \ decreases steadily while the flux at 24\mum \ remains roughly constant. This indicates that the very inner parts of the disk are dissipated while the outer disk remains essentially intact, i.e. an inner hole may form. Going from \fig\ref{fig:SED-sample}(e) to \fig\ref{fig:SED-sample}(f) we see that the excess emission at 8.0\mum \ is significantly reduced, while the excess at 24\mum \ has more than doubled. This may be understood in terms of an inner disk that has essentially dissipated, i.e. an inner gap of order 1~AU in radius, combined with a dusty outer disk whose inner edge is no longer partially shielded by the inner disk but rather is fully exposed to the stellar radiation field, and has a temperature of $\lesssim$200\,K making it very bright at 24\mum \ but very faint at $\lesssim$8\mum.

The star shown in \fig\ref{fig:SED-sample}(g) displays no excess emission at any of the Spitzer bands, indicating it has essentially no circumstellar material within $\sim$20\,AU. In \fig\ref{fig:SED-sample}(h) we show a star that has essentially photospheric emission levels at $\lambda \leq$8\mum \ but shows a weak excess at 24\mum. {\rev Such an SED is typical of ``debris disks'', which possess relatively small amounts of circumstellar material originating in the collisional grind-down of larger bodies (``planetesimals''). While the source shown in \fig\ref{fig:SED-sample}(g) does show a debris disk-like SED, it is unclear whether it is of the same physical nature, given its young age of $\lesssim$4\,Myrs.}

\begin{table}
\caption{\label{tab:excess_flux} Excess emission fluxes for YSOs in \fig\ref{fig:SED-sample}}
 \centering
\begin{tabular}{ccccccc}
\hline\hline
sub-  & ID&  [3.6]  & [4.5] & [5.8] &[8.0] &[24]\\
panel & (L1641) & (mag)   & (mag) & (mag) & (mag) & (mag)\\
 \hline
(a) &     166 &      1.39  &     2.07  &     2.46 &     3.34  &     6.45\\
(b) &     239 &      1.49  &     1.89  &     2.15 &     2.81  &     5.71\\
(c) &     177 &      0.84  &     1.24  &     1.57 &     2.12  &     4.12\\
(d) &     205 &      0.56  &     0.82  &     1.13 &     1.83  &     4.38\\
(e) &      64 &    \nodata &     0.05  &     0.34 &     1.22  &     4.20\\
(f) &     204 &      0.06  &     0.23  &     0.29 &     0.72  &     5.17\\
(g) &      44 &     \nodata&    \nodata&   \nodata&    \nodata&    \nodata\\
(h) &     192 &     \nodata&    \nodata&   \nodata&    \nodata&     0.75\\
 \hline
\end{tabular}
\end{table}

\begin{figure}
\centering
\includegraphics[width=\columnwidth]{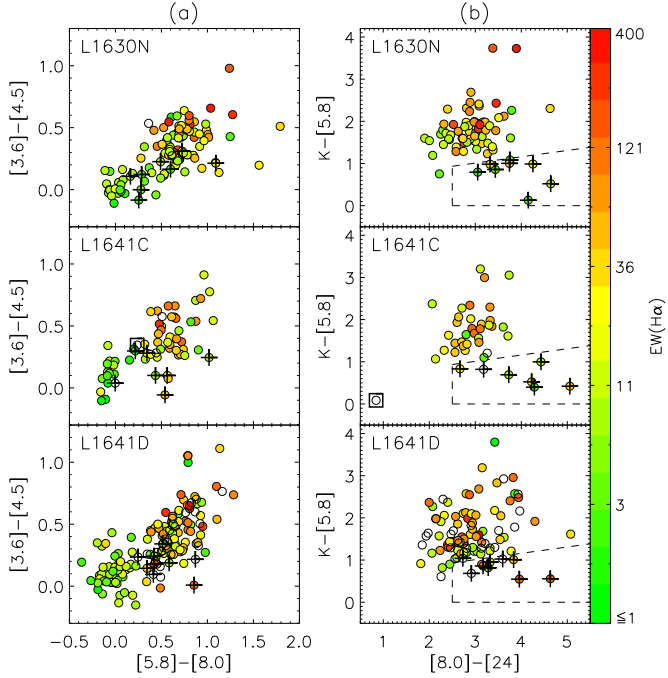}
\caption{\label{fig:ccmap_IRAC} Spitzer [3.6]-[4.5] vs. [5.8]-[8.0] (left) and [8.0]-[24] vs. K$_{s}$-[5.8] (right) color-color diagrams for our YSO sample in L1630N, L1641C, and L1641. The pluses mark YSOs with transition disks. The debris disk candidate is marked with a square box. All points are color-coded according to the H$\alpha$  equivalent width (blank circles represent sources where our VIMOS spectra did not cover the H$\alpha$ spectral region)}.
\end{figure}

\vspace{0.3cm}

\noindent
\noindent \textbf{(b) Disk frequency} \\
\vspace{-0.3cm}

\noindent In \fig\ref{fig:ccmap_IRAC}(a) we show the ``classical Spitzer IRAC color-color'' diagram of all YSOs in our sample, i.e. the [3.6]-[4.5] vs. [5.8]-[8.0] colors. A number of stars cluster around the origin, indicating approximately photospheric near-infrared colors and the absence of an optically thick inner disk. Most stars are located towards the top-right of the origin due to infrared excess emission indicative of hot dust in the inner disk. Note that in \fig\ref{fig:ccmap_IRAC}, the H$\alpha$ equivalent width of each star is indicated using a color coding, this will be further discussed in \sect\ref{sec:results:accretion_signatures}.

\newcommand{\alphairac}{$\alpha_{\rm{\scriptscriptstyle{IRAC}}}$}

\begin{figure}
\centering
\includegraphics[width=\columnwidth]{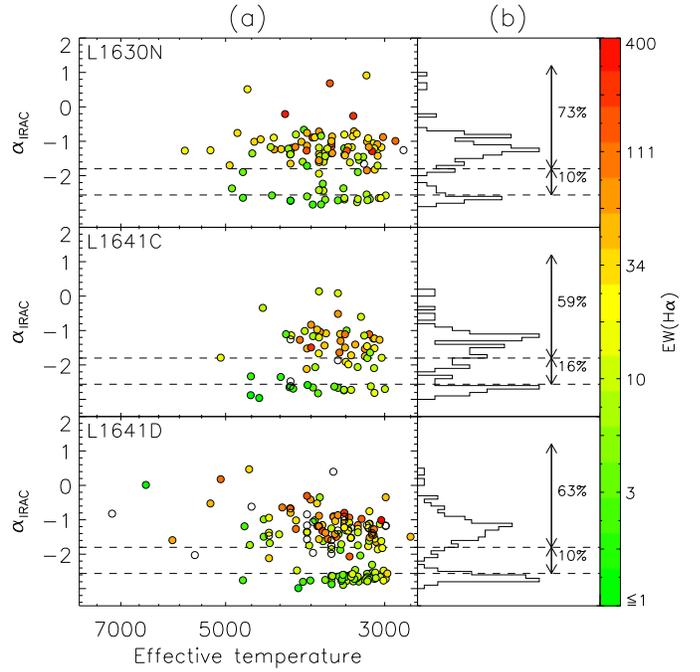}
\caption{\label{fig:alpha_irac} (a): The dereddened power-law slope of the IRAC SEDs (\alphairac)  for YSOs in L1630N, L1641C, and L1641D, plotted as a   function of  effective temperature. (b): Distributions of \alphairac for  YSOs in L1630N, L1641C, and L1641D.}
\end{figure}

The average slope \alphairac \ of the SED in the IRAC spectral range is often used as a proxy for the nature of the inner disk. \fig\ref{fig:alpha_irac} shows this slope as a function of stellar effective tempareture, again with symbols color-coded according to H$\alpha$ equivalent width. Diskless M-type stars show a spectral slope of \alphairac\,$\lesssim$\,$-$2.56, while objects with an optically thick inner disk have \alphairac\,$\geq$\,-1.8. An intermediate spectral slope of $-$2.56\,$\leq$\,\alphairac\,$<$\,$-$1.8 points at a partially dissipated ``evolved'' inner disks. Using these criteria, we find the following disk statistics in our sample of YSOs \emph{with detections in all four IRAC bands}. In L1630N, the YSOs with optically thick disks and evolved disks account for 73$\pm$9\% and 10$\pm$3\% of all sources, respectively. In L1641C, 59$\pm$11\% of sources have an optically thick disk and 16$\pm$6\% have an evolved disk. In L1641D, these fractions are 63$\pm$7\% and
  10$\pm$3\%, respectively.

Some YSOs, preferentially those without strong excess emission, did not have solid detections in all four IRAC bands. For these sources, we visually inspected the SEDs for signs of infrared excess emission, by comparing the observed IRAC fluxes with the reddened model atmospheres fitted to the optical and near-infrared data. Including these stars leads to somewhat lower total disk frequencies (optically thick + evolved) compared to the numbers quoted above: 70\% in L1630N, 67\% in L1641C (aggregate/clutser population in L1641), and 59\% in L1641D (distributed population in L1641). We remind the reader that these values likely still slightly overestimate the true disk frequencies in an abolute sense, due to the bias introduced by our target selection (see \sect\ref{sec:observations:biases}).

\vspace{0.3cm}

\noindent
\noindent \textbf{(c) Disk flaring \& dust settling} \\
\vspace{-0.3cm}

\begin{figure}
\centering
\includegraphics[width=\columnwidth]{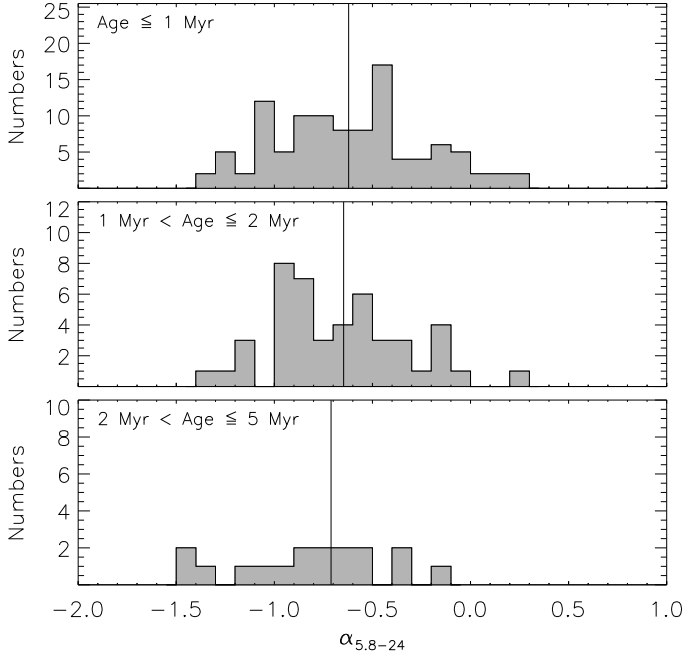}
\caption{\label{fig:geometry} The spectral slope \alfasixtwentyfour \ between the IRAC 5.8 and MIPS 24 bands in different age bins. This slope is a measure for the disk geometry on scales of a few AU, see \sect\ref{sec:results:disk_geometry}(c).}
\end{figure}

\noindent 
In \fig\ref{fig:geometry} we show histograms of the spectral slope between the IRAC 5.8 and MIPS 24 bands. As discussed in \sect\ref{sec:analysis:disk_structure_IR_SED}, this slope is a measure of the disk geometry on scales of a few AU. We have split our sample into three age bins, and have excluded sources with an SED indicative of a transition disk. Since the special geometry of transition systems causes the 5.8\mum \ radiation to mostly not arise in the disk whereas the stellar photons are converted into 24\mum \ radiation with above average efficiency, their \alfasixtwentyfour \ as a proxy for the disk geometry cannot be compared directly with that of normal disks.

From \fig\ref{fig:geometry} it becomes immediately apparent that there is a large range of observed spectral slopes \alfasixtwentyfour \ in each of the age bins. The median slope decreases slightly when going to larger ages, but this effect is much smaller than the scatter within the individual age bins. Thus, while our data do suggest the occurence of ``flat'' disks to increase with progressing age, they also imply that age is not the main factor determining whether a disk has a flared or a flat geometry. Assuming that all disks start out with a flared geometry, the rate at which dust settling turns them into flat disks varies strongly from source to source. The uncertainties in the ages of the individual objects are too small to explain the observed spread in the rates at which the disks evolve, and the spread must instead be intrinsic. A prime candidate mechanism for influencing this process is binarity \citep[e.g.][]{2006ApJ...653L..57B}, a property which we do not assess in this work.

\begin{figure}
\centering
\includegraphics[width=\columnwidth]{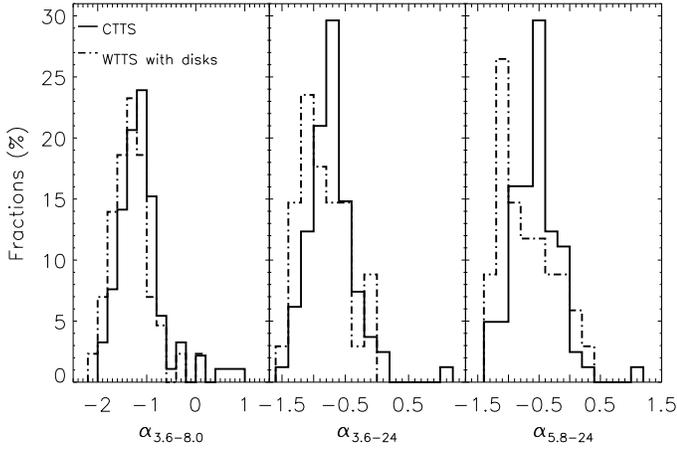}
\caption{\label{fig:geometry_multislopes} The distributions of the spectral slopes  between the IRAC 3.6 and 8.0 bands, the IRAC 3.6 and MIPS 24 bands, and the  IRAC 5.8 and MIPS 24 bands for CTTSs and WTTSs with disks.}
\end{figure}

In \fig\ref{fig:geometry_multislopes} we show distributions of the spectral slopes  between the IRAC 3.6 and 8.0 bands (\alfathreeeight), the IRAC 3.6  and MIPS 24 bands (\alfathreetwentyfour), and the  IRAC 5.8 and MIPS 24 bands(\alfasixtwentyfour), for CTTSs and WTTSs with disks. For the same reason as discussed above, we {\lrev have excluded} transition disk objects. Though a large dispersion in the spectral slopes is present for both populations, the CTTSs show on average ``redder'' slopes than the WTTSs with disks in all the spectral indices considered, especially for \alfathreetwentyfour, and \alfasixtwentyfour. A KS test toward the distributions of the specral slopes hints that \alfathreeeight \ is different for both populations, with a probability of $P$\,$\sim$0.14 for CTTSs and WTTSs with disks to be drawn randomly from the same parent distribution. The difference in the other two indices \alfathreetwentyfour \ and \alfasixtwentyfour \ is more evident, with probabilities of $P$\,$\sim$\,0.02 and $P$\,$\sim$\,0.03 for both sets to be drawn randomly from the same population, respetively. Since the infrared spectral slopes trace the disk geometry, our observations provide evidence for a correlation between disk geometry evolution and accretion evolution \citep[see also e.g.][]{2006AJ....132.2135S}. This does, however, not necessarily imply a causal connection between the two.

\vspace{0.3cm}

\noindent \textbf{(d) Transition disks} \\
\vspace{-0.3cm}

\begin{figure*}
\centering
\includegraphics[width=18cm]{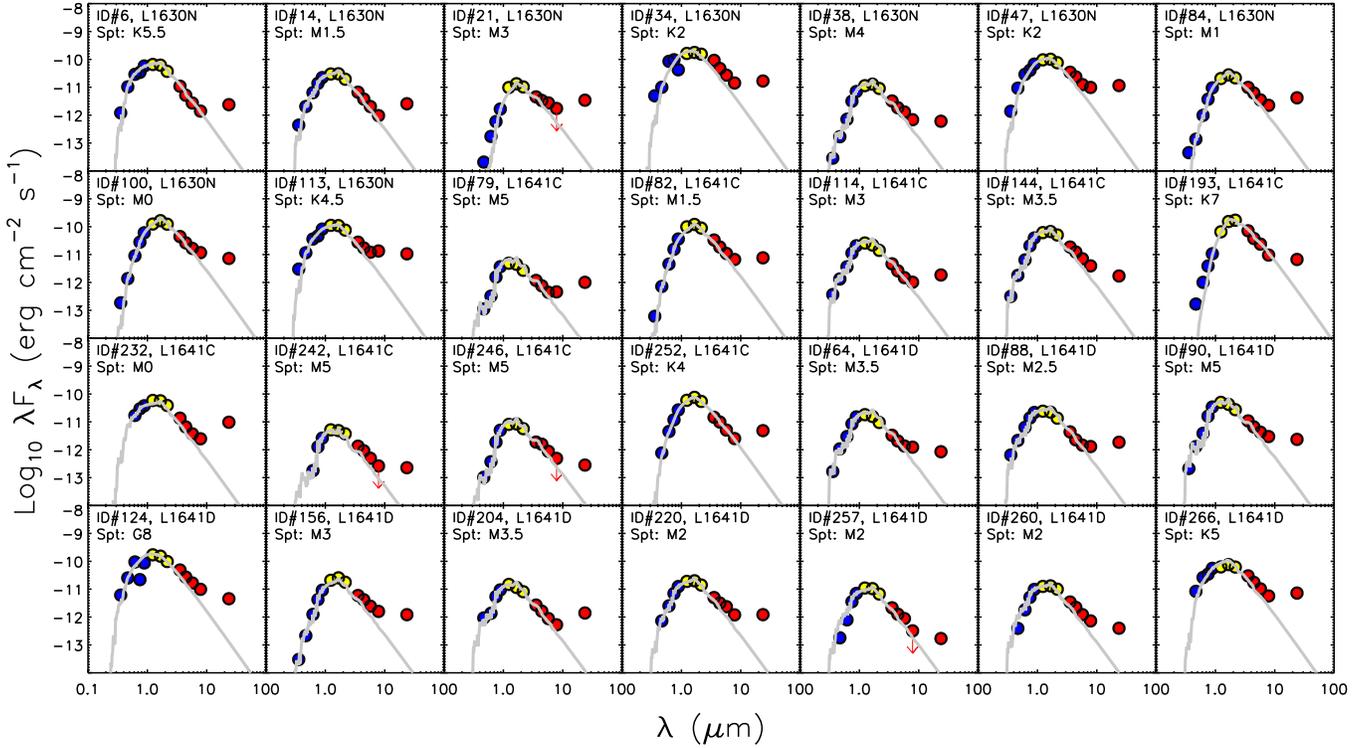}
\caption{\label{fig:TD_SED} The SEDs of YSOs with transition disks in L1630N and L1641. The photospheric emission level is indiacted with a grey curve in each panel.}
\end{figure*}

\noindent
The SEDs of all sample stars were visually inspected to select those that comply with the description of a transition disk given by \cite{Muzerolle_Spitzer}. We include all 3 flavors of transition disks discussed by \cite{Muzerolle_Spitzer}, see also \sect\ref{sec:analysis:disk_structure_IR_SED}. The SEDs are shown in \fig\ref{fig:TD_SED}, in which we also show the best fit (reddened) model atmospheres. In \fig\ref{fig:ccmap_IRAC}(b) we show a Ks-[5.8] vs. [8.0]-[24] color-color diagram incorporating 2MASS and Spitzer data. In this diagram, the transition disks have been marked with $+$ symbols. They occupy a well defined area in the lower right part of the diagram, outlined by the folowing borders:
\begin{eqnarray}
    [8.0]-[24]\geq2.5\\
    K_{s}-[5.8]\leq0.56+([8.0]-[24])\times0.15
\end{eqnarray}

Note that the transition objects are much more clearly separated from the rest of the population in this diagram than in the ``classical'' IRAC [3.6]-[4.5] vs. [5.8]-[8.0] color-color diagram shown in \fig\ref{fig:ccmap_IRAC}(a). In the latter diagram, the transition objects cover on average intermediate positions between ``diskless'' stars and normal disks, but show a signifiant overlap with both distributions. Stars with little or no excess emission at $\lambda$\,$\leq$\,8\mum \ but strong excess at 24\mum \ are naturally properly identified only in the Ks-[5.8] vs. [8.0]-[24] diagram {\rev (\fig\ref{fig:ccmap_IRAC}(b))}. Among the 256 stars we identify to possess a circumstellar disk there are 28 transition disk objects, i.e. the transition disks constitute $\sim$11\% of the disk population. Note that, Using the criteria described above, we indentify 47 additional transition disk candidates based on their photometry. Their photometric magnitudes at optical and infrared wavelengths are listed in \tab\ref{tab:TD_candidates}.

\subsubsection{Accretion signatures}
\label{sec:results:accretion_signatures}
Accretion characteristics are obtained from observations of optical emission lines, in particular the hydrogen Balmer series and the \HeI~5876\AA \ line.

\newcommand{\FWten}{$FW_{10\%}$}

\vspace{0.3cm}

\noindent
\noindent \textbf{(a) H$\alpha$ emission profiles} \\
\vspace{-0.3cm}

\noindent 
We observed H$\alpha$ emission in 86 young stars in L1630N, and 178 objects in L1641. The vast majority of these stars, 78\% in L1630N and 88\% in L1641, show symmetrical profiles with a single peak in our observations. These have a median full width at 10\% of the peak intensity of \FWten$\sim$270\kms. If we divide the sample into CTTSs and WTTSs according to their H$\alpha$ EWs (see Appendix~\ref{boundary_sec}), we find that the former have broader lines with a median \FWten$\sim$320\kms \ and the latter are narrower with a median \FWten$\sim$230\kms. The limited spectral resolution of our observations would yield a \FWten$\sim$180\kms \ for a line with zero intrinsic width. Therefore, the majority of the observed H$\alpha$ lines are spectrally resolved, even if in some cases only marginally so.

\begin{figure*}
\centering
\includegraphics[width=18cm]{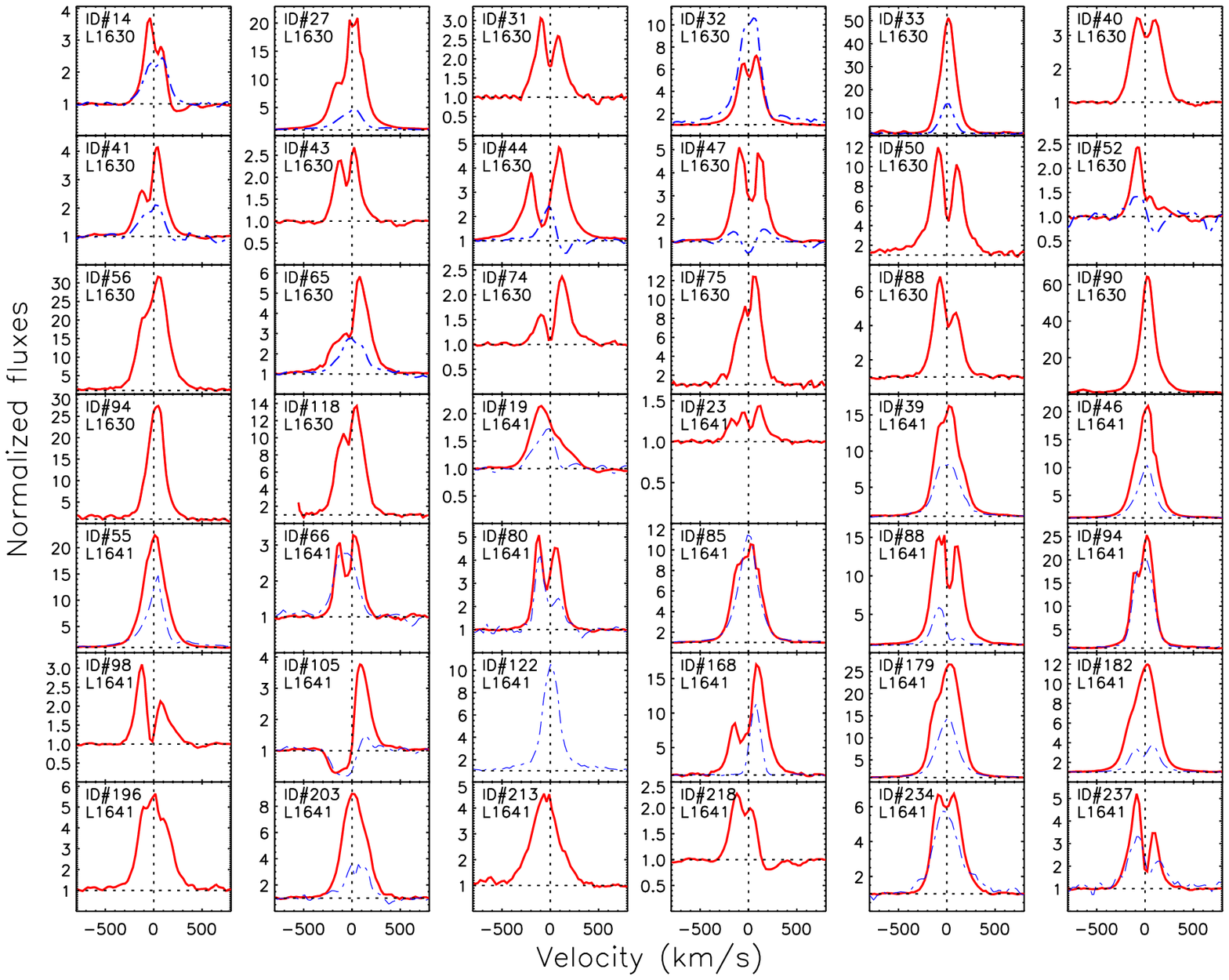}
\caption{\label{fig:profile} Normalized H$\alpha$ (solid lines) and H$\beta$ (dashed lines) emission lines of sources that show particularly broad and complex line profiles. Note that the majority of sources show single-peaked, much more symmetric profiles.}
\end{figure*}

About 22\% and 12\% of YSOs in L1630N and L1641, respectively, show broad asymmetrical H$\alpha$ lines that are easily resolved in our observations. We show normalized H$\alpha$ profiles for most of these in \fig\ref{fig:profile}. The median \FWten \ of these stars is approximately 500\kms, i.e. their H$\alpha$ lines are much broader than those of the CTTSs with symmetrical line profiles. We generally see clear substructure in the line, mostly in the form of double-peaked emission. For comparison, we have overplotted their H$\beta$ profiles, which are usually similar but on occasions strikingly different from the H$\alpha$ lines. They illustrate a complex accretion geometry in which the line emission is at least partially optically thick.

\vspace{0.3cm}

\noindent
\noindent \textbf{(b) H$\alpha$ equivalent widths} \\
\vspace{-0.3cm}

\noindent
The equivalent widths of the H$\alpha$ lines we {\lrev observed} in our sample range from less than 1\,\AA \ to over 300\,\AA. Equivalent widths of more than 3 to 20\,\AA \ (depending on spectral type) point at active accretion and our measurements reveal a range of accretion rates spanning several orders of magnitude (see Appendix~\ref{boundary_sec} for an exact description of the H$\alpha$ equivalent width threshold used to distinguish between CTTSs and WTTSs, as a function of spectral type). H$\alpha$ emission with an equivalent width below the CTTS/WTTS threshold cannot be unambiguously attributed to accretion activity, as it may also be caused by stellar coronal activity. {\rev While H$\alpha$ emission below the threshold may in some cases be caused by weak accretion \citep{2006AJ....132.2135S}, our purpose here is to define the EW limit above which we can be sure that the H$\alpha$ emission is dominantly accretion induced}.

\vspace{0.3cm}

\noindent
\noindent \textbf{(c) Accretion rates derived from H$\alpha$, H$\beta$, and \HeI} \\
\vspace{-0.3cm}

\begin{figure}
\centering
\includegraphics[width=\columnwidth]{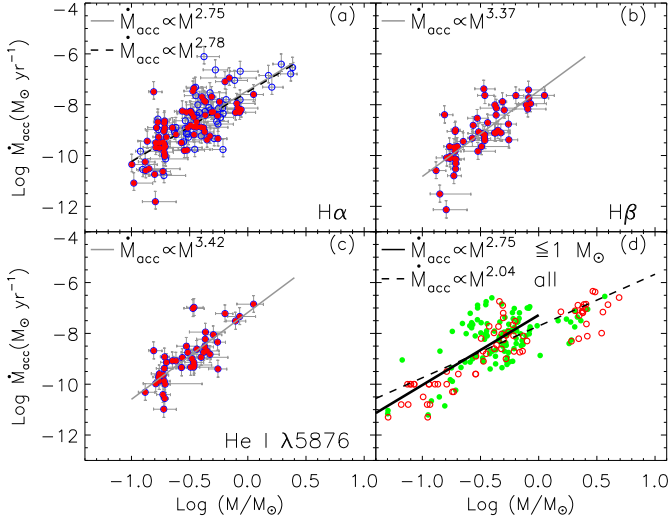}
\caption{\label{fig:acc_rate_mass_all} (a)(b)(c): The relation between accretion rates deduced from the H$\alpha$(a), H$\beta$ (b), or \HeI\,$\lambda=5876$ (c) emission line luminosity and stellar mass for the YSOs in our spectroscopic sample in L1630N and L1641. In panel (a), the filled circles represent the YSOs observed with VIMOS, and the open circles show the YSOs from literature.
The solid line in each figure represents the best fit power law to the observed distribution for all sources in the VIMOS sample. In panel (a), the dashed line shows the best fit if also literature values are included. (d): the relation between accretion rates and stellar masses for YSOs in the literature, i.e. explicitly excluding our results (see \sect\ref{sec:discussion_Mdot_Mstar}). The filled circles represent YSOs with accretion rates estimated from U-band excess emission or veiling in UV/optical spectra, the open circles represent stars for which accretion rates were estimated from emission lines (H$\alpha$, \CaII$\lambda$8662, Br$\gamma$, Pa$\beta$, and Pa$\gamma$).}
\end{figure}

\noindent
In \fig\ref{fig:acc_rate_mass_all}(a-c) we show the accretion rates as derived from the H$\alpha$, H$\beta$, and \HeI \ equivalent widths using the methods described in \sect\ref{sec:analysis:accretion_rates} and Appendix~\ref{relation-acc-line}, as a function of stellar mass. The uncertainties in the accretion rates were calculated from the uncertainties in the equivalent width measurements, and assuming a 10\% error in the continuum flux when converting equivalent widths to line luminosities. For the H$\alpha$  EWs from the literature that did not have published uncertainties, we assigned an error of 10\% of emission-line flux. Since these diagrams are mostly populated with the same stars, we can use them to investigate whether the derived mass accretion rate of an ensemble of stars depends on which line is used as a diagnostic. We {\lrev have excluded} stars with extinctions above \AV$=$5\,mag.

The relation between the mass accretion rate as derived from the H$\alpha$ emission line and the stellar mass is shown in \fig\ref{fig:acc_rate_mass_all}(a). A linear regression in log-log space of the entire sample investigated (including spectral types and H$\alpha$ EWs from the literature), taking into account the uncertainty in the mass accretion rate of individual objects, yields the following best fit:
\begin{equation}\label{Halpha_mass1}
Log \dot M_{\rm{acc}}=-7.44(\pm0.11)+2.78(\pm0.22)\times log M_*
\end{equation}
\noindent
where \Mdotacc \ is the mass accretion rate in \Msunyr, and $M_*$ is the stellar mass in \Msun. If we include only those stars that we observed with VIMOS, the best fit remains essentially unchanged:
\begin{equation}\label{Halpha_mass2}
log \dot M_{\rm{acc}}=-7.50(\pm0.17)+2.7(\pm0.30)\times log M_*
\end{equation}

\noindent
If we use the H$\beta$ emission line to estimate the accretion rate, a linear regression gives a best fit of:
\begin{equation}\label{Hbeta_mass}
log \dot M_{\rm{acc}}=-7.46(\pm0.18)+3.36(\pm0.38)\times log M_*
\end{equation}
\noindent
and in case of the \HeI\,$\lambda$\,5876\,$\AA$  emission line, we get:
\begin{equation}\label{He_mass}
log \dot M_{\rm{acc}}=-7.18(\pm0.21)+3.42(\pm0.38)\times log M_*
\end{equation}

Thus, within uncertainties, we find the same behavior of the accretion rate with stellar mass from each of the three emission lines investigated. Taking simply a weighted average of the exponents in the above relations, we find: \Mdotacc\,$\propto$\,$M_*^{3.13\pm0.34}$, where the standard deviation in the distribution was adopted as the uncertainty in the exponent.

\vspace{0.3cm}

\noindent
\noindent \textbf{(d) Accretion rates as a function of stellar mass and age} \\

\vspace{-0.3cm}

\noindent
As outlined above, we find that the mass accretion rate in our sample shows an approximate power law relation with the stellar mass, with an exponent of 2.8$-$3.4. This is only an average relation, and individual objects scatter around this relation by at least an order of magnitude. Interestingly, the relation that we find is steeper than those found in previous work, where exponents in the range of 1.0$-$2.1 were found \citep{2003ApJ...582.1109W,2003ApJ...592..266M,2004AJ....128.1294C,2005ApJ...625..906M,2005ApJ...626..498M,2006A&A...452..245N,2006A&A...459..837G,2008ApJ...681..594H,2008A&A...481..423G}. We will discuss this further in Sect.\,\ref{sec:discussion_Mdot_Mstar}.

\begin{figure}
\centering
\includegraphics[width=\columnwidth]{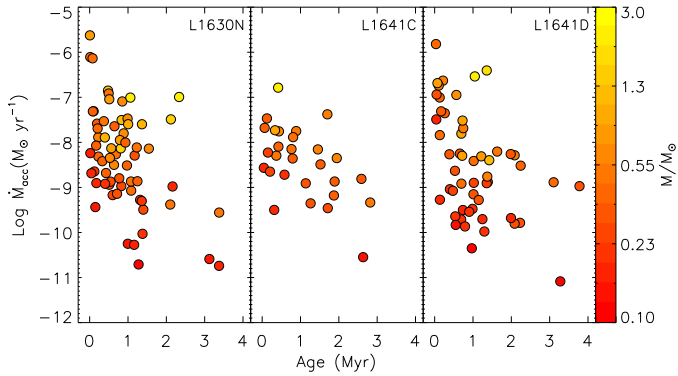}
\caption{\label{fig:acc_rate_age} The relations between accretion rates from H$\alpha$ emission luminosity and stellar ages for YSOs in L1630N, L1641C, and L1641D. {\newrev The symbol colors correspond to different stellar masses.}}
\end{figure}

In \fig\ref{fig:acc_rate_age} we show the accretion rates as a function of stellar age. In this figure, each star is color-coded according to its mass. On the whole, we observe a clear trend of decreasing accretion rate with increasing age. This trend is more obvious for stars below 0.5\Msun \ than for the more massive members, though. For very low mass stars ($M_*$$<$0.3\Msun), the accretion rate evolves from $\sim$10$^{-8}$\,$-$\,$10^{-9}$\Msunyr \ at ages below 1\,Myr to $\sim$10$^{-9}$\,$-$\,$10^{-11}$\Msunyr \ at ages above 3\,Myr. Our ensemble of stars with measured accretion rates contains only 11 stars with a mass above 1.0\Msun. These all are younger than 3\,Myr and have accretion rates of order 10$^{-9}$\,$-$\,$10^{-11}$\Msunyr. Within this sample there is no clear evidence for an evolution of accretion rate with age.

\subsubsection{Other emission lines}
\label{sec:results:emission_lines}
In addition to the hydrogen and helium emission lines, which were discussed in the previous section, a number of other emission lines are frequently seen in the optical spectra of our sample stars. These may trace accreting material, the disk surface, jets, or stellar/disk winds. In the following paragraph we briefly recall existing knowledge concerning these lines.

Forbidden line emission in the \OI\,6300\,\AA \ line often shows a low- and a high-velocity component. The low-velocity component may trace the disk surface {\rev \citep[e.g.][]{2006A&A...449..267A} or in a poorly collimated disk wind \citep[e.g.][]{1995ApJ...454..382K}}, whereas the high-velocity component arises in a jet close to the star \citep{1995ApJ...452..736H}. Forbidden line emission in the \OI\,5577\AA \ line appears primarily as a low-velocity component \citep{1994ApJS...93..485H,1995ApJ...452..736H}. Note that this line may suffer from terrestrial atmospheric contamination in some cases. Forbidden line emission in the \NII~6583\,\AA \ and 6678\,\AA \ lines usually shows up only in the high-velocity component, tracing the unresolved stellar jet \citep{1995ApJ...452..736H}. The \CaII \ near-infrared triplet (8498, 8542, and 8662~\AA) often shows anomalous intensity ratios in T~Tauri stars, strongly deviating from the nominal 1:9:5 line ratio for optically thin media, with the 8498\AA \ line being stronger than the 8542\AA \ line in some sources \citep{1994ApJS...93..485H}. The \CaII \ triplet often shows a broad emission profile, and the line fluxes correlate strongly with the accretion rate, suggesting these lines are formed in the magnetospheric infall flow \citep{1998ApJ...492..743M}. Forbidden emission in the \SII \ line is rarely observed towards T~Tauri stars due to the low abundance of sulfur in the interstellar medium \citep{1995ApJ...452..736H}.

\newcommand{\Lacc}{$L_{\rm{acc}}$}

\begin{figure}
\centering
\includegraphics[width=\columnwidth]{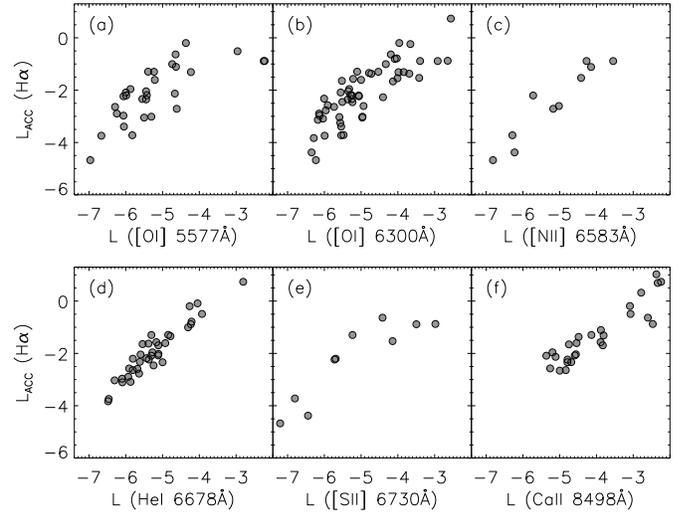}
\caption{\label{fig:other_lines} The correlation between accretion luminosities and the emission luminosities of different emission lines: (a) \OI\,5577\,$\AA$, (b) \OI 6300\,$\AA$, (c) \NII\,6583\,$\AA$, (d) \HeI\,6678\,$\AA$, (e) \SII\,6730\,$\AA$, and (f) \CaII\,8498\,$\AA$.  The accretion luminosities are calculated from H$\alpha$ emission luminosity using the formulae in Appendix~\ref{relation-acc-line}.}
\end{figure}

We examined all young stars observed with VIMOS for the abovementioned emission lines. In \tabs\ref{tab:emission_lines_L1630N} and \ref{tab:emission_lines_L1641} we give an inventory of the lines detected in the individual stars, listing their equivalent widths. In \fig\ref{fig:other_lines}, we compare the observed luminosities in six emission lines with the accretion luminosities (\Lacc) as derived from  H$\alpha$. All lines are strongly correlated with the accretion luminosity, but some lines show a larger dispersion around the average correlation than others. In particular, the \OI \ 5577 and 6300\,\AA \ lines show a rather large scatter. This agrees with the notion that these lines are not directly related to the accretion process but instead arise in, e.g., the disk surface \citep[see e.g.][]{2006A&A...449..267A,2008A&A...491..809F,2008A&A...485..487V}. For these lines, the correlation between line luminosity and accretion luminosity may simply be explained by the scaling of \Lacc \ with stellar mass, and the fact that more massive, more luminous stars will induce more \OI \ emission in their vicinity. The \HeI~6678\,\AA \ line is very strongly correlated with \Lacc, showing only a very small dispersion. This confirms that it is a good tracer for accretion. The \CaII~8498\,\AA \ line only appears in objects with \Lacc$>$3$\times$10$^{-4}$\Lsun \ and also shows a relatively small dispersion around the average relation. The \NII\,6583\,\AA \ and \SII\,6730\,\AA \ lines are detected in too few sources ($\sim$10) to properly assess how tightly they correlate with \Lacc. These lines are thought to arise in jets emerging from accreting systems and therefore their luminosities are expected to roughly scale with \Lacc. Based on our data we cannot distinguish between such an accretion-jet relation and a simple scaling of the line luminosities with the stellar luminosity. Note also that line emission in jets depends on shocks being present, i.e. the jet impinging on an ambient medium or having varying outflow rates and velocities, causing jet-internal shocks. Thus, a relation between the luminosity of line emission from jets and the accretion rate may be dilluted by temporal variations in both quantities, as well as by source to source differences in the ambient medium.

Eight of the young stars we observed show both the \SII~6716 and 6731~\AA \ lines. The flux ratio of these lines provides a direct measure of the electron density. The relative intensities (6716/6731) are identical within uncertainties, with a median ratio of 0.53, suggesting that both \SII \ lines are saturated with electron density $>$7$\times$10$^{3}$\,cm$^{-3}$ \citep{1994ApJS...93..485H}.

\subsection{Exotic objects}
\label{sec:discussion:exotic_objects}
A natural consequence of conducting a survey in which large numbers of sources are studied, is the discovery of objects with extreme properties. We will briefly discuss several objects we considered worth special attention.

\subsubsection{Subluminous objects with high emission line equivalent widths}
\label{sec:discussion:high_Halpha_EW}

\begin{figure}
\centering
\includegraphics[width=\columnwidth]{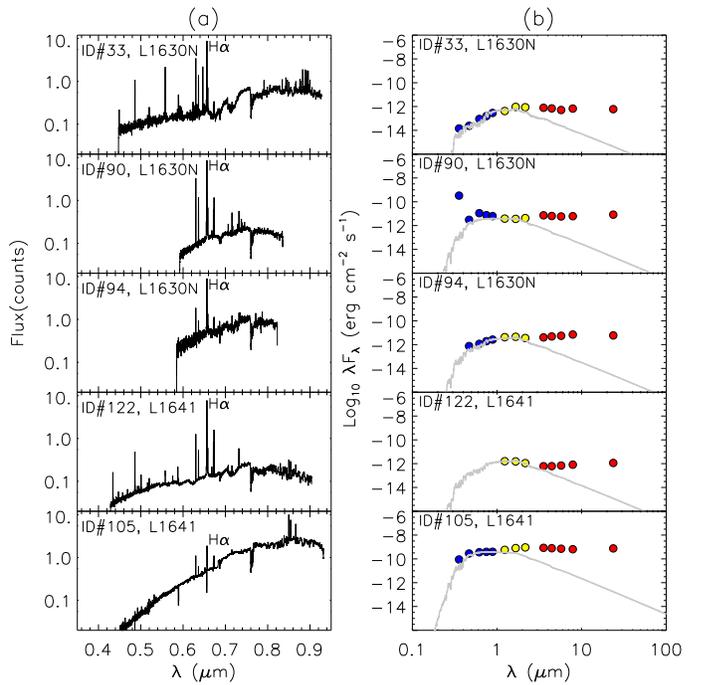}
\caption{\label{fig:exotic} The optical spectra (left panels) and dereddened SEDs (right panels) for the  four exotic  stars in L1630 (see \sect\ref{sec:discussion:high_Halpha_EW}), and L1641, and a new the FU Orionis object (ID\#105 in L1641, see \sect\ref{sec:discussion:fuor}). The photospheric emission level is indiacted with a grey curve in each panel. {\rev The blue points represent the optical photometry from SDSS or LAICA, the yellow points indicate 2MASS photometry, and the red points indicate Spitzer photometry.}}
\end{figure}

Three stars in L1630 (\#33, \#90, and \#94) and one star in L1641 (\#122) show exceptionally rich emission line spectra, with the emission lines exhibiting very high equivalent widths. In addition to the hydrogen balmer series, emission lines of He\,I, N\,II, S\,II, O\,I, O\,II, Ca\,II, Fe\,I, Fe\,II, and Na\,I are identified. All these stars have strong infrared excesses in the IRAC and MIPS bands, but generally show little or no excess in the near-infrared 2MASS bands. Their optical spectra and SEDs are shown in \fig\ref{fig:exotic}.

The high equivalent widths of e.g. H$\alpha$ are not due to abnormally high line luminosities, but instead seem to be caused by reduced photospheric continuum levels, making the lines appear bright in contrast. We have marked these sources in the HR diagram (\fig\ref{fig:HR_diagrams}), and they appear subluminous by a factor $\sim$10 with respect to the bulk of objects of similar spectral type.

\cite{2003A&A...406.1001C} have previously identified sources with similar properties in the Lupus~3 dark cloud. They investigated 3 possible explanations: (1) the sources are embedded, strongly accreting Class~I sources seen in scattered light; (2) they are sources with edge-on disks of which we see the photosphere only in scattered light but the emission lines (more) directly, making the latter appear bright in contrast; (3) they are sources in which strong sustained accretion alters the pre-main sequence evolution making objects appear subluminous for their spectral type \citep[][]{1997ApJ...475..770H,1997A&A...326.1001S}.

Option (1) is excluded by \cite{2003A&A...406.1001C} since Class~I sources seen in scattered light would have easily been spatially resolved in their observations, which instead revealed point sources. In our VIMOS pre-imaging observations, stars \#33 and \#90 in L1630N appear as single point sources at the resolution limit of our observations of $fwhm$$\sim$0\farcs7 (315\,AU). Our image of star \#94 in L1630N is well reproduced by the sum of two point sources, suggesting that this star is a binary at a separation of 0\farcs66 ($\sim$300\,AU). The fainter component (in the R-band) is located south-east of the brighter component at a position angle of $\sim$109\degree \ E of N. We do not think that this object is an edge-on disk seen in scattered light, since in that case the individual ``blobs'' of scattered light would be expected to be spatially resolved, given their large separation. Star \#122 in L1641 appears as a point source, but has a faint, spatially extended ``blob'' of emission about 0\farcs8 north of the star. This may be shock-induced H$\alpha$ emission from a jet emerging from this object. We conclude that we do not find spatially extended emission from the central sources, in agreement with \cite{2003A&A...406.1001C}.

 Option (2) was considered to be unlikely by \cite{2003A&A...406.1001C} since there is strong evidence from the H$\alpha$ line strengths and line widths that they are formed in magnetospheric accretion columns very close to the star. Thus, an hypothetical edge-on disk would obscure the emission lines as much as the photospheric continuum and the equivalent widths would remain unaltered, though the apparent under-luminosity can be explained in such a scenario. Also in our observations the balmer lines are clearly spectrally resolved and likely arise close to the central star. Thus, we agree with \cite{2003A&A...406.1001C} that edge-on disks do not provide a satisfactory explanation of the nature of these objects.

\cite{2003A&A...406.1001C} consider option (3) the most likely scenario for their exotic objects. Even though the magnitude of the ``under-luminosity'' in their objects is clearly larger than predicted by theory, the models do provide a good qualitative match to the observations. \cite{2003A&A...406.1001C} argue, though, that the effect will be stronger for their sources because of the lower mass compared to the range studied by \cite{1997ApJ...475..770H}.

\subsubsection{A new FU Orionis object}
\label{sec:discussion:fuor}
Star \#105 in L1641 shows a clear P-cygni profile in both the H$\alpha$ and H$\beta$ lines. Blue-shifted absorption is observed at a range of velocities up to approximately -300\kms, the peak of the H$\alpha$ emission is redshifted to $\sim$100\kms \ (see \fig\ref{fig:profile}), a typical profile of H$\alpha$ emission line from FU Orionis type stars \citep{1996A&AS..120..229R,2007A&A...472..207F}. In \fig\ref{fig:exotic}, its optical spectrum and SED are shown.  The IR excess of this source is extraordinarily strong and the IR luminosity exceeds the estimated stellar luminosity by a factor of $\gtrsim$5. Emission in forbidden transitions of \OI, \SII, and \NII, as well as permitted transitions of Fe\,I and Fe\,II  are detected. We {\lrev derived} a spectral type of K5.5 for this star, which compares reasonably well with an earlier determination of K3 by \cite{1995PhDT..........A}.

The observed properties of this object provide a perfect match to those of FUOR stars, a relatively rare class of young stars thought to undergo an accretion outburst, and whose system luminosity is dominated by the release of gravitational energy of accreting material rather than stellar photospheric emission. The IR excess emission constitutes an estimated luminosity of 20\Lsun. Given the estimated stellar mass and radius, an accretion rate of a few times 10$^{-6}$\Msunyr \ is implied. The forbidden emission points at strong outflow activity.

\subsection{Extinction}
\label{sec:results:extinction}
\begin{figure}
\centering
\includegraphics[width=\columnwidth]{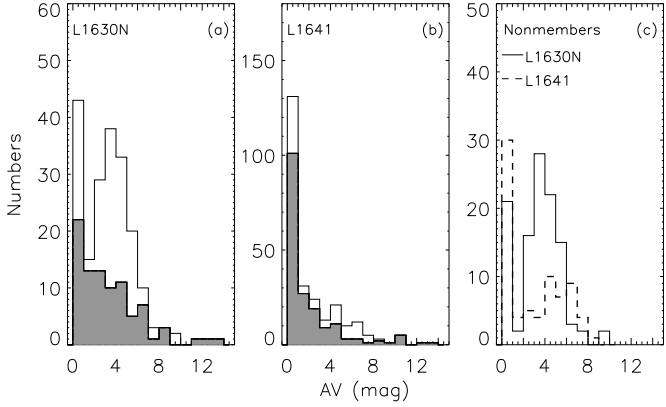}
\caption{\label{fig:AV_dis} (a)(b): The visual extinction distribution of our  sample observed with VIMOS in L1630N and L1641. The open histograms show the  distribution of all the stars with reliable spectral types. The filled  histograms display the YSO distribution among them. (c): The visual  extinction distribution for non-members in L1630N (solid-line histogram) and  L1641 (dashed-line histogram).}
\end{figure}

Robust extinction estimates were made for each individual object using the method outlined in \sect\ref{sec:analysis:optical_sed_fitting}. In \fig\ref{fig:AV_dis}(a,b) we show the resulting A$_{V}$ distributions, distinguishing between cloud members (shaded grey) and field stars. In \fig\ref{fig:AV_dis}(c) we zoom in on the field population.

The YSOs show a similar distribution of extinction values in both clouds: a peak in the distribution at very low extinction (\AV$<$1\,mag) and a gradual decrease in number counts towards higher \AV. The fraction of sources with very high optical extinctions (\AV$>$5\,mag) is somewhat higher in L1630 than in L1641, the opposite is true for sources with \AV$<$1. The distribution of extinction values may reflect the distribution of the young stars within the molecular cloud: stars located ``on the near side'' of the cloud will on average have low extinction, whereas objects located ``on the far side'' have higher extinctions. However, ``local'' density variations within the cloud may affect the distribution equally well: some sources will be located behind very dense clumps whereas others are not. Additionally, some objects may have a disk inclination such that our sight line goes through the outer part of their flared disks, causing additional local ``circumstellar'' extinction. Remnant envelopes may have the same effect.

The field population shows a different distribution of extinctions, best seen in \fig\ref{fig:AV_dis}(c). Also here there is a peak at \AV$<$1\,mag, consisting of foreground objects and background objects at relatively ``clean'' sight lines. In addition, there is a distribution between 1 and $\sim$8 magnitudes of \AV. This distribution is relatively well defined in L1630N with a clear peak around \AV=3$-$4\,mag, but is somewhat broader in L1641. It reflects the total column density of dust in the entire cloud in the sight lines towards the respective stars. Note that in both the YSO and field population there is a bias \emph{against} objects with very high extinctions, since such objects are too faint for optical spectroscopy.

\subsection{Spatial distribution of YSOs in L1630N and L1641}
\label{sec:results:spatial_distribution}
The spatial distributions of young stars in the L1630N and L1641 clouds are shown in \fig\ref{fig:YSODIS}, overplotted on $^{13}$CO maps by \cite{1987ApJ...312L..45B} and \cite{1994ApJ...429..645M}. The positions of (Lada) Class~0/I and Class~II sources, identified on the basis of their infrared colors using the criteria in \cite{2008ApJ...674..336G}, have been indicated with yellow {\newrev asterisks} and red $+$\,signs, respectively. The young stars included in our spectroscopic (VIMOS) sample have been marked with blue dots. Note that these include both Class~II and Class~III sources.

As noted before by numerous authors \citep[e.g.][]{2002ApJ...578..914H,2005IAUS..227..383M,2009ApJS..181..321E}, the distribution of Class~0/I sources closely follows the densest parts of the molecular cloud as traced by the $^{13}$CO data, and shows a filamentary structure. The Class~II sources are distributed more evenly, though they also show concentrations in the densest cloud parts. The YSOs in L1630N are mainly distributed in two clusters, which are associated with two dense regions of $^{13}$CO molecular clouds. In L1641 the young stars are more dispersed across the whole region, with several local concentrations. Based on their near-infrared image survey, \citet{1993ApJ...412..233S} have identified 8 aggregates in this region. Since their observations do not cover the entire L1641 region, we {\lrev applied} the nearest neighbor method to the spitzer data to find the new aggregates/clusters. With the NNM, we find 9 aggregates in L1641, named groups 1-9. Among these, group 3 and 4 are newly identified. Groups 1, 2, 5, 6, 7, 8 and 9  correspond to the L1641 south cluster, the CK group, L1641~C, KMS~36, KMS~35, the V380~Ori group, and L1641N, respectively, as identified by \citet{1993ApJ...412..233S}. We do not recover the aggregate HH~34/KMS~12 due to the small number of stars ($\sim$5) it possesses.

\section{Discussion}
\label{sec:discussion}
A general goal of surveys like ours is to gain insight in the physics of disk evolution by finding empirical relations between observable properties. In particular, we may seek for correlations of disk geometry and accretion rate with stellar mass, age, and formation environment. An evolution of the disk geometry and the accretion rate with age is certainly expected, as well as a dependence of the accretion rate on stellar mass. Disk evolution is a function of stellar mass as well, in the sense that disks around high-mass stars (approximately B type and earlier on the main sequence) evolve faster than those around low-mass stars \citep[e.g.][]{2000prpl.conf..559N,2009A&A...497..117A}. Whether the evolution of the disk depends on stellar mass within the low-mass regime probed here, is not clear. The influence of the formation environmment, i.e. whether a star formed in a clustered environment or in isolation, is a matter of debate. One may consider how the disk evolution is affected by gravitational interactions with other cluster members and by the strongly enhanced (far-) UV radiation field compared to interstellar space, causing photo-evaporation of the outer disk regions. When making quantitave theoretical assessments of these effects, one finds that the disk regions we see at $\lambda$\,$\lesssim$\,24\mum \ are not noticably affected, even in clusters of 100-1000 members \citep{2008PhST..130a4029A}. The number of members in the aggregates or clusters that we identify is usually considerably smaller (N$<$100), and thus, from a theoretical vantage point, we do not expect to see a significant effect of the formation environment on the disk properties.

\subsection{Disk frequency: trends with stellar mass and age}
\label{sec:discussion:disk_frequency}
\begin{figure}
\centering
\includegraphics[width=\columnwidth]{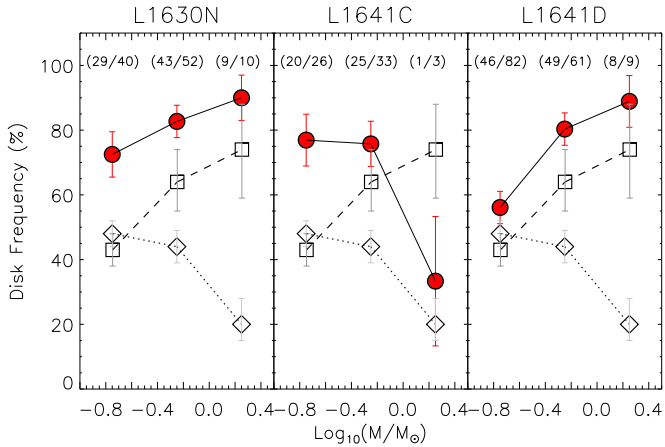}
\caption{\label{fig:df_mass} The disk frequency  as a function of stellar mass  for YSOs in L1630N (filled circles) and L1641C (filled circles, cluster/aggregate population), and L1641D (filled circles, distributed population), IC~348 \citep[open diamonds, ][]{2006AJ....131.1574L}, Chamaeleon~I \citep[open squares, ][]{2008ApJ...675.1375L}. Absolute number counts are given at the top of each panel, for each mass bin.}
\end{figure}

In \fig\ref{fig:df_mass} we display the fraction of YSOs which show clear indications of circumstellar material in the form of infrared excess emission, as a function of stellar mass\footnote{The mass bins used are, in log($M_*$/\Msun), -1 to -0.5, -0.5 to 0, and 0 to 0.5.}. For reasons outlined in \sect\ref{sec:observations:biases}, these numbers may somewhat over-estimate the true disk frequency, but trends with stellar mass and age remain unaffected. In \fig\ref{fig:df_mass} we also show the disk frequencies for IC~348 as found by \cite{2006AJ....131.1574L} and Chamaeleon~I as found by \cite{2008ApJ...675.1375L}. For L1630N and L1641D, we find some evidence for a disk frequency that \emph{increases} with stellar mass: the increase between $\sim$0.2\Msun \ and $\sim$1.5\Msun \ is $\sim$2$\sigma$ and $\sim$3$\sigma$ for these populations, respectively. For the clustered population in L1641, our data are consistent with a disk fraction that is constant with stellar mass (the highest mass bin contains just 3 stars, of which 1 has a disk, and the apparent decrease is not significant). The trend of increasing disk frequency with increasing stellar mass that we tentatively identify in L1630N and L1641D is remarkably similar to that seen by \citep{2008ApJ...675.1375L} in the $\sim$2\,Myr old Chamaeleon~I cloud. In the somewhat older IC~348 \citep{2006AJ....131.1574L} and $\sigma$\,Orionis \citep{2007ApJ...662.1067H} regions, the disk frequency is lower for the stars in the highest mass bin (see \fig\ref{fig:df_mass}). If the behavior of increasing disk frequency with increasing stellar mass at young ages, and a reduced disk frequency for the higher mass stars at later times is universal, it may be naturally explained if the higher mass stars form somewhat later than the low mass stars but subsequently dissipate their disks faster.

\begin{figure}
\centering
\includegraphics[width=\columnwidth]{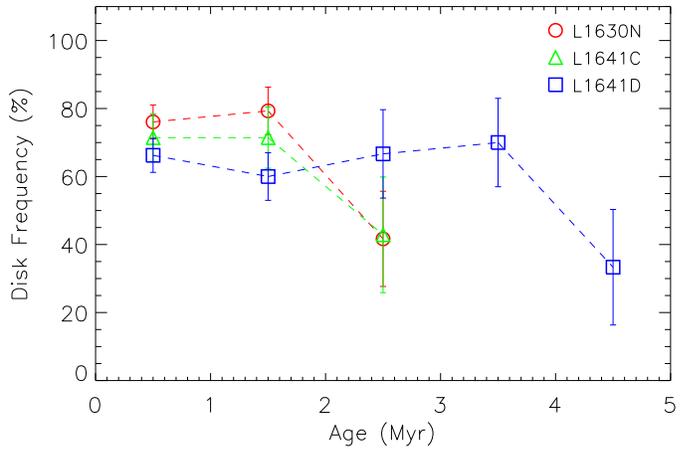}
\caption{\label{fig:df_age} The disk frequency  as a function of age for YSOs in L1630N ({\newrev open circles}) and L1641 ({\newrev open triangles} for cluster/aggregate population, and {\newrev open squares} for distributed population).  }
\end{figure}

\vspace{0.15cm}

The disk frequency as a function of stellar age is shown in \fig\ref{fig:df_age} (top panel). For the distributed population in L1641 we see a constant disk frequency for the first 3$-$4\,Myr, after which we see a $\sim$2$\sigma$ decrease in the last age bin. In the clustered populations L1630N and L1641C we observe a quantitatively different behavior: in the first 1$-$2\,Myr, the disk frequency is approximately constant and consistent with that observed in L1641D, within uncertainties. In the age bin 2$-$3\,Myr the disk frequency is reduced by $\sim$2$\sigma$ in both the L1630N and L1641C populations. Both the magnitude and the timing of the decrease in disk frequency are strikingly similar in L1630N and L1641C, underlining the significance of the result. The age bins above 3\,Myrs contain insufficient stars in the clustered populations to give meaningful disk frequencies. Our results strongly support the notion that disk dissipation proceeds faster in objects that form in a clustered environments than in objects that form in isolation. Qualitatively, this behavior is expected from theory. However, given the relatively low stellar densities in the clustered environments that we consider here, compared to dense clusters such as the Orion nebula cluster, the magnitude of the observed effect is higher than expected \citep{2008PhST..130a4029A}.

\subsection{Accretion rate as a function of stellar mass}
\label{sec:discussion_Mdot_Mstar}
How the rate at which material accretes depends on the mass of the central star has been a topic of considerable attention. Power law fits of the form \Mdotacc\,$\propto$\,$M_*^{\alpha}$ have yielded values in the range 1.0$-$2.1 for the exponent $\alpha$ \citep{2003ApJ...582.1109W,2003ApJ...592..266M,2004AJ....128.1294C,2005ApJ...625..906M,2005ApJ...626..498M,2006A&A...452..245N,2006A&A...459..837G,2008ApJ...681..594H,2008A&A...481..423G}. As shown in \sect\ref{sec:results:accretion_signatures} and \fig\ref{fig:acc_rate_mass_all}(a-c), we consistently find a steeper exponent of $\alpha$\,$=$\,2.8$-$3.4 from each of the three emission lines we use as a tracer of \Mdotacc \ (H$\alpha$, H$\beta$, \HeI). The sample we investigated contains numerous stars in the mass range of 0.1$-$1.0\Msun \ but only very few stars more massive than the sun.

To investigate this further we gathered a large set of accretion rate estimates from the literature \citep{1998ApJ...492..323G,2003ApJ...582.1109W,2003ApJ...592..266M,2003ApJ...583..334H,2004AJ....128.1294C,2005ApJ...625..906M,2006A&A...452..245N,2005ApJ...626..498M,2006A&A...459..837G,2008ApJ...681..594H,2008AJ....136..521D}, of stars in a range of star forming regions including Taurus-Aurigae, IC\,348, $\rho$\,Oph, $\lambda$\,Ori, Orion OB and upper Scorpius. We estimate their masses using their luminosities and effective temperatures, which were taken directly from the literature, with the PMS evolutionary tracks from \cite{2008ApJS..178...89D} (for M$\geq$0.1\,M$_{\odot}$), and \cite{1998A&A...337..403B} (for M$<$0.1\,M$_{\odot}$). From the stars discussed in \cite{2006A&A...452..245N} we included only those with confirmed spectral types and adopted their spectral types and luminosities from \cite{1999ApJ...525..440L}. 

The result is shown in \fig\ref{fig:acc_rate_mass_all}(d). The red open circles in this figure represent measurements of the mass accretion rate derived from emission line strengths, the filled green circles represent those derived from UV excess emission and veiling of optical spectra. The usual picture emerges: a clear trend of increasing accretion rate with increasing stellar mass, and a large scatter of individual sources around the average relation. If we make a power law fit of the form \Mdotacc\,$\propto$\,$M_*^{\alpha}$, and we restrict the fit to the mass range $M_*$\,$\leq$\,1.0\Msun, we find a value of $\alpha$\,$\sim$2.8 for the exponent. This is very similar to the result we have obtained from the sample studied in this work, which covers the same mass range. If we perform the same fit over the whole mass range plotted in \fig\ref{fig:acc_rate_mass_all}(d), which includes a number of HAeBe stars, we find an exponent of $\alpha$\,$\sim$2.0. The latter value agrees with that found by previous authors. Our results indicate that the dependence of accretion rate on stellar mass is not uniform over the entire mass regime, but instead is considerably steeper below $\sim$1\,\Msun \ than at higher masses \citep[see also][]{2006ApJ...648..484H}.

Due to the dependence of the mass accretion rate on the system age, it is important to consider this aspect. If the stars with a mass above 1\,\Msun \ would be significantly older than those of lower mass, the the difference between the \Mdotacc($R_*$) in the subsolar and super-solar mass regime as seen in \fig\ref{fig:acc_rate_mass_all}(d) could be an artifact. We therefore checked what would happen if we include only stars from Taurus-Aurigae, IC\,348, $\rho$\,Oph, and $\lambda$\,Ori, i.e. from regions that all have an age of $\lesssim$3\,Myr. This yielded results that are consistent with those described above\footnote{In a fit over the whole mass regime in which only stars from the four mentioned star forming regions were included, we find $\alpha$\,$=$\,2.27, i.e. only slightly larger than in the fit including all objects in \fig\ref{fig:acc_rate_mass_all}(d). This small difference is due to the fact that the abovementioned star forming regions contain relatively few $M_*$\,$>$\,1.0\Msun \ stars, thus biasing the fit towards the slope of the low-mass regime.}, and thus the mass dependence of the accretion rate that we derived is robust.

\subsection{Ages of the different populations}
In the traditional view of low-mass star formation, CTTSs evolve into WTTSs when their circumstellar disks dissipate, and WTTSs should therefore on average be older than CTTSs. Observations of various star-forming regions, however, did usually \emph{not} yield evidence for an age difference between the CTTS and WTTS population \citep[e.g.][]{1998ApJ...497..736H,2001AJ....121.1030H,2002AJ....123..304H,2005AJ....129..829D}, though in some cases the WTTS are indeed found to be older than the CTTSs \citep{1995ApJ...452..736H,2007A&A...473L..21B}.

\begin{figure}
\centering
\includegraphics[width=\columnwidth]{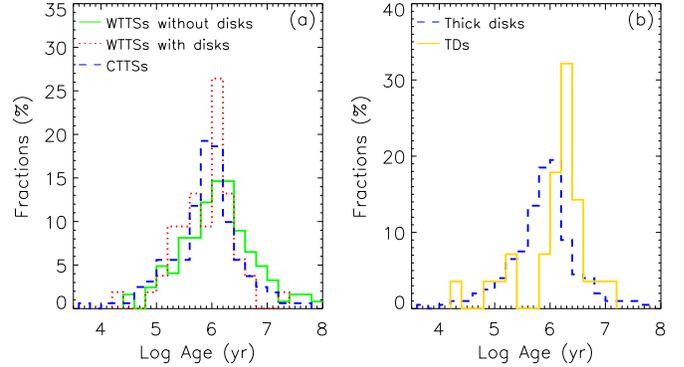}
\caption{\label{fig:TTS_age_dis} Left panel: The histograms showing the  distribution of ages for the WTTSs without disks, WTTSs with disks, and CTTSs. Right panel: The distributions of ages for  YSOs with ``normal'' optically thick disks and transition disks.}
\end{figure}

The age distribution of both classes of objects is shown in \fig\ref{fig:TTS_age_dis}(a). This figure encompasses our entire sample of YSOs, i.e. the sources in L1630N and L1641 have been put together. The median ages of the CTTSs and WTTSs are $\sim$0.8\,Myr and  $\sim$1.2\,Myr, respectively. A KS test reveals a  relatively low  probability ($P$\,$\sim$0.03) for the WTTSs and CTTSs to be drawn randomly from the same age distribution. The WTTSs are further divided into disk harboring objects with clear IR excess emission, and those who appear diskless. The probability that the WTTSs with and without disks are drawn randomly from the same population is low ($P$\,$\sim$0.05), whereas the age distributions of WTTSs with disks and CTTSs are indistinguishable ($P$\,$\sim$0.94).

In \fig\ref{fig:TTS_age_dis}(b) we show the age distributions for YSOs with optically thick disks and those with transition disks. The median ages for both populations are 0.8\,Myr and 1.9\,Myr, respectively, and the transition disks thus are on average clearly older. A KS test yields a very low probability ($P$\,$\sim$0.003) for the stars with normal optically thick disks and the transition objects to be randomly drawn from the same population. In contrast, the age distributions of the WTTSs without disks and the transition objects are indistinguishable ($P$\,$\sim$0.53 in a KS test).

Transition disk objects are on average older than the normal disk population. This suggests that disk-binary interaction and gravitational instabilities in the disk are not the dominant mechanisms causing a transition disk appearance, as these mechanisms would take effect from the earliest stages and would not lead to an ``old'' transition disk population.

\vspace{0.5cm}

\subsection{Median SEDs}
\label{sec:discussion:median_SEDs}

\begin{figure*}
\centering
\includegraphics[width=17cm]{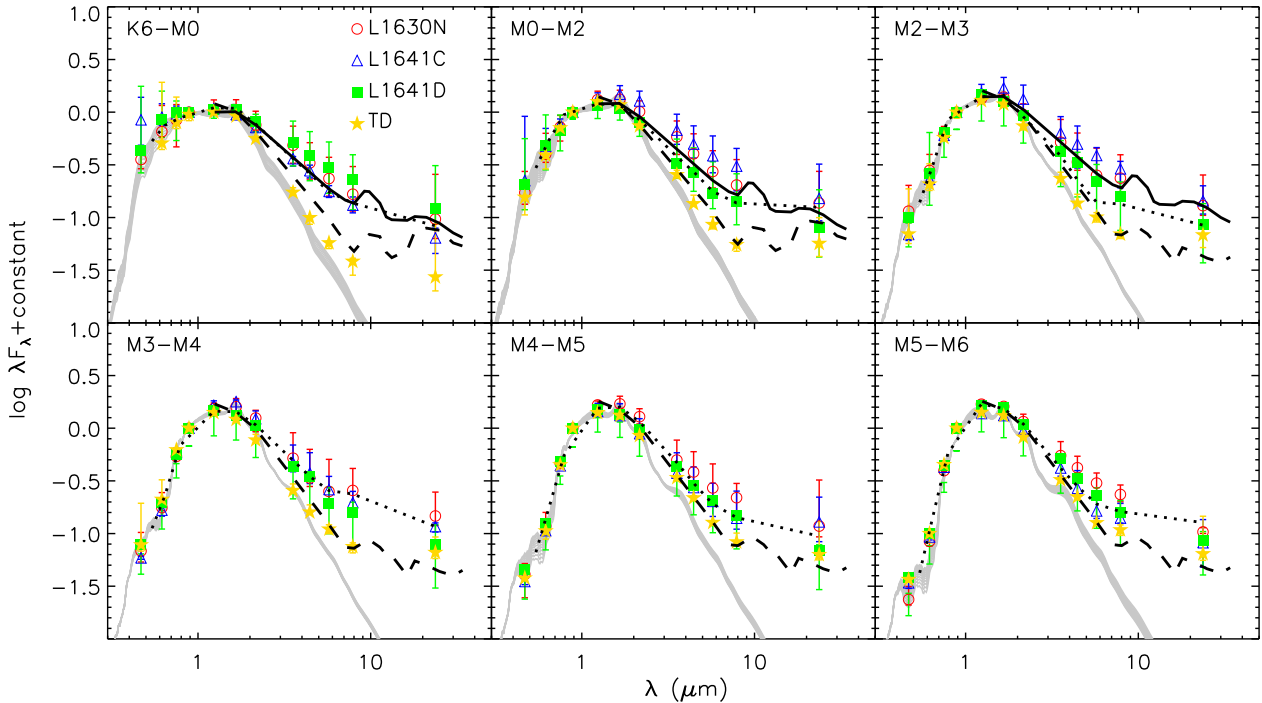}
\caption{\label{fig:median_seds} Median SEDs of the stars with optically thick disks in L1630N, L1641C and L1641D. We also plot the median SED of the transition disks, as well as those of the IC~348 \citep[dotted line, 2-3\,Myr,][]{2006AJ....131.1574L}, Taurus \citep[solid line, 1-2\,Myr,][]{2006ApJS..165..568F}, and upper Scorpius \citep[dashed line, $\sim$5\,Myr,][]{2009arXiv0901.4120D} regions. The photospheric emission level is indicated with a grey curve in each panel.}
\end{figure*}

In \fig\ref{fig:median_seds} we show the median SEDs of the YSOs in our sample that are surrounded by ``normal'' optically thick disks, defined as those having \alphairac$\geq$-1.8 \citep{2006AJ....131.1574L}. We also show the SEDs of sources classified as having transition disks. Following \cite{2006AJ....131.1574L}, we divide the sample in six spectral type bins. We also show the median SEDs of the $\sim$2-3\,Myr old cluster IC~348 from \cite{2006AJ....131.1574L} with dotted curves, of the  $\sim$1-2\,Myr old distributed population in Taurus \citep{2006ApJS..165..568F} with solid curves, and the $\sim$5\,Myr old population in upper Scorpius \citep{2009arXiv0901.4120D} with dashed curves. For the cluster IC~348, \cite{2006AJ....131.1574L} only presents 'observed' SEDs. We determine reddening by fitting model atmospheres of corresponding spectral types to the observations in optical bands. The derived visual extinctions are around 1-3\,mag for different spectral types. Note that the Taurus median SED covers spectraly types K4-M2 without distinguishing between various sub types, the upper Scorpius SEDs distinguish only between spectral types earlier and later than M2.

The median SEDs of the ``normal'' optically thick disks in L1630N, L1641C and L1641D are identical within errors at all wavelengths and for all spectral types, except for a marginal ($\sim$2\,$\sigma$) difference at 24\mum \ for the K6-M0 spectral bin. These SEDs are also indistinguishable from the population in IC~348 and Taurus. The transition disk SEDs are clearly different, and closely resemble the median SED of upper Scorpius.

The excess emission of the optically thick disks in the IRAC bands is a function of spectral type, with the excess being strongest for the K-type stars and becoming progressively weaker towards later spectral types. This trend is fully consistent with that found by \cite{2006AJ....131.1574L} in IC~348. The transition disk sources show an opposite trend: the IRAC excess is lowest at the early spectral types but becomes noticably stronger at spectral types later than M4. The combined effect is that for spectral types later than M4, the discrimination between stars classified as having optically thick disks and those having transition disks is no longer obvious. The MIPS 24\mum \ excess of the transition disks is only slightly smaller than that of the optically thick disks for the early spectral types and equal within uncertainties for the later spectral types.

\subsection{Star formation modes in L1641}
With the Spitzer data, we have found 9 aggregates/clusters in L1641, which include 31$\pm$3\%  of YSOs identified by four IRAC bands. We have classified the YSOs identified by Spitzer into class 0/I or class II types. Among them, 58$\pm$5\% of class 0/I sources and 79$\pm$6\% of class II sources  are formed in isolation. The high fraction of class\,0/I sources in the distributed population suggests that most stars in L1641 are formed in isolation, since these class 0/I sources are very young \citep[less than 0.5~Myr,][]{2009ApJS..181..321E}, and cannot leave their parental molecular cloud given the typical stellar velocity dispersion of in the Orion molecular cloud complex of $\sim$1~km~s$^{-1}$ \citep{1993ApJ...412..233S}.

In our sample, YSOs formed in aggregates/clusters account for 31$\pm$3\% of all YSOs in L1641, consistent with previous estimates of, e.g., $\sim$30\% \citep{1993ApJ...412..233S}, or 25-50\% \citep{1995PhDT..........A}. The median age for YSOs in aggregates/clusters is 1.1~Myr, which is similar to the median age ($\sim$ 1.2~Myr) for the distributed populations.  A Kolmogorov-Smirnov (KS) test yields no statistically significant difference betweeen the age distributions of both populations ($P$\,$\sim$0.3) and we thus find no evidence for a different star formation history between the clustered and distributed populations.

\begin{figure}
\centering
\includegraphics[width=\columnwidth]{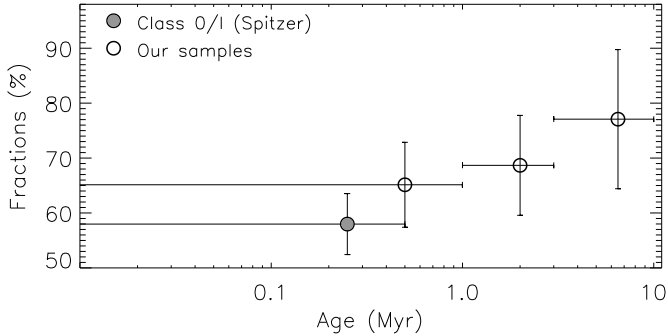}
\caption{\label{fig:L1641_SFH} The fractions of distributed population in
  L1641 at different age bins.}
\end{figure}

We divide the ages $\tau$ of our sample into three bins: $\tau$\,$\leq$\,1~Myr, 1~Myr$\leq$\,$\tau$\,$<$3~Myr, and 3~Myr$\leq$\,$\tau$\,$<$10~Myr. For each age bin, we compute the fraction of stars that resides in the distributed population, and plot these in \fig\ref{fig:L1641_SFH}. We obtain an additional point in this diagram at the youngest ages by including the class\,0/I sources identified from the Spitzer data and assigning an age of less than 0.5\,Myr to these \citep{2009ApJS..181..321E}. {\rev The resulting fractions of YSOs in the distributed population are 58$\pm6$\% for objects younger than $\sim$0.5~Myr (the class~0/I sources), 65$\pm$8\% for stars younger than 1~Myr (class~II sources), 69$\pm$9\% in the interval 1$-$3~Myr, and 77$\pm$13\% in the interval 3$-$10~Myr}. These fractions are equal within the uncertainties of the individual points. However, a slight trend of an increasing fraction in the distributed population with increasing age is seen. This suggests that the older, unbound aggregate/cluster population is slowly spreading into the distributed population. We note, however, that this effect is too small to produce significant differences in the average age of the aggregate/cluser and distributed populations, and that most of the stars that we see in isolation were formed in isolation.

\subsection{Accretion in transition disks}
\label{sec:discussion:transition_disks}
\citet{2007MNRAS.378..369N} have compiled the properties of a dozen transition disks in the Taurus-Aurigae star foming region from the literature, studying in particular the accretion rate and the disk mass. They find that, in comparison to other stars of similar age, objects harboring transition disks have a $\sim$10 times \emph{lower} median accretion rate and a $\sim$4 times \emph{higher} median disk mass. These results suggest an important role for forming jovian planets in the evolution from optically thick disks to transition objects: if a planet is massive enough it will open a gap in the disk, after which the disk region interior to the planet is quickly drained by viscous evolution and the accretion rate onto the central star drops. Since massive giant planets may form more easily within massive disks, this scenario would explain both the lower median accretion rates and the higher median disk masses of transition objects. Note, though, that based on SMA observations of a  sample of transition disks, \citet{2008ApJ...686L.115C} find that the inner disks only start to be cleared out after the outer disk has been significantly dissipated. These observations therefore argue for photo-evaporation to be the dominant processes driving disk evolution.

\begin{figure}
\centering
\includegraphics[width=\columnwidth]{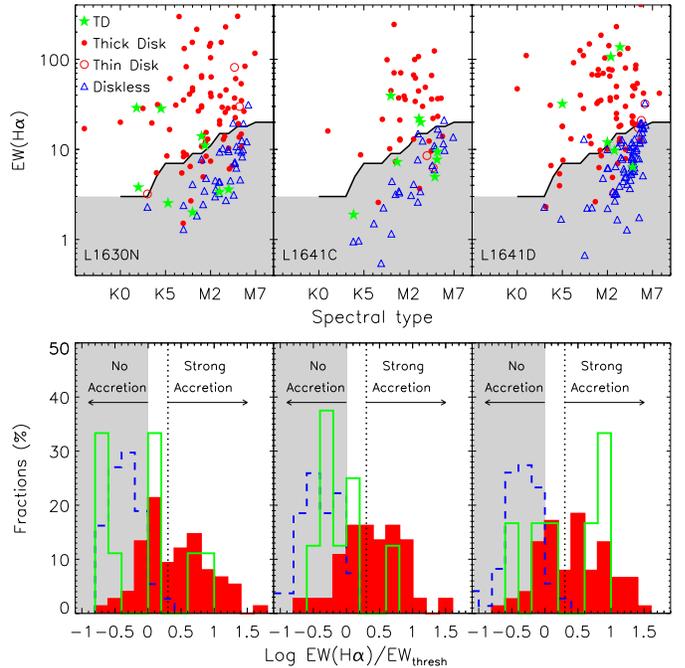}
\caption{\label{fig:TD_acc} The top panels: The H$\alpha$ EWs  vs. spectral  type for YSOs in {\newrev the three populations discussed in this work}. The filled circles are for YSOs with optically thick disks, open circles for YSOs with optically thin disks, asterisks for YSOs with transition disks, and {\newrev triangles} for YSOs  without disks. The solid line is the boundary between WTTSs and CTTSs   (see~Appendix~\ref{boundary_sec}). The bottom panels: the distribution of   logarithmic ratio between the observed H$\alpha$ EW and the EW threshold, which used to classify the YSOs into CTTSs or WTTSs for the corresponding spectral type, for  three population: YSOs with optically thick disks (filled histogram),  YSOs with  transition disks (solid-line histogram), and YSOs without disks (dashed-line histogram).}
\end{figure}

In our spectroscopic sample there are 28 transition disks distributed over the L1630N and L1641 clouds, of which 23 have estimated H$\alpha$ equivalent widths. We compare their H$\alpha$ EWs with those of the other YSOs in \fig\ref{fig:TD_acc}. Among the transition disks, 26$\pm$11\% (6 out of 23) show ``strong'' accretion activity, here defined as having an H$\alpha$ equivalent width of more than 2 times the EW threshold used to distinguish between CTTSs and WTTSs (see Appendix~\ref{boundary_sec}). In the stars surrounded by optically thick disks this fraction is 57$\pm$6\%, and thus there are significantly fewer strong accretors among the transition disks. However, among the transition disks that do show active accretion (H$\alpha$ EW above the WTTS/CTTS threshold\footnote{We remind the reader that H$\alpha$ emission with an EW below the CTTS/WTTS threshold cannot be unambiguously attributed to accretion, and we therefore exclude the WTTSs from the current comparison.}), the median accretion rate is 3.0$\times$10$^{-9}$\Msunyr, which is very similar to the median accretion rate of the CTTSs with optically thick disks in our sample: $\sim$4.0$\times$10$^{-9}$\Msunyr. Note that \citet{2007MNRAS.378..369N} find a median accretion rate of 3.2$\times$10$^{-9}$\Msunyr \ for their transition disks, which is consistent with the value we find. However, \cite{2007MNRAS.378..369N} find a much higher median accretion rate of $\sim$2.5$\times$10$^{-8}$\Msunyr \ for the optically thick disk-harboring CTTSs in the Taurus-Aurigae region than we find for our Orion sample ($\sim$4.0$\times$10$^{-9}$\Msunyr). The latter value is consistent with the ``typical'' accretion rate of young T~Tauri stars of $\sim$3$\times$10$^{-9}$\Msunyr \ \citep{1998ApJ...492..323G,2003ApJ...582.1109W,2003ApJ...592..266M}, and is also in perfect agreement with the findings of Sicilia-Aguilar et al. (2009, in prep.) who find a median accretion rate of 2-3$\times$10$^{-9}$\Msunyr \ for both the young stars with optically thick disks and those with transition disks in Trumpler~37.

In summary, we find that the fequency of accreting stars is lower among transition disk objects than among stars with optically thick accretion disks, but the median accretion rate among the actively accreting stars is similar in both populations.

\subsection{Clumpiness of the molecular clouds}
\label{sec:discussion:clumpiness}
Our observations of numerous background objects yield accurate ``pencil-beam'' estimates of the integrated extinction through the whole cloud. The extinction can be converted into estimates of the dust column density along each sight line. The angular scale probed by each measurement is simply the angular diameter of the respective background star, and will typically be of order 10$^{-2}$ milli-arcseconds. The millimeter $^{13}$CO data yield estimates of the gas surface density at a spatial resolution of 1\farcm7  \citep{1994ApJ...429..645M}. Thus, we get two measures of the surface density of the molecular cloud $\Sigma_{\rm{cloud}}$, at vastly different spatial resolutions. The comparison of both yields information on the clumpiness of the regions probed: both column density measures will correlate well in the case of a homogeneous medium, whereas they will correlate poorly if the medium shows strong substructure on scales below the resolution of the millimeter data. A number of authors have studied similar correlations between pencil beam extinction estimates, usually based on near-infrared photometry, and tracers of the gas density \citep[C$^{18}$O, $^{13}$CO, CS, e.g.][]{1994ApJ...429..694L,1999A&A...342..257K,1999ApJ...515..265A}, and found evidence for a clumpy cloud structure.

The direct comparison of both measures of $\Sigma_{\rm{cloud}}$ depends on a number of assumptions: (1) the dust properties are homogeneous across the molecular cloud; (2) the gas to dust ratio is uniform; (3) CO freeze-out is not important in the regions probed; (4) the molecular tracer is optically thin in the regions probed. Since our sample contains only sight lines with relatively low extinction (\AV$\lesssim$10\,mag, with only a handful of objects with \AV$>$7\,mag), the $^{13}$CO line remains optically thin, and we do not cover dense cores in which CO freeze-out occurs. Rather than the dense regions, which are known to show strong density enhancements on small scales \citep[e.g.][]{2006A&A...454..781L}, we trace the low density regions that surround the cores.

\begin{figure}
\centering
\includegraphics[width=\columnwidth]{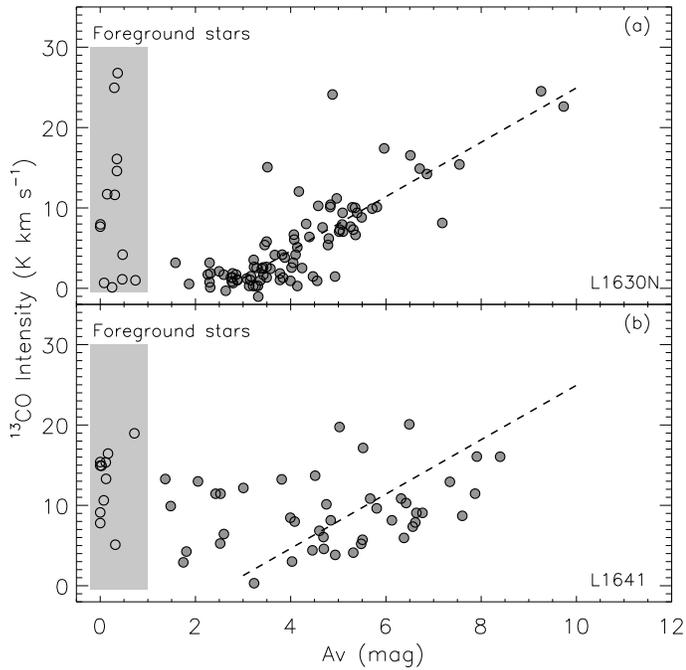}
\caption{\label{fig:AV_CO} The correlation between $^{13}$CO intensity and extinction estimated from the non-members in L1630N and L1641. The dashed line shows the best linear fit in L1630N, in the regime \AV$>$2\,mag where CO is detectable. In L1641 the scatter in the distribution is much larger, and we only overplot the relation found in L1630N for reference.}
\end{figure}

In \fig\ref{fig:AV_CO} we show the relation between our pencil beam \AV  \ measurements and the $^{13}$CO intensities from \cite{1987ApJ...312L..45B,1994ApJ...429..645M} interpolated at the same positions for L1630N and L1641. Apart from a few obvious outliers, we find a fairly tight correlation between both quantities in L1630N. In L1641, the correlation is much worse. A linear fit to the $^{13}$CO intensity as a function of \AV, within the range 3\,mag\,$\leq$\,\AV\,$\leq$\,10\,mag, yields the following relation in L1630N:
\begin{equation}
I_{^{13}CO}=(-8.88\pm1.50)+(3.38\pm0.31)\times A_{\rm V}
\end{equation}
This fit yields zero $^{13}$CO intensity for \AV$\approx$2\,mag, which is expected since at extinctions below this value CO is dissociated by the interstellar UV radiation field.

Since the L1630N and L1641 are at approximately the same distance, the millimeter data have the same physical resolution of $\sim$0.2\,pc. Our observations therefore suggest that the low density medium in L1630 shows relatively little substructure at these scales, whereas the L1641 medium has a substantial amount of such inhomogeneities.

\section{Summary}
\label{sec:summary}

We have performed a large optical spectroscopic and imaging survey of the L1630N and L1641 star-forming clouds in Orion. We {\lrev combined} our data with optical and infrared imaging data from the literature. The optical spectroscopy and photometry allow accurate determination of the stellar effective temperature, luminosity, and line of sight extinction through model atmosphere fitting. Mass and age estimates of individual objects {\lrev were} obtained by placement in the Hertzsprung-Russell diagram. Accretion rates {\lrev were} estimated from optical emission lines (H$\alpha$, H$\beta$, \HeI). The infrared photometry provides good measures of the IR excess emission, from which the evolutionary state of the disk can be deduced. In total we {\lrev investigated} 132 YSOs in L1630N and 267 in L1641, of which 65 and 117 were newly identified in the respective clouds.

Our survey products are threefold: (1) a list of identified YSOs with spectral types, optical and infrared magnitudes, equivalent widths of a number of emission lines, estimates of the stellar mass/age, an SED based disk classification, estimates of the accretion rate, and line of sight extinction (399 sources); (2) a list of unrelated field objects, mostly background stars, with photometric magnitudes, spectral types, and line of sight extinction (179 sources); (3) a photometric catalog with optical and infrared magnitudes of all sources detected in both wavelength regimes (21694 sources).

The young stars in L1630N predominantly formed in 2 clusters, whereas in L1641 some stars live in clusters (or rather ``aggregates'') and some in a distributed population. We {\lrev distinguished} between these populations using the nearest neighbor method, and identify two previously unknown aggregates in L1641.

Based on the infrared SEDs we {\lrev divided} the young stars into objects with still full blown optically thick disks, optically thin disks, transition objects, and diskless stars. We find that the disk frequency does not strongly depend on the mass of the star, with some evidence for an increasing disk frequency with increasing stellar mass. The disk frequency decreases with age, and does so earlier in clustered environments than in the distributed population. The latter is not expected from a theoretical vantage point, given the low densities and small sizes of the clusters/aggregates under consideration. {\rev Our data yield no evidence for a substantial age difference between the aggregate and distributed populations, and the possible difference in disk dissipation time scales between both populations requires further investigation.} In our spectroscopic sample we {\lrev identified} 28 transition disks, of which 20 were previously unknown. Moreover, we {\lrev identified} 47 additional transition disk candidates based on photometric measurements only.

 We {\lrev estimated} accretion rates from the equivalent widths of the H$\alpha$, H$\beta$, and \HeI\,5876\AA \ lines. The mass accretion rate show a clear dependence on stellar mass, with individual objects showing a large scatter around the average relation. When making power law fits of the form \Mdotacc\,$\propto$\,$M_*^{\alpha}$ we find values in the range of 2.8$-$3.4 for the exponent $\alpha$, which is steeper than the range 1.0$-$2.1 found by previous authors. We compiled an independent set of mass and accretion rate estimates from the literature. We find that over the whole mass range of $\sim$0.05$-$5.0\Msun \ we obtain $\alpha$$\sim$2.0, consistent with previous work. If we limit the fit to the mass range $M_*$\,$\lesssim$\,1.0\Msun, i.e. the range occupied by our sample, the literature data yield $\alpha$$\sim$2.8, which is consistent with our results. Thus, we show that the dependence of the accretion rate on stellar mass is not uniform, but instead is steeper at subsolar masses than above $\sim$1\,\Msun. We find a general decrease of accretion rate with age, which is less evident for the stars in our highest mass bin ($M_*$$>$1.0\Msun), possibly due to low number statistics.

The WTTSs are on average older than the CTTSs in our sample, but both populations show a large intrinsic scatter in their age distributions. The WTTSs with IR excess, emission indicative of a circumstelar disk, have the same average age as the CTTSs whereas apparently diskless WTTSs are on average older. The age distributions of transition disk objects and WTTSs without disks are indistinguishable, and they are on average older by a factor of $\gtrsim$2 than the CTTSs.

For L1641 we {\lrev investigated} what fraction of stars exist in the distributed and clustered populations. Our data suggest that the fraction of stars in the distributed population increases with age, from $\sim$60\% at $\leq$0.5\,Myr to $\sim$80\% at $>$3\,Myr, though at our current statistics this effect is $\lesssim$2$\sigma$ and should be further investigated.

We find that among the stars with transition disks there are significantly fewer strong accretors (26$\pm$9\%) than among stars with optically thick disks  (57$\pm$6\%). We find that the median accretion rate of those transitional objects that are actively accreting is similar to that of accreting stars with normal optically thick disks. This agrees with recent results in Trumpler~37 (Sicilia-Aguilar et al. 2009, in prep.), but is in contrast to earlier results for the Taurus-Aurigae region, in which transition disks were found to have a $\sim$10 times lower accretion rate than objects with normal optically thick disks \citep{2007MNRAS.378..369N}. Note that the median accretion rate we {\lrev derived} for the Orion TDs is consistent with that found in the Taurus TDs, but that the median accretion rate derived for stars with normal optically thick disks is much higher in Taurus, and that this causes the aformentioned discrepancy.

We find one previously unknown object whose characteristics stongly suggests it is an FU~Orionis star. Its bolometric (infrared) luminosity exceeds the estimated stellar luminosity by a factor $\gtrsim$5, and it shows P-Cygni profiles in both H$\alpha$ and H$\beta$. Furthermore we {\lrev identified} four objects with apparently under-luminous photospheres for their spectral types which show very rich emission line spectra with very high equivalent widths. Such objects were found before \citep[][]{2003A&A...406.1001C}, who suggested that these objects are pre-main sequence stars whose evolution has been altered by strong accretion. Such stars would be less massive and less luminous than other stars of similar age and spectral type. The strong accretion and associated outflow activity yields the rich emission line spectrum.

By comparing molecular cloud column density estimates from \AV \ measurements of background stars (spatial resolution $\sim$0.005\,AU) and $^{13}$CO intensity maps (spatial resolution $\sim$45000\,AU or 0.22\,pc) we find that the low density medium in which the star forming cores are embedded shows significant substructure on scales of $\sim$0.2\,pc in L1641, but little substructure in L1630.

\begin{acknowledgements}
Doug Finkbeiner and Craig Loomis are kindly acknowledged for their generous
help with the unpublished SDSS data of the north-west half of L1641. {\rev Many thanks to J. Hern\'{a}ndez for providing his spectral classification code on which our code is based, and to Carlo Izzo for providing the new VIMOS pipeline.} 

We thank Calar Alto Observatory for allocation of director's discretionary time to this programme. This research has made use of the SIMBAD database, operated at CDS, Strasbourg, France. This publication makes use of data products from the Two Micron All Sky Survey, which is a joint project of the University of Massachusetts and the Infrared Processing and Analysis Center/California Institute of Technology, funded by the National Aeronautics and Space Administration and the National Science Foundation. This work is in part  based  on observations made with the Spitzer Space Telescope, which is operated by the Jet Propulsion Laboratory, California Institute of Technology under a contract with NASA. This publication is in part based on data from the Sloan Digital Sky Survey project, whose many contributors can be found at the following web-page: http://www.sdss.org/collaboration/credits.html. We thank an anonymous referee for a constructive report and many comments and suggestions which improved the clarity of this paper.
\end{acknowledgements}

\bibliographystyle{aa}
\bibliography{references}

\newpage

\begin{appendix}

\section{Classifying  CTTS and WTTS}
\label{boundary_sec}

In this appendix we present the criteria we apply to separate Classical T-Tauri Stars (CTTS) from Weak-Line T-Tauri Stars (WTTS). Our classification scheme is a refined version of that of \citet{2003ApJ...582.1109W}. Our basic method is very simple: we select a large number of young stars from the literature that show no evidence for a significant near-infrared excess, i.e. their (inner) disks have already dissipated. The vast majority of these will \emph{not} be actively accreting, though the sample will contain some objects that do still accrete. We then inspect the H$\alpha$ equivalent widths of all objects and, as a function of spectral type, define the region that contains the bulk of objects as the WTTS domain.

CTTSs are distinguished from WTTSs based on the equivalent width of the H$\alpha$ emission line. CTTSs show relatively strong, often broad H$\alpha$ emission lines that are attributed to active accretion. WTTSs exhibit weaker and narrower H$\alpha$ lines which do not point at active accretion, but instead appear to have their origin in chromospheric activity. \citet{2003ApJ...582.1109W} use accretion-induced veiling of optical spectra to distinguish CTTSs and WTTSs, and classify a T Tauri star as a CTTS if 
   EW(H$\alpha$)$\geq3\AA$ for K0-K5 stars,
   EW(H$\alpha$)$\geq10\AA$ for K7-M2.5 stars,
   EW(H$\alpha$)$\geq20\AA$ for M3-M5.5 stars,
 and EW(H$\alpha$)$\geq40\AA$ for M6-M7.5 stars.
The EW(H$\alpha$) threshold significantly increases with spectral type due to the substantial decrease of the photospheric continuum level near the H$\alpha$ line in cool stars.

We collected PMS stars from the literature and distributed in different star-formation regions, i.e. Ophiuchus, Lupus \citep{2007ApJ...667..308C}, Taurus (Luhman et al. 2006,Gudel et al. 2007),Tr37, NGC7160 \citep{2005AJ....130..188S,2006ApJ...638..897S}, L1630N, L1641 (this paper), NGC2264 \citep{2005AJ....129..829D,2007ApJ...671..605C}, NGC2362 \citep{2005AJ....130.1805D,2007AJ....133.2072D}, $\lambda$ Orionis cluster \citep{2007ApJ...664..481B}, Orion OB1 association \citep{2005AJ....129..907B,2007ApJ...661.1119B,2007ApJ...671.1784H}, Chamaeleon II \citep{2008ApJ...676..427A,2008ApJ...680.1295S}, $\sigma$\,Ori cluster \citep{1999ApJ...521..671B,2003A&A...404..171B,2008A&A...488..167S,2007ApJ...662.1067H}, Coronet cluster \citep{2008ApJ...687.1145S}, IC\,348 \citep{2003ApJ...593.1093L,2006AJ....131.1574L}, Serpens cloud core, and NGC\,1333 \citep{2007ApJ...669..493W,2009arXiv0904.1244W}. In Taurus, some PMS stars have multiple measurements of the H$\alpha$ EW, in which case we used the average value.  

\begin{figure}
\centering
\includegraphics[width=\columnwidth]{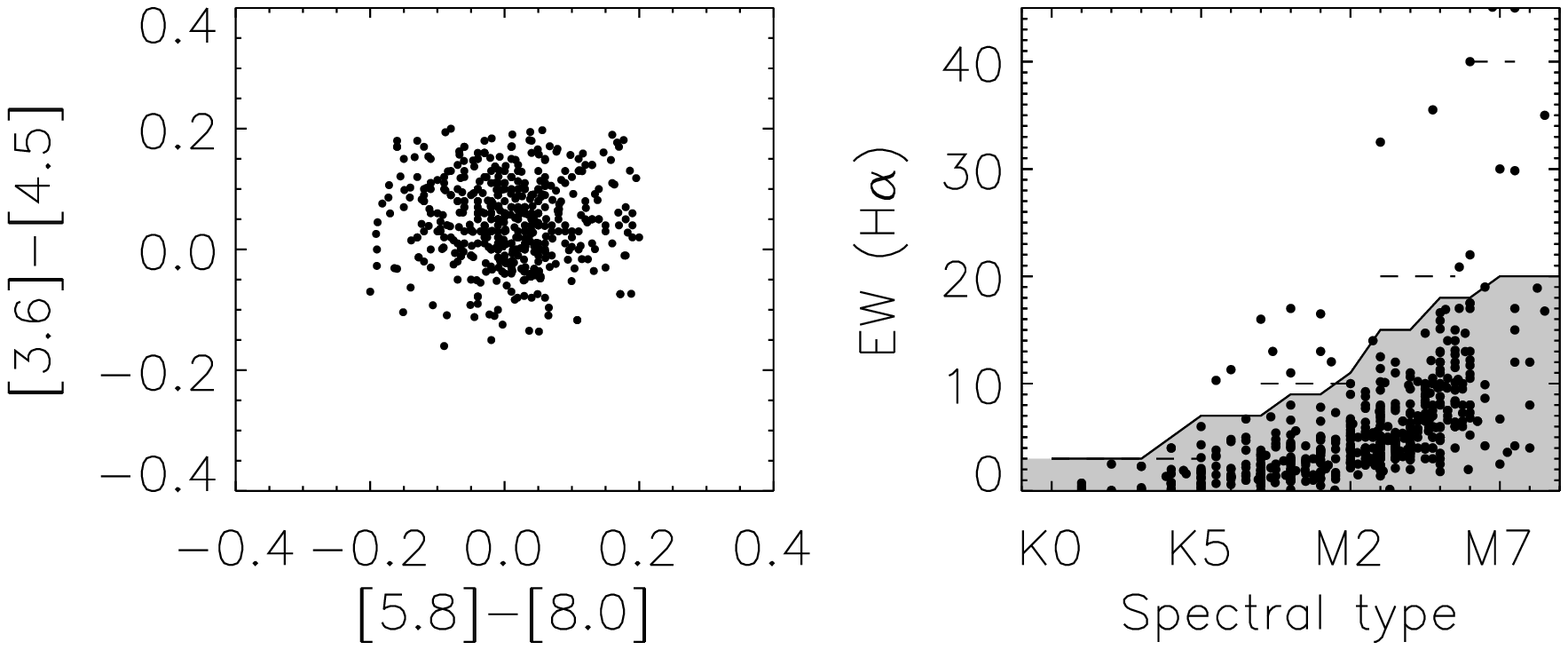}
\caption{\label{fig:boundary} \emph{Left panel:} Spitzer [5.8]-[8.0] vs. [3.6]-[4.5] color-color diagrams for YSOs with no or little infrared   excess in different regions,  i.e. Ophiuchus, Lupus, Perseus   \citep{2007ApJ...667..308C}, Tr37, NGC7160   \citep{2005AJ....130..188S,2006ApJ...638..897S}, L1630N, L1641 (this paper),   NGC2264 \citep{2005AJ....129..829D,2007ApJ...671..605C}, NGC2362   \citep{2005AJ....130.1805D,2007AJ....133.2072D}, $\lambda$ Orionis cluster   \citep{2007ApJ...664..481B}, Orion OB1 association   \citep{2005AJ....129..907B,2007ApJ...661.1119B,2007ApJ...671.1784H},  Chamaeleon II \citep{2008ApJ...676..427A,2008ApJ...680.1295S}, $\sigma$\,Ori cluster \citep{1999ApJ...521..671B,2003A&A...404..171B,2008A&A...488..167S,2007ApJ...662.1067H}, Coronet cluster \citep{2008ApJ...687.1145S}, IC\,348 \citep{2003ApJ...593.1093L,2006AJ....131.1574L}, Serpens cloud core, and NGC\,1333 \citep{2007ApJ...669..493W,2009arXiv0904.1244W}.. \emph{Right panel:} the relation between spectral types and EW(H$\alpha$) for stars in the left panel. The dashed lines show the criteria that \citet{2003ApJ...582.1109W} used to distinguish CTTSs and WTTSs. The grey-shaded area outlined by the solid line represents our refined criteria to classify CTTSs and WTTSs.}
\end{figure}

In total, we obtained a sample of 536 stars with no or weak infrared excess by constraining $|$[3.6]-[4.5]$|$$\leq$0.2,and $|$[5.8]-[8.0]$|$$\leq$0.2. The inner disks around these stars have been mostly cleared and the bulk of these objects will no longer be actively accreting. In \fig\ref{fig:boundary}, we show their [5.8]-[8.0] vs. [3.6]-[4.5] color-color diagram (left panel) and the distribution of H$\alpha$ equivalent widths as a function of spectral type. We find that there is a quite clearly defined border between the ``bulk'' of sources and the ``outliers'' with high EW. We interpret the former sources to be WTTS where the H$\alpha$ emission arises in an active chromosphere, whereas the latter sources are still actively accreting, despite there weak or absent near-IR excess. These sources are rare, and most or all of them will be transition disk objects.

In the right panel we also plot the commonly used criteria for distinguishing between CTTSs and WTTSs by \citet{2003ApJ...582.1109W}, with dashed horizontal lines. We can see that these criteria are not stringent enough to describe the boundary between the ``bulk'' and ``outlier'' objects, in particular many of the ``outlier'' objects of late-M spectral type would be classified as WTTSs according to these criteria. We propose more stringent limits to separate WTTSs from CTTSs: we consider a star to be a CTTS if EW(H$\alpha$)$\geq3\AA$ for K0-K3 stars, EW(H$\alpha$)$\geq5\AA$ for K4 stars, EW(H$\alpha$)$\geq7\AA$ for K5-K7 stars, EW(H$\alpha$)$\geq9\AA$ for M0-M1 stars, EW(H$\alpha$)$\geq11\AA$ for M2 stars, EW(H$\alpha$)$\geq15\AA$ for M3-M4 stars, EW(H$\alpha$)$\geq18\AA$ for M5-M6 stars, and EW(H$\alpha$)$\geq20\AA$ for M7-M8 stars. Our newly defined boundary is illustrated in the right panel of \fig\ref{fig:boundary} by the solid line and grey-shaded area.

\section{Relations between emission line luminosity and accretion luminosity}
\label{relation-acc-line}

In this appendix we derive the relations between the luminosity in several optical emision lines to the total accretion luminosity as measured using diagnostics that are independent of those lines. Both the line luminosities and the total accretion luminosities are adopted from the quoted papers.

\subsection{Hydrogen emission lines}
\begin{figure}
\centering
\includegraphics[width=\columnwidth]{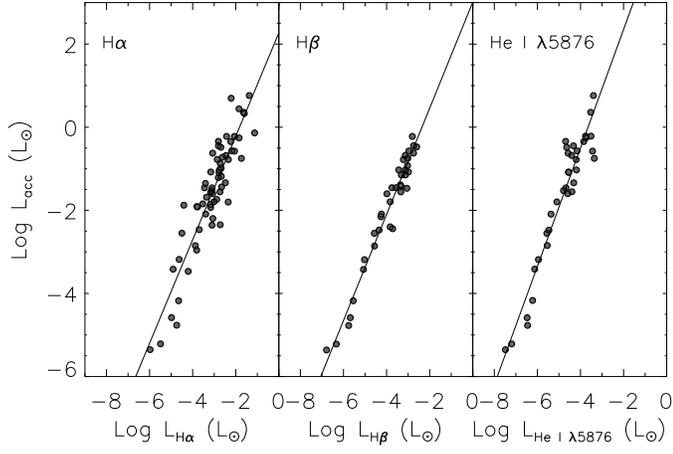}
\caption{\label{fig:lineluminosity_vs_accretionluminosity} The relations between accretion luminosity and emission line luminosity. The solid lines show the best power law fits to the respective distributions.}
\end{figure}

We collected YSOs with measured H$\alpha$ emission line luminosities and accretion luminosities from the literature \citep[][]{1998ApJ...492..323G,2008AJ....136..521D,2008ApJ...681..594H} and plotted these in \fig\ref{fig:lineluminosity_vs_accretionluminosity}. A resulting least squares fit to this distribution yields the following relation:
\begin{equation}
log(L_{\rm{acc}}/L_{\odot})=(2.27\pm0.23)+(1.25\pm0.07)\times log (L_{\rm{H}\alpha}/L_{\odot})
\end{equation}

Similarly, we collected YSOs with measured H$\beta$ emission line luminosities and accretion luminosities from the literature \citep[][]{1998ApJ...492..323G,2008ApJ...681..594H}. The best fit relation in this case becomes:
\begin{equation}
log(L_{\rm{acc}}/L_{\odot})=(3.01\pm0.19)+(1.28\pm0.05)\times log (L_{\rm{H}\beta}/L_{\odot})
\end{equation}

\subsection{Helium emission line}
We collected YSOs with measured \HeI \ emission line luminosities and accretion luminosities from the literature \citep[][]{2008AJ....136..521D,2008ApJ...681..594H}. In this case, the best fit reads:
\begin{equation}
log(L_{\rm{acc}}/L_{\odot})=(5.20\pm0.38)+(1.42\pm0.08)\times log (L_{\rm{He\,I}\,\lambda \rm{5876}}/L_{\odot})
\end{equation}

\end{appendix}

\clearpage

\scriptsize
\onltab{4}{\longtab{4}{
\begin{landscape}
\begin{longtable}{lcccccccccccccccccccccc}
\caption{\label{tab:YSO_parameters_L1630N} Parameters for YSOs in L1630N. Column 4: $^{a}$ the values estimated from our VIMOS spectra; $^{b}$ the values from \citet{2008AJ....135..966F}. Column 6: stellar bolometric luminosity derived from optical SED fitting (see Sect.\,\ref{sec:analysis:optical_sed_fitting}). Column 9: classification based on the criteria in Appendix~\ref{boundary_sec}. Column 11: aggregate/cluster population or distributed population. Column 13-14: stellar masses and ages estimated using PMS evolutionary tracks from \citet{2008ApJS..178...89D}. Column 16-18: accretion rates estimated with  H$\alpha$, H$\beta$, and \HeI\,5876\AA \ emission line luminosities, respectively.}\\
\hline\hline
(1) &(2) &(3) &(4)&(5)&(6)&(7)&(8)&(9)&(10)&(11)&(12)&(13)&(14)&(15)&(16)&(17)&(18)\\
&  RA     & DEC     &              &adopted      & Lum & EW(H$_{\alpha}$) &EW(Li)   & TTS        & Av &   & &Mass&Age&Disk&Log $\dot{M}_{acc}$&Log $\dot{M}_{acc}$&Log $\dot{M}_{acc}$\\
ID& (J2000) & (J2000) & Spectral type&Spectral type& (L$_{\odot}$) & ($\AA$)       & ($\AA$) &type & (mag) &clu  &$\alpha(3.6-8.0 \mu m)$&(M$_{\odot}$) & (Myr) &property &(M$_{\odot} $\,yr$^{-1}$) &(M$_{\odot} $\,yr$^{-1}$) & (M$_{\odot} $\,yr$^{-1}$) \\ 
\hline
\endfirsthead
\caption{continued.}\\
\hline\hline
\hline\hline
(1) &(2) &(3)
&(4)&(5)&(6)&(7)&(8)&(9)&(10)&(11)&(12)&(13)&(14)&(15)&(16)&(17)&(18)\\
&  RA     & DEC     &              &adopted      & Lum & EW(H$_{\alpha}$) &EW(Li)   & TTS        & Av &   & &Mass$^{c}$&Age$^{c}$&Disk&Log $\dot{M}_{acc}$&Log $\dot{M}_{acc}$&Log $\dot{M}_{acc}$\\
D& (J2000) & (J2000) & Spectral type&Spectral type& (L$_{\odot}$) & ($\AA$)       & ($\AA$) &type & (mag) &clu  &$\alpha(3.6-8.0 \mu m)$&(M$_{\odot}$) & (Myr) &property &(M$_{\odot} $\,yr$^{-1}$) &(M$_{\odot} $\,yr$^{-1}$) & (M$_{\odot} $\,yr$^{-1}$) \\ 
\hline
\endhead
\hline
\endfoot
1& 05 45 21.38& $-$00 00 45.9&M3$^{b}$&M3&0.298&$-$4.4$^{b}$&0.2$^{b}$&WTTS&0.9&yes&$-$2.69&0.30$^{+0.10}_{-0.06}$&0.84$^{+1.11}_{-0.39}$&No disk&...&...&...\\
2& 05 45 22.68& $-$00 14 27.5&M6$^{b}$&M6&0.128&$-$12.2$^{b}$&0.6$^{b}$&WTTS&0.0&yes&$-$2.58&0.13$^{+0.06}_{-0.01}$&0.38$^{+0.70}_{-0.12}$&No disk&...&...&...\\
3& 05 45 26.16& +00 06 37.9&M3$^{b}$&M3&0.393&$-$20.8$^{b}$&0.3$^{b}$&CTTS&0.4&yes&$-$1.39&0.29$^{+0.09}_{-0.06}$&0.51$^{+0.51}_{-0.22}$&Thick&$-$8.94&...&...\\
4& 05 45 38.26& $-$00 08 11.0&M0$^{b}$&M0&0.552&$-$4.1$^{b}$&0.4$^{b}$&WTTS&2.1&yes&$-$2.60&0.55$^{+0.21}_{-0.11}$&1.36$^{+3.15}_{-0.61}$&No disk&...&...&...\\
5& 05 45 41.68& $-$00 04 02.4&M4$^{a}$,M3$^{b}$&M4&0.273&$-$10.0$^{a}$,$-$5.4$^{b}$&0.4$^{b}$&WTTS&0.2&yes&$-$2.77&0.24$^{+0.03}_{-0.03}$&0.56$^{+0.20}_{-0.15}$&No disk&...&...&...\\
6& 05 45 41.94& $-$00 12 05.3&K5.5$^{a}$,K4$^{b}$&K5.5&0.994&$-$2.5$^{a}$,$-$1.2$^{b}$&0.5$^{a}$,0.5$^{b}$&WTTS&1.3&yes&$-$2.57&0.92$^{+0.23}_{-0.23}$&2.00$^{+4.04}_{-0.93}$&Transition disk&...&...&...\\
7& 05 45 42.80& $-$00 01 01.8&M5.5$^{a}$&M5.5&0.093&$-$14.8$^{a}$&0.4$^{a}$&WTTS&0.3&yes&$-$1.24&0.15$^{+0.04}_{-0.03}$&1.07$^{+0.89}_{-0.40}$&Thick&...&...&...\\
8& 05 45 44.37& +00 22 58.2&M1$^{b}$&M1&0.608&$-$29.4$^{b}$&0.6$^{b}$&CTTS&0.8&yes&$-$1.37&0.43$^{+0.12}_{-0.08}$&0.68$^{+0.82}_{-0.28}$&Thick&$-$8.26&...&...\\
\end{longtable}
\end{landscape}
} }

\onltab{5}{\longtab{5}{
\begin{landscape}
\begin{longtable}{lcccccccccccccccccccccc}
\caption{\label{tab:YSO_parameters_L1641} Parameters for YSOs in L1641. Column 4: $^a$ the values estimated from our VIMOS spectra; $^b$ the values from \citet{1995PhDT..........A}; $^c$ the values from \citet{2008A&A...489.1409G}. Column 6: stellar bolometric luminosity derived from optical SED fitting (see Sect.\,\ref{sec:analysis:optical_sed_fitting}). Column 9: classification based on the criteria in Appendix~\ref{boundary_sec}. Column 11: aggregate/cluster population or distributed population. Column 13-14: stellar masses and ages estimated using PMS evolutionary tracks from \citet{2008ApJS..178...89D}. Column 16-18: accretion rates estimated with  H$\alpha$, H$\beta$, and \HeI\,5876\AA \ emission line luminosities, respectively.}\\
\hline\hline
(1) &(2) &(3) &(4)&(5)&(6)&(7)&(8)&(9)&(10)&(11)&(12)&(13)&(14)&(15)&(16)&(17)&(18)\\
&  RA     & DEC     &              &adopted      & Lum & EW(H$_{\alpha}$) &EW(Li)   & TTS        & Av &   & &Mass&Age&Disk&Log $\dot{M}_{acc}$&Log $\dot{M}_{acc}$&Log $\dot{M}_{acc}$\\
ID& (J2000) & (J2000) & Spectral type&Spectral type& (L$_{\odot}$) & ($\AA$)       & ($\AA$) &type & (mag) &clu  &$\alpha(3.6-8.0 \mu m)$&(M$_{\odot}$) & (Myr) &property &(M$_{\odot} $\,yr$^{-1}$) &(M$_{\odot} $\,yr$^{-1}$) & (M$_{\odot} $\,yr$^{-1}$) \\ 
\hline
\endfirsthead
\caption{continued.}\\
\hline\hline
1) &(2) &(3) &(4)&(5)&(6)&(7)&(8)&(9)&(10)&(11)&(12)&(13)&(14)&(15)&(16)&(17)&(18)\\
&  RA     & DEC     &              &adopted      & Lum & EW(H$_{\alpha}$) &EW(Li)   & TTS        & Av &   & &Mass&Age&Disk&Log $\dot{M}_{acc}$&Log $\dot{M}_{acc}$&Log $\dot{M}_{acc}$\\
ID& (J2000) & (J2000) & Spectral type&Spectral type& (L$_{\odot}$) & ($\AA$)       & ($\AA$) &type & (mag) &clu  &$\alpha(3.6-8.0 \mu m)$&(M$_{\odot}$) & (Myr) &property &(M$_{\odot} $\,yr$^{-1}$) &(M$_{\odot} $\,yr$^{-1}$) & (M$_{\odot} $\,yr$^{-1}$) \\ 
\hline
\endhead
\hline
\endfoot
1& 05 35 14.56& -06 15 12.7&M1$^{a}$&M1&0.569&$-$9.8$^{a}$&...&CTTS&2.3&no&$-$1.20&0.42$^{+0.06}_{-0.05}$&0.70$^{+0.52}_{-0.22}$&Thick&$-$8.92&$-$9.11&$-$8.74\\
2& 05 35 14.67& $-$06 15 07.3&M5$^{a}$&M5&0.168&$-$23.2$^{a}$&...&CTTS&2.6&no&$-$0.75&0.19$^{+0.04}_{-0.03}$&0.69$^{+0.53}_{-0.24}$&Thick&$-$9.71&...&...\\
3& 05 35 14.72& $-$06 13 39.9&M5.5$^{a}$&M5.5&0.292&$-$19.6$^{a}$&...&CTTS&0.8&no&$-$2.60&0.15$^{+0.03}_{-0.02}$&0.14$^{+0.12}_{-0.06}$&No disk&$-$9.48&$-$9.86&$-$9.79\\
4& 05 35 15.83& $-$06 24 45.7&M5$^{a}$&M5&0.126&$-$10.9$^{a}$&0.9$^{a}$&WTTS&0.0&no&$-$2.56&0.18$^{+0.03}_{-0.02}$&0.99$^{+0.47}_{-0.27}$&Thin&...&...&...\\
5& 05 35 17.80& $-$06 24 38.4&M1$^{a}$,M2$^{b}$&M1&0.964&$-$6.3$^{a}$,$-$24.0$^{b}$&0.7$^{a}$&WTTS&0.3&no&$-$0.35&0.39$^{+0.03}_{-0.03}$&0.28$^{+0.07}_{-0.06}$&Thick&...&...&...\\
6& 05 35 18.26& $-$06 24 30.3&M6$^{a}$,M3.5$^{b}$&M6&0.116&$-$12.7$^{a}$&...&WTTS&0.0&no&$-$2.48&0.13$^{+0.06}_{-0.02}$&0.51$^{+1.06}_{-0.16}$&Thin&...&...&...\\
7& 05 35 18.95& $-$06 27 25.6&M4$^{b}$&M4&0.187&$-$80.0$^{b}$&...&CTTS&0.0&no&$-$1.16&0.24$^{+0.09}_{-0.06}$&1.02$^{+1.17}_{-0.47}$&Thick&$-$8.90&...&...\\
8& 05 35 19.34& $-$06 24 14.5&M2.5$^{a}$,M2.5$^{b}$&M2.5&0.262&$-$3.4$^{a}$&...&WTTS&0.0&no&$-$2.65&0.34$^{+0.04}_{-0.03}$&1.34$^{+0.48}_{-0.26}$&No disk&...&...&...\\
\end{longtable}
\end{landscape}
} }

\renewcommand{\tabcolsep}{0.07cm}
\onltab{6}{\longtab{6}{
\begin{landscape}
\begin{longtable}{lccccccccccccccc}
\caption{\label{tab:YSO_magnitudes_L1630N}Photometric magnitudes of YSOs in L1630N.}\\
\hline\hline
& RA & DEC & u$'$ & g$'$ & r$'$ & i$'$ & z$'$ & J & H &K$_{s}$ & [3.6] &[4.5] &[5.8] &[8.0] &[24] \\
ID& (J2000) & (J2000) & (mag) & (mag) & (mag) & (mag) &(mag)& (mag) & (mag) &(mag)  & (mag)  & (mag)  & (mag)  & (mag)\\
\hline
\endfirsthead
\caption{continued.}\\
\hline\hline
& RA & DEC & u$'$ & g$'$ & r$'$ & i$'$ & z$'$ & J & H &K$_{s}$ & [3.6] &[4.5] &[5.8] &[8.0] &[24]\\
ID& (J2000) & (J2000) & (mag) & (mag) & (mag) & (mag) &(mag)& (mag) & (mag) &(mag) & (mag)  & (mag)  & (mag)  & (mag)\\
\hline
\endhead
\hline
\endfoot
 ﻿     1 &05 45 21.38 &$-$00 00 45.9   &21.51$\pm$ 0.06   &18.71$\pm$ 0.01   &17.09$\pm$ 0.01   &15.46$\pm$ 0.01   &14.55$\pm$ 0.01 &12.96$\pm$ 0.02 &12.22$\pm$ 0.02 &11.93$\pm$ 0.02 &11.62$\pm$ 0.06 &11.64$\pm$ 0.06 &11.44$\pm$ 0.06 &11.52$\pm$ 0.06  & ...          \\ 
     2 &05 45 22.68 &$-$00 14 27.5   &22.36$\pm$ 0.11   &19.91$\pm$ 0.01   &18.35$\pm$ 0.01   &16.39$\pm$ 0.01   &15.25$\pm$ 0.01 &13.50$\pm$ 0.02 &12.87$\pm$ 0.03 &12.54$\pm$ 0.03 &12.21$\pm$ 0.06 &12.05$\pm$ 0.06 &12.03$\pm$ 0.06 &11.96$\pm$ 0.07  & ...          \\ 
     3 &05 45 26.16 &+00 06 37.9   &19.24$\pm$ 0.01   &17.62$\pm$ 0.01   &16.18$\pm$ 0.01   &14.87$\pm$ 0.01   &13.93$\pm$ 0.01 &12.37$\pm$ 0.02 &11.61$\pm$ 0.02 &11.21$\pm$ 0.02 &10.50$\pm$ 0.06 &10.30$\pm$ 0.06 & 9.91$\pm$ 0.06 & 9.25$\pm$ 0.06  & ...          \\ 
     4 &05 45 38.26 &$-$00 08 11.0   &21.16$\pm$ 0.05   &18.10$\pm$ 0.01   &16.30$\pm$ 0.01   &15.13$\pm$ 0.01   &14.37$\pm$ 0.01 &12.65$\pm$ 0.02 &11.82$\pm$ 0.02 &11.53$\pm$ 0.02 &11.38$\pm$ 0.06 &11.38$\pm$ 0.06 &11.20$\pm$ 0.06 &11.17$\pm$ 0.06  & ...          \\ 
     5 &05 45 41.68 &$-$00 04 02.4   &21.03$\pm$ 0.05   &18.37$\pm$ 0.01   &16.79$\pm$ 0.01   &15.26$\pm$ 0.01   &14.41$\pm$ 0.01 &12.86$\pm$ 0.03 &12.09$\pm$ 0.03 &11.85$\pm$ 0.03 &11.50$\pm$ 0.06 &11.46$\pm$ 0.06 &11.36$\pm$ 0.06 &11.45$\pm$ 0.06  & ...          \\ 
     6 &05 45 41.94 &$-$00 12 05.3   &18.55$\pm$ 0.01   &15.85$\pm$ 0.01   &14.44$\pm$ 0.01   &14.05$\pm$ 0.01   &13.32$\pm$ 0.01 &11.93$\pm$ 0.02 &11.18$\pm$ 0.02 &10.99$\pm$ 0.02 &10.84$\pm$ 0.06 &10.93$\pm$ 0.06 &10.86$\pm$ 0.06 &10.61$\pm$ 0.06  &6.45$\pm$ 0.06\\ 
     7 &05 45 42.80 &$-$00 01 01.8   &22.97$\pm$ 0.22   &20.59$\pm$ 0.01   &18.98$\pm$ 0.01   &16.89$\pm$ 0.01   &15.67$\pm$ 0.01 &13.85$\pm$ 0.02 &13.18$\pm$ 0.03 &12.75$\pm$ 0.03 &11.93$\pm$ 0.06 &11.68$\pm$ 0.06 &11.25$\pm$ 0.19 &10.55$\pm$ 0.06  & ...          \\
\end{longtable}
\end{landscape}
}}

\tiny 
\onltab{7}{\longtab{7}{
\begin{landscape}
\begin{longtable}{lccccccccccccccc}
\caption{\label{tab:YSO_magnitudes_L1641}Photometric magnitudes of YSOs in L1641.}\\
\hline\hline
& RA & DEC & u$'$ & g$'$ & r$'$ & i$'$ & z$'$ & J & H &K$_{s}$ & [3.6] &[4.5] &[5.8] &[8.0] &[24] \\
ID& (J2000) & (J2000) & (mag) & (mag) & (mag) & (mag) &(mag)& (mag) & (mag) &(mag)  & (mag)  & (mag)  & (mag)  & (mag)\\
\hline
\endfirsthead
\caption{continued.}\\
\hline\hline
& RA & DEC & u$'$ & g$'$ & r$'$ & i$'$ & z$'$ & J & H &K$_{s}$ & [3.6] &[4.5] &[5.8] &[8.0] &[24]\\
ID& (J2000) & (J2000) & (mag) & (mag) & (mag) & (mag) &(mag)& (mag) & (mag) &(mag) & (mag)  & (mag)  & (mag)  & (mag)\\
\hline
\endhead
\hline
\endfoot
﻿  1 &05 35 14.56 &$-$06 15 12.7   &21.82$\pm$ 0.22   &18.89$\pm$ 0.01   &16.78$\pm$ 0.01   &15.42$\pm$ 0.01   &14.48$\pm$ 0.01 &12.49$\pm$ 0.04 &11.38$\pm$ 0.04 &10.86$\pm$ 0.03 &10.02$\pm$ 0.06 & 9.35$\pm$ 0.06 & 9.32$\pm$ 0.06 & 8.43$\pm$ 0.06  &4.80$\pm$ 0.06\\
     2 &05 35 14.67 &$-$06 15 07.3   & ...              &22.04$\pm$ 0.09   &19.86$\pm$ 0.02   &17.65$\pm$ 0.01   &16.08$\pm$ 0.01 &13.58$\pm$ 0.04 &12.43$\pm$ 0.04 &11.79$\pm$ 0.02 &10.93$\pm$ 0.06 &10.36$\pm$ 0.06 & 9.96$\pm$ 0.06 & 9.03$\pm$ 0.06  & ...          \\
     3 &05 35 14.72 &$-$06 13 39.9   &21.89$\pm$ 0.24   &20.14$\pm$ 0.02   &18.29$\pm$ 0.01   &16.03$\pm$ 0.01   &14.70$\pm$ 0.01 &12.67$\pm$ 0.03 &11.85$\pm$ 0.03 &11.49$\pm$ 0.02 &11.03$\pm$ 0.06 &10.92$\pm$ 0.06 &10.87$\pm$ 0.06 &10.80$\pm$ 0.08  & ...          \\
     4 &05 35 15.83 &$-$06 24 45.7   &21.43$\pm$ 0.14   &19.19$\pm$ 0.01   &17.76$\pm$ 0.01   &15.99$\pm$ 0.01   &15.02$\pm$ 0.01 &13.42$\pm$ 0.02 &12.84$\pm$ 0.03 &12.55$\pm$ 0.02 &12.20$\pm$ 0.06 &12.07$\pm$ 0.06 &12.05$\pm$ 0.07 &11.93$\pm$ 0.17  & ...          \\
     5 &05 35 17.80 &$-$06 24 38.4   &17.85$\pm$ 0.01   &15.78$\pm$ 0.01   &14.38$\pm$ 0.01   &15.25$\pm$ 0.01   &12.94$\pm$ 0.01 &11.31$\pm$ 0.02 &10.52$\pm$ 0.03 &10.22$\pm$ 0.02 & 9.21$\pm$ 0.06 & 8.53$\pm$ 0.06 & 7.95$\pm$ 0.06 & 7.01$\pm$ 0.06  &4.94$\pm$ 0.06\\
     6 &05 35 18.26 &$-$06 24 30.3   &22.61$\pm$ 0.72   &19.95$\pm$ 0.02   &18.51$\pm$ 0.01   &16.51$\pm$ 0.01   &15.37$\pm$ 0.01 &13.69$\pm$ 0.03 &13.11$\pm$ 0.03 &12.80$\pm$ 0.03 &12.36$\pm$ 0.06 &12.21$\pm$ 0.06 &12.17$\pm$ 0.07 &12.02$\pm$ 0.22  & ...          \\
     7 &05 35 18.95 &$-$06 27 25.6   &20.04$\pm$ 0.06   &18.59$\pm$ 0.01   &17.20$\pm$ 0.01   &15.64$\pm$ 0.01   &14.73$\pm$ 0.01 &13.14$\pm$ 0.03 &12.39$\pm$ 0.03 &11.89$\pm$ 0.02 &11.19$\pm$ 0.06 &10.89$\pm$ 0.06 &10.46$\pm$ 0.06 & 9.73$\pm$ 0.06  &6.59$\pm$ 0.06\\
\end{longtable}
\end{landscape}
}}

\renewcommand{\tabcolsep}{0.212cm}
\scriptsize \longtab{8}{
\begin{landscape}
\begin{longtable}{lccccccccccc}
\caption{\label{tab:L1630_mass_age} Mass and age estimate for YSOs in L1630N
  with different PMS evolutionary tracks.  A complete version of this table,
  containing 132 stars, is available at the CDS. Here we only show the first 10 entries.}\\
\hline\hline
 &  RA     & DEC     &   mass        & age    & mass  & age & mass    & age  & mass & age\\
ID& (J2000) & (J2000) & (M$_{\odot}$, DM97) & (Myr, DM97)  &  ( M$_{\odot}$, D08) & (Myr, D08)  &  (M$_{\odot}$, S00) & (Myr, S00)  &  (M$_{\odot}$, B98) & (Myr, B98) \\
\hline
\endfirsthead
\caption{continued.}\\
\hline\hline
  &  RA     & DEC     &   mass        & age    & mass  & age & mass    & age  & mass & age\\
ID& (J2000) & (J2000) & (M$_{\odot}$, DM97) & (Myr, DM97)  &  ( M$_{\odot}$, D08) & (Myr, D08)  &  (M$_{\odot}$, S00) & (Myr, S00)  &  (M$_{\odot}$, B98) & (Myr, B98) \\
\hline
\endhead
\hline
\endfoot
1& 05 45 21.38& $-$00 00 45.9&0.24$^{+0.10}_{-0.05}$&0.92$^{+0.65}_{-0.47}$&0.30$^{+0.10}_{-0.06}$&0.84$^{+1.11}_{-0.39}$&0.28$^{+0.09}_{-0.03}$&1.92$^{+0.75}_{-0.53}$&0.41$^{+0.15}_{-0.06}$&1.76$^{+2.08}_{-0.41}$\\
2& 05 45 22.68& $-$00 14 27.5&0.13$^{+0.03}_{-0.02}$&0.82$^{+0.55}_{-0.45}$&0.13$^{+0.06}_{-0.01}$&0.38$^{+0.70}_{-0.12}$&0.13$^{+0.04}_{-0.02}$&0.61$^{+2.06}_{-0.16}$&...&...\\
3& 05 45 26.16& +00 06 37.9&0.23$^{+0.08}_{-0.05}$&0.68$^{+0.22}_{-0.42}$&0.29$^{+0.09}_{-0.06}$&0.51$^{+0.51}_{-0.22}$&0.32$^{+0.07}_{-0.05}$&1.54$^{+0.41}_{-0.51}$&0.45$^{+0.14}_{-0.03}$&1.33$^{+1.23}_{-0.19}$\\
4& 05 45 38.26& $-$00 08 11.0&0.43$^{+0.14}_{-0.08}$&0.90$^{+1.56}_{-0.31}$&0.55$^{+0.21}_{-0.11}$&1.36$^{+3.15}_{-0.61}$&0.57$^{+0.22}_{-0.10}$&2.14$^{+3.60}_{-0.68}$&0.96$^{+0.12}_{-0.17}$&4.41$^{+4.86}_{-1.82}$\\
5& 05 45 41.68& $-$00 04 02.4&0.18$^{+0.03}_{-0.02}$&0.72$^{+0.18}_{-0.36}$&0.24$^{+0.03}_{-0.03}$&0.56$^{+0.20}_{-0.15}$&0.25$^{+0.02}_{-0.02}$&2.04$^{+0.06}_{-0.13}$&0.31$^{+0.05}_{-0.02}$&1.15$^{+0.37}_{-0.08}$\\
6& 05 45 41.94& $-$00 12 05.3&0.61$^{+0.27}_{-0.13}$&0.93$^{+2.30}_{-0.33}$&0.92$^{+0.23}_{-0.23}$&2.00$^{+4.04}_{-0.93}$&1.07$^{+0.15}_{-0.26}$&3.52$^{+5.77}_{-1.42}$&1.35$^{+0.03}_{-0.16}$&6.09$^{+8.46}_{-2.48}$\\
7& 05 45 42.80& $-$00 01 01.8&0.15$^{+0.02}_{-0.02}$&1.74$^{+0.35}_{-0.34}$&0.15$^{+0.04}_{-0.03}$&1.07$^{+0.89}_{-0.40}$&0.15$^{+0.04}_{-0.02}$&3.77$^{+0.45}_{-2.02}$&0.15$^{+0.06}_{-0.02}$&1.53$^{+4.80}_{-0.19}$\\
8& 05 45 44.37& +00 22 58.2&0.34$^{+0.08}_{-0.07}$&0.60$^{+0.38}_{-0.23}$&0.43$^{+0.12}_{-0.08}$&0.68$^{+0.82}_{-0.28}$&0.46$^{+0.11}_{-0.08}$&1.40$^{+0.94}_{-0.25}$&0.78$^{+0.16}_{-0.13}$&2.33$^{+2.08}_{-0.76}$\\
\end{longtable}
\end{landscape}
}

\clearpage \scriptsize \longtab{9}{
\begin{landscape}
\begin{longtable}{lccccccccccc}
\caption{\label{tab:L1641_mass_age} Mass and age estimate for YSOs in L1641
  with different PMS evolutionary tracks. A complete version of this table,
  containing 267 stars, is available at the CDS. Here we only show the first 10 entries.}\\
\hline\hline
  &  RA     & DEC     &   Mass        & Age    & Mass  & Age & Mass    & Age  & Mass & Age\\
ID& (J2000) & (J2000) & (M$_{\odot}$, DM97) & (Myr, DM97)  &  ( M$_{\odot}$, D08) & (Myr, D08)  &  (M$_{\odot}$, S00) & (Myr, S00)  &  (M$_{\odot}$, B98) & (Myr, B98) \\
\hline
\endfirsthead
\caption{continued.}\\
\hline\hline
 &  RA     & DEC     &   Mass        & Age    & Mass  & Age & Mass    & Age  & Mass & Age\\
ID& (J2000) & (J2000) & (M$_{\odot}$, DM97) & (Myr, DM97)  &  ( M$_{\odot}$, D08) & (Myr, D08)  &  (M$_{\odot}$, S00) & (Myr, S00)  &  (M$_{\odot}$, B98) & (Myr, B98) \\
\hline
\endhead
\hline
\endfoot
1& 05 35 14.56& $-$06 15 12.7&0.34$^{+0.05}_{-0.04}$&0.63$^{+0.26}_{-0.15}$&0.42$^{+0.06}_{-0.05}$&0.70$^{+0.52}_{-0.22}$&0.45$^{+0.05}_{-0.04}$&1.44$^{+0.47}_{-0.22}$&0.74$^{+0.10}_{-0.08}$&2.29$^{+1.18}_{-0.63}$\\
2& 05 35 14.67& $-$06 15 07.3&0.16$^{+0.02}_{-0.02}$&1.11$^{+0.28}_{-0.51}$&0.19$^{+0.04}_{-0.03}$&0.69$^{+0.53}_{-0.24}$&0.20$^{+0.03}_{-0.03}$&2.55$^{+0.46}_{-0.94}$&...&...&\\
3& 05 35 14.72& $-$06 13 39.9&0.13$^{+0.01}_{-0.01}$&0.12$^{+0.09}_{-0.03}$&0.15$^{+0.03}_{-0.02}$&0.14$^{+0.12}_{-0.06}$&0.19$^{+0.03}_{-0.02}$&0.37$^{+1.03}_{-0.11}$&...&...&\\
4& 05 35 15.83& $-$06 24 45.7&0.16$^{+0.02}_{-0.01}$&1.42$^{+0.19}_{-0.21}$&0.18$^{+0.03}_{-0.02}$&0.99$^{+0.47}_{-0.27}$&0.17$^{+0.03}_{-0.01}$&3.17$^{+0.28}_{-1.00}$&0.19$^{+0.04}_{-0.02}$&1.45$^{+2.29}_{-0.19}$\\
5& 05 35 17.80& $-$06 24 38.4&0.29$^{+0.02}_{-0.02}$&0.19$^{+0.15}_{-0.06}$&0.39$^{+0.03}_{-0.03}$&0.28$^{+0.07}_{-0.06}$&0.46$^{+0.03}_{-0.03}$&0.93$^{+0.06}_{-0.04}$&0.82$^{+0.06}_{-0.04}$&1.17$^{+0.27}_{-0.10}$\\
6& 05 35 18.26& $-$06 24 30.3&0.14$^{+0.04}_{-0.02}$&1.32$^{+0.42}_{-0.70}$&0.13$^{+0.06}_{-0.02}$&0.51$^{+1.06}_{-0.16}$&0.14$^{+0.05}_{-0.02}$&0.74$^{+2.46}_{-0.18}$&...&...\\
7& 05 35 18.95& $-$06 27 25.6&0.19$^{+0.07}_{-0.04}$&1.15$^{+0.53}_{-0.51}$&0.24$^{+0.09}_{-0.06}$&1.02$^{+1.17}_{-0.47}$&0.25$^{+0.06}_{-0.05}$&2.68$^{+0.69}_{-0.83}$&0.28$^{+0.11}_{-0.04}$&1.64$^{+3.22}_{-0.33}$\\
8& 05 35 19.34& $-$06 24 14.5&0.27$^{+0.04}_{-0.03}$&1.16$^{+0.27}_{-0.12}$&0.34$^{+0.04}_{-0.03}$&1.34$^{+0.48}_{-0.26}$&0.34$^{+0.02}_{-0.03}$&2.35$^{+0.33}_{-0.18}$&0.44$^{+0.06}_{-0.04}$&2.57$^{+0.92}_{-0.48}$\\
\end{longtable}
\end{landscape}
}

\renewcommand{\tabcolsep}{0.05cm}
\scriptsize \onltab{10}{\longtab{10}{
\begin{longtable}{lccccccccccccc}
\caption{\label{tab:emission_lines_L1630N} The equivalent widths of emission lines on the optical spectra for YSOs in L1630N.}\\
\hline\hline
  &  RA     & DEC     &    [O I]    &    [O I]          & [O I] & [N II]         & [N II]         & He I         & [S II]       & [S II]  & Ca II & Ca II & Ca II\\
ID& (J2000) & (J2000) &(5577\,$\AA$)&    (6300\,$\AA$)  &  (6363\,$\AA$) & (6548\,$\AA$)  &  (6583\,$\AA$) &(6678\,$\AA$) &(6716\,$\AA$) & (6731\,$\AA$) & (8498\,$\AA$) & (8542\,$\AA$) & (8662\,$\AA$)\\
\hline
\endfirsthead
\caption{continued.}\\
\hline\hline
  &  RA     & DEC     &    [O I]    &    [O I]          & [N II] &         & [N II]         & He I         & [S II]       & [S II]  & Ca II & Ca II & Ca II\\
ID& (J2000) & (J2000) &(5577\,$\AA$)&    (6300\,$\AA$)  &  (6363\,$\AA$) & (6548\,$\AA$)  &  (6583\,$\AA$) &(6678\,$\AA$) &(6716\,$\AA$) & (6731\,$\AA$) & (8498\,$\AA$) & (8542\,$\AA$) & (8662\,$\AA$)\\
\hline
\hline
\endhead
\hline
\endfoot
﻿    14 &05 45 53.11 &$-$00 13 24.9 &   ...            &   $-$0.6$\pm$   0.01&    ...            &    ...            &    ...            &   $-$0.5$\pm$   0.01&    ...            &    ...            &    ...            &    ...            &    ...            \\
    20 &05 45 56.31 &+00 07 08.6 &   ...            &   $-$1.3$\pm$   0.01&    ...            &    ...            &    ...            &    ...            &    ...            &    ...            &    ...            &    ...            &    ...            \\
    23 &05 45 57.62 &+00 07 21.1 &  $-$1.0$\pm$   0.02&   $-$5.1$\pm$   0.15&    ...            &    ...            &    ...            &   $-$1.3$\pm$   0.02&    ...            &    ...            &    ...            &    ...            &    ...            \\
    24 &05 45 57.93 &+00 02 48.6 &  $-$0.8$\pm$   0.01&   $-$3.2$\pm$   0.04&    ...            &    ...            &    ...            &   $-$2.0$\pm$   0.02&    ...            &    ...            &   $-$3.6$\pm$   0.03&   $-$4.6$\pm$   0.04&   $-$4.6$\pm$   0.04\\
    25 &05 46 00.18 &+00 03 07.0 &   ...            &   $-$0.4$\pm$   0.01&    ...            &    ...            &    ...            &   $-$0.8$\pm$   0.01&    ...            &    ...            &    ...            &    ...            &    ...            \\
    27 &05 46 04.64 &+00 04 58.1 &  $-$0.6$\pm$   0.01&   $-$1.5$\pm$   0.01&   $-$0.6$\pm$   0.01&    ...            &    ...            &   $-$0.7$\pm$   0.01&    ...            &    ...            &    ...            &    ...            &    ...            \\
    32 &05 46 10.31 &$-$00 00 06.7 &  $-$0.9$\pm$   0.01&   $-$1.9$\pm$   0.02&   $-$0.8$\pm$   0.01&    ...            &    ...            &   $-$2.7$\pm$   0.02&    ...            &    ...            &   $-$2.0$\pm$   0.02&   $-$2.2$\pm$   0.02&   $-$1.6$\pm$   0.01\\
\end{longtable}
}}

\clearpage \scriptsize \onltab{11}{\longtab{11}{
\begin{longtable}{lccccccccccccc}
\caption{\label{tab:emission_lines_L1641} The equivalent widths of emission lines on the optical spectra for YSOs in L1641.}\\
\hline\hline
  &  RA     & DEC     &    [O I]    &    [O I]          & [O I] & [N II]         & [N II]         & He I         & [S II]       & [S II]  & Ca II & Ca II & Ca II\\
ID& (J2000) & (J2000) &(5577\,$\AA$)&    (6300\,$\AA$)  &  (6363\,$\AA$) & (6548\,$\AA$)  &  (6583\,$\AA$) &(6678\,$\AA$) &(6716\,$\AA$) & (6731\,$\AA$) & (8498\,$\AA$) & (8542\,$\AA$) & (8662\,$\AA$)\\
\hline
\endfirsthead
\caption{continued.}\\
\hline\hline
  &  RA     & DEC     &    [O I]    &    [O I]          & [O I] & [N II]        & [N II]         & He I          & [S II]       & [S II]  & Ca II & Ca II & Ca II\\
ID& (J2000) & (J2000) &(5577\,$\AA$)&    (6300\,$\AA$)  &  (6363\,$\AA$) & (6548\,$\AA$)  &  (6583\,$\AA$) &(6678\,$\AA$) &(6716\,$\AA$) & (6731\,$\AA$) & (8498\,$\AA$) & (8542\,$\AA$) & (8662\,$\AA$)\\
\hline
\hline
\endhead
\hline
\endfoot
﻿     5 &05 35 17.80 &$-$06 24 38.4&    ...            &    ...            &    ...            &    ...            &    ...            &    ...            &    ...            &    ...            &   $-$0.6$\pm$   0.01&   $-$0.6$\pm$   0.01&   $-$0.6$\pm$    ...\\
    20 &05 35 27.35 &$-$06 19 31.3&    ...            &   $-$3.5$\pm$   0.12&    ...            &    ...            &    ...            &    ...            &    ...            &    ...            &    ...            &    ...            &    ...            \\
    28 &05 35 33.93 &$-$06 14 32.8&    ...            &   $-$1.2$\pm$   0.01&    ...            &    ...            &    ...            &   $-$0.9$\pm$   0.01&    ...            &    ...            &    ...            &    ...            &    ...            \\
    34 &05 35 41.02 &$-$06 22 45.4&   $-$0.7$\pm$   0.01&   $-$1.6$\pm$   0.02&    ...            &    ...            &    ...            &    ...            &    ...            &    ...            &    ...            &    ...            &    ...            \\
    39 &05 35 43.37 &$-$06 22 19.6&    ...            &   $-$4.6$\pm$   0.05&   $-$1.6$\pm$   0.01&    ...            &    ...            &   $-$2.4$\pm$   0.02&    ...            &    ...            &    ...            &    ...            &    ...            \\
    40 &05 35 44.18 &$-$06 28 16.4&    ...            &    ...            &    ...            &    ...            &    ...            &   $-$4.3$\pm$   0.06&    ...            &    ...            &    ...            &    ...            &    ...            \\
    45 &05 35 50.19 &$-$06 50 40.2& $-$283.0$\pm$  28.22&  $-$56.5$\pm$   6.12&    ...            &    ...            &  $-$12.2$\pm$   1.46&    ...            &    ...            &    ...            &    ...            &    ...            &    ...            \\
\end{longtable}
}}

\renewcommand{\tabcolsep}{0.05cm}
\scriptsize \longtab{12}{
\begin{landscape}
\begin{longtable}{lcccccccccccccccrc}
\caption{\label{tab:background_stars_L1630N} Unrelated field stars with
  photometric magnitudes in L1630N.  A complete version of this table,
  containing 108 stars, is available at the CDS. Here we only show the first 10 entries.}\\
\hline\hline
& RA & DEC & u$'$ & g$'$ & r$'$ & i$'$ & z$'$ & J & H &K$_{s}$ & [3.6] &[4.5] &[5.8] &[8.0] &[24] &  &Av\\
ID& (J2000) & (J2000) & (mag) & (mag) & (mag) & (mag) &(mag)& (mag) & (mag)
&(mag)  & (mag)  & (mag)  & (mag)  & (mag) & (mag)& Spt & (mag)\\
\hline
\endfirsthead
\caption{continued.}\\
\hline\hline
& RA & DEC & u$'$ & g$'$ & r$'$ & i$'$ & z$'$ & J & H &K$_{s}$ & [3.6] &[4.5] &[5.8] &[8.0] &[24] & &Av\\
ID& (J2000) & (J2000) & (mag) & (mag) & (mag) & (mag) &(mag)& (mag) & (mag) &(mag) & (mag)  & (mag)  & (mag)  & (mag) &Spt &(mag)\\
\hline
\endhead
\hline
\endfoot
     1 &05 45 29.13 &$-$00 13 49.4   &24.24$\pm$ 0.68   &20.82$\pm$ 0.02   &18.82$\pm$ 0.01   &17.88$\pm$ 0.01   &17.21$\pm$ 0.01 &15.43$\pm$ 0.05 &14.73$\pm$ 0.05 &14.29$\pm$ 0.09 &14.21$\pm$ 0.06 &14.24$\pm$ 0.06 &14.19$\pm$ 0.10 &14.23$\pm$ 0.26   &...       &  K6  &2.7\\ 
     2 &05 45 29.45 &$-$00 12 12.9   & ...              &21.09$\pm$ 0.02   &19.27$\pm$ 0.01   &18.02$\pm$ 0.01   &17.35$\pm$ 0.01 &15.87$\pm$ 0.08 &15.48$\pm$ 0.14 &15.18$\pm$ 0.17 &14.87$\pm$ 0.06 &14.84$\pm$ 0.06 &  ...           &  ...             &...       &M3.5  &0.1\\
     3 &05 45 30.81 &$-$00 09 43.1   &23.43$\pm$ 0.36   &20.69$\pm$ 0.01   &19.22$\pm$ 0.01   &18.41$\pm$ 0.01   &17.83$\pm$ 0.01 &16.20$\pm$ 0.11 &15.87$\pm$ 0.17 &15.36$\pm$ 0.20 &15.29$\pm$ 0.06 &15.40$\pm$ 0.06 &  ...           &  ...             &...       &  F8  &3.8\\
     4 &05 45 31.70 &$-$00 08 33.5   &21.99$\pm$ 0.10   &19.37$\pm$ 0.01   &17.71$\pm$ 0.01   &16.83$\pm$ 0.01   &16.21$\pm$ 0.01 &14.60$\pm$ 0.03 &13.86$\pm$ 0.03 &13.67$\pm$ 0.05 &13.40$\pm$ 0.06 &13.40$\pm$ 0.06 &13.45$\pm$ 0.07 &13.25$\pm$ 0.13   &...       &G1.5  &4.0\\
     5 &05 45 33.01 &$-$00 04 55.0   &23.72$\pm$ 0.46   &21.73$\pm$ 0.03   &20.11$\pm$ 0.01   &19.21$\pm$ 0.01   &18.60$\pm$ 0.02 &  ....          &  ...           &  ...           &15.96$\pm$ 0.06 &15.91$\pm$ 0.07 &  ...           &  ...             &...       &  G0  &4.1\\
     6 &05 45 33.72 &$-$00 04 49.0   &21.28$\pm$ 0.06   &18.28$\pm$ 0.01   &16.68$\pm$ 0.01   &15.90$\pm$ 0.01   &15.37$\pm$ 0.01 &13.91$\pm$ 0.02 &13.11$\pm$ 0.02 &12.93$\pm$ 0.03 &12.75$\pm$ 0.06 &12.78$\pm$ 0.06 &12.77$\pm$ 0.06 &12.83$\pm$ 0.10   &...       &  K4  &2.3\\
     7 &05 45 34.32 &$-$00 04 27.5   &23.46$\pm$ 0.42   &21.19$\pm$ 0.02   &18.91$\pm$ 0.01   &17.39$\pm$ 0.01   &16.45$\pm$ 0.01 &14.69$\pm$ 0.03 &13.63$\pm$ 0.04 &13.36$\pm$ 0.03 &13.20$\pm$ 0.06 &13.07$\pm$ 0.06 &12.98$\pm$ 0.06 &12.91$\pm$ 0.10   &...       &M2.5  &1.6\\
\end{longtable}
\end{landscape}
 }

\scriptsize \longtab{13}{
\begin{landscape}
\begin{longtable}{lcccccccccccccccrc}
\caption{\label{tab:background_stars_L1641}  Unrelated field stars with
  photometric magnitudes in L1641. A complete version of this table,
  containing 71 stars, is available at the CDS. Here we only show the first 10 entries.}\\
\hline\hline
& RA & DEC & u$'$ & g$'$ & r$'$ & i$'$ & z$'$ & J & H &K$_{s}$ & [3.6] &[4.5] &[5.8] &[8.0] &[24] &  &Av\\
ID& (J2000) & (J2000) & (mag) & (mag) & (mag) & (mag) &(mag)& (mag) & (mag)
&(mag)  & (mag)  & (mag)  & (mag)  & (mag) & (mag) &Spt & (mag)\\
\hline
\endfirsthead
\caption{continued.}\\
\hline\hline
& RA & DEC & u$'$ & g$'$ & r$'$ & i$'$ & z$'$ & J & H &K$_{s}$ & [3.6] &[4.5] &[5.8] &[8.0] &[24] & &Av\\
ID& (J2000) & (J2000) & (mag) & (mag) & (mag) & (mag) &(mag)& (mag) & (mag) &(mag) & (mag)  & (mag)  & (mag)  & (mag) &Spt &(mag)\\
\hline
\endhead
\hline
\endfoot
﻿     1 &05 35 18.65 &$-$06 25 09.7   &21.12$\pm$ 0.16   &18.42$\pm$ 0.01   &17.01$\pm$ 0.01   &15.76$\pm$ 0.01   &15.10$\pm$ 0.01 &13.72$\pm$ 0.02 &13.01$\pm$ 0.03 &12.82$\pm$ 0.02 &12.51$\pm$ 0.06 &12.47$\pm$ 0.06 &12.33$\pm$ 0.07 &12.02$\pm$ 0.18  & ...            &M3.5  &0.1\\ 
     2 &05 35 28.58 &$-$06 17 00.0   &22.40$\pm$ 0.59   &22.67$\pm$ 0.17   &19.77$\pm$ 0.02   &18.06$\pm$ 0.01   &16.83$\pm$ 0.01 &14.48$\pm$ 0.03 &13.42$\pm$ 0.03 &12.86$\pm$ 0.02 &12.52$\pm$ 0.06 &12.38$\pm$ 0.06 &12.39$\pm$ 0.07 &12.48$\pm$ 0.25  & ...            &  K5  &5.7\\
     3 &05 35 32.41 &$-$06 16 48.0   &21.02$\pm$ 0.18   &20.81$\pm$ 0.04   &16.86$\pm$ 0.01   &13.63$\pm$ 0.01   &11.75$\pm$ 0.01 & 7.93$\pm$ 0.03 & 6.24$\pm$ 0.06 & 5.43$\pm$ 0.01 & 5.59$\pm$ 0.06 & 4.94$\pm$ 0.06 & 4.79$\pm$ 0.06 & 4.64$\pm$ 0.06  &4.05$\pm$ 0.06  &  M6  &7.3\\
     4 &05 35 34.81 &$-$06 21 51.3   &13.71$\pm$ 0.01   &10.91$\pm$ 0.01   &10.44$\pm$ 0.01   &10.35$\pm$ 0.01   &12.83$\pm$ 0.01 & 9.50$\pm$ 0.02 & 9.25$\pm$ 0.03 & 9.17$\pm$ 0.01 & 9.11$\pm$ 0.06 & 9.12$\pm$ 0.06 & 8.99$\pm$ 0.06 & 9.17$\pm$ 0.06  & ...            &B8.5  &2.1\\
     5 &05 35 39.41 &$-$06 16 02.7   &19.87$\pm$ 0.07   &17.15$\pm$ 0.01   &15.71$\pm$ 0.01   &15.14$\pm$ 0.01   &14.54$\pm$ 0.01 &13.27$\pm$ 0.03 &12.61$\pm$ 0.03 &12.39$\pm$ 0.02 &12.09$\pm$ 0.06 &12.27$\pm$ 0.06 &12.23$\pm$ 0.07 &12.23$\pm$ 0.20  & ...            &M1.5  &0.1\\
     6 &05 35 40.03 &$-$06 23 54.3   & ...              &21.44$\pm$ 0.06   &18.99$\pm$ 0.01   &17.58$\pm$ 0.01   &16.54$\pm$ 0.01 &14.48$\pm$ 0.03 &13.51$\pm$ 0.03 &13.10$\pm$ 0.02 &12.78$\pm$ 0.06 &12.65$\pm$ 0.06 &12.71$\pm$ 0.07 &12.67$\pm$ 0.23  & ...            &  K5  &4.5\\
     7 &05 35 46.67 &$-$06 39 55.1   &22.35$\pm$ 0.75   &20.62$\pm$ 0.04   &19.03$\pm$ 0.01   &18.22$\pm$ 0.01   &17.59$\pm$ 0.01 &16.03$\pm$ 0.09 &15.46$\pm$ 0.12 &15.36$\pm$ 0.16 &15.12$\pm$ 0.08 &15.06$\pm$ 0.07 &  ...           &  ...            & ...            &F6.5  &4.0\\
\end{longtable}
\end{landscape}
 }

\scriptsize \longtab{14}{
\begin{landscape}
\begin{longtable}{lccccccccccccccc}
\caption{\label{tab:all_photometry_L1630N} Photometric magnitudes for stars in
  L1630N. A complete version of this table, containing  12199 stars, is
  available at the CDS. Here we only show the first 10 entries.}\\
\hline\hline
& RA & DEC & u$'$ & g$'$ & r$'$ & i$'$ & z$'$ & J & H &K$_{s}$ & [3.6] &[4.5] &[5.8] &[8.0] &[24] \\
ID& (J2000) & (J2000) & (mag) & (mag) & (mag) & (mag) &(mag)& (mag) & (mag) &(mag)  & (mag)  & (mag)  & (mag)  & (mag) & (mag)\\
\hline
\endfirsthead
\caption{continued.}\\
\hline\hline
& RA & DEC & u$'$ & g$'$ & r$'$ & i$'$ & z$'$ & J & H &K$_{s}$ & [3.6] &[4.5] &[5.8] &[8.0] &[24]\\
ID& (J2000) & (J2000) & (mag) & (mag) & (mag) & (mag) &(mag)& (mag) & (mag) &(mag) & (mag)  & (mag)  & (mag)  & (mag) & (mag)\\
\hline
\endhead
\hline
\endfoot
﻿     1 &05 45 21.05 &$+$00 00 52.7   & ...              & ...              &23.79$\pm$ 0.30   &22.20$\pm$ 0.11   &21.18$\pm$ 0.21 &  ...           &  ...           &  ...           &17.08$\pm$ 0.08 &17.03$\pm$ 0.10 &  ...           &  ...          &   ... \\
     2 &05 45 21.07 &$+$00 49 19.8   &21.10$\pm$ 0.06   &19.60$\pm$ 0.01   &18.79$\pm$ 0.01   &18.43$\pm$ 0.01   &18.25$\pm$ 0.02 &  ...           &  ...           &  ...           &16.51$\pm$ 0.07 &16.21$\pm$ 0.07 &  ...           &  ...          &   ... \\
     3 &05 45 21.11 &$+$00 38 36.9   & ...              & ...              &23.15$\pm$ 0.39   &21.93$\pm$ 0.21   &21.21$\pm$ 0.38 &  ...           &  ...           &  ...           &16.63$\pm$ 0.08 &16.66$\pm$ 0.08 &  ...           &  ...          &   ... \\
     4 &05 45 21.14 &$+$00 37 48.4   &23.91$\pm$ 0.44   &21.20$\pm$ 0.02   &19.82$\pm$ 0.01   &19.20$\pm$ 0.01   &18.77$\pm$ 0.02 &  ...           &  ...           &  ...           &16.30$\pm$ 0.07 &16.46$\pm$ 0.07 &  ...           &  ...          &   ... \\
     5 &05 45 21.15 &$-$00 11 05.2   &19.82$\pm$ 0.02   &17.44$\pm$ 0.01   &16.03$\pm$ 0.01   &15.31$\pm$ 0.01   &14.80$\pm$ 0.01 &13.38$\pm$ 0.02 &12.82$\pm$ 0.04 &12.57$\pm$ 0.02 &12.45$\pm$ 0.06 &12.46$\pm$ 0.06 &12.39$\pm$ 0.06 &12.04$\pm$ 0.07&   ... \\
     6 &05 45 21.15 &$+$00 18 48.0   & ...              &24.52$\pm$ 0.38   &22.44$\pm$ 0.10   &20.86$\pm$ 0.04   &20.00$\pm$ 0.08 &  ...           &  ...           &  ...           &  ...           &16.97$\pm$ 0.10 &  ...           &  ...          &   ... \\
     7 &05 45 21.16 &$+$00 47 15.0   & ...              & ...              &22.93$\pm$ 0.35   &21.50$\pm$ 0.16   &20.06$\pm$ 0.14 &  ...           &  ...           &  ...           &16.96$\pm$ 0.09 &  ...           &  ...           &  ...          &   ... \\
\end{longtable}
\end{landscape}
 }

\scriptsize \longtab{15}{
\begin{landscape}
\begin{longtable}{lccccccccccccccc}
\caption{\label{tab:all_photometry_L1641} Photometric magnitudes for stars in
  L1641. A complete version of this table, containing  9495 stars, is
  available at the CDS. Here we only show the first 10 entries.}\\
\hline\hline
& RA & DEC & u$'$ & g$'$ & r$'$ & i$'$ & z$'$ & J & H &K$_{s}$ & [3.6] &[4.5] &[5.8] &[8.0] &[24] \\
ID& (J2000) & (J2000) & (mag) & (mag) & (mag) & (mag) &(mag)& (mag) & (mag) &(mag)  & (mag)  & (mag)  & (mag)  & (mag) & (mag)\\
\hline
\endfirsthead
\caption{continued.}\\
\hline\hline
& RA & DEC & u$'$ & g$'$ & r$'$ & i$'$ & z$'$ & J & H &K$_{s}$ & [3.6] &[4.5] &[5.8] &[8.0] &[24]\\
ID& (J2000) & (J2000) & (mag) & (mag) & (mag) & (mag) &(mag)& (mag) & (mag) &(mag) & (mag)  & (mag)  & (mag)  & (mag) & (mag)\\
\hline
\endhead
\hline
\endfoot
﻿    1 &05 34 04.47 &$-$06 31 37.9   &22.43$\pm$ 0.56   &19.53$\pm$ 0.02   &18.10$\pm$ 0.01   &16.47$\pm$ 0.01   &15.52$\pm$ 0.01 &13.90$\pm$ 0.03 &13.13$\pm$ 0.04 &12.68$\pm$ 0.02 &  ...           &  ...           &  ...           &11.40$\pm$ 0.08  & ...          \\ 
     2 &05 34 05.73 &$-$06 28 51.7   &21.67$\pm$ 0.18   &17.90$\pm$ 0.01   &15.61$\pm$ 0.01   &14.46$\pm$ 0.01   &13.63$\pm$ 0.01 &11.77$\pm$ 0.02 &10.84$\pm$ 0.03 &10.47$\pm$ 0.02 &  ...           &  ...           &  ...           &10.03$\pm$ 0.06  & ...          \\
     3 &05 34 06.23 &$-$06 30 45.1   &19.71$\pm$ 0.06   &16.74$\pm$ 0.01   &15.24$\pm$ 0.01   &16.37$\pm$ 0.01   &13.14$\pm$ 0.01 &11.59$\pm$ 0.02 &10.84$\pm$ 0.03 &10.59$\pm$ 0.02 &  ...           &  ...           &  ...           &10.18$\pm$ 0.06  &5.70$\pm$ 0.06\\
     4 &05 34 06.28 &$-$06 28 39.8   &23.47$\pm$ 0.91   &19.76$\pm$ 0.01   &17.69$\pm$ 0.01   &16.57$\pm$ 0.01   &15.79$\pm$ 0.01 &14.04$\pm$ 0.03 &13.10$\pm$ 0.03 &12.73$\pm$ 0.02 &  ...           &  ...           &  ...           &12.26$\pm$ 0.13  & ...          \\
     5 &05 34 06.96 &$-$06 32 07.9   &20.87$\pm$ 0.15   &18.22$\pm$ 0.01   &16.80$\pm$ 0.01   &15.45$\pm$ 0.01   &14.71$\pm$ 0.01 &13.29$\pm$ 0.02 &12.59$\pm$ 0.03 &12.26$\pm$ 0.02 &  ...           &  ...           &  ...           & 9.34$\pm$ 0.06  &5.67$\pm$ 0.06\\
     6 &05 34 09.07 &$-$06 33 56.7   &21.05$\pm$ 0.18   &19.26$\pm$ 0.01   &17.91$\pm$ 0.01   &16.30$\pm$ 0.01   &15.38$\pm$ 0.01 &13.84$\pm$ 0.02 &13.20$\pm$ 0.03 &12.99$\pm$ 0.02 &  ...           &  ...           &  ...           &12.43$\pm$ 0.15  & ...          \\
     7 &05 34 09.57 &$-$06 25 41.4   & ...              &22.92$\pm$ 0.16   &21.33$\pm$ 0.05   &20.54$\pm$ 0.04   &19.99$\pm$ 0.07 &  ...           &  ...           &  ...           &17.40$\pm$ 0.16 &17.12$\pm$ 0.13 &  ...           &  ...            & ...          \\
\end{longtable}
\end{landscape}
 }

\scriptsize \onltab{16}{\longtab{16}{
\begin{landscape}
\begin{longtable}{lccccccccccccccc}
\caption{\label{tab:TD_candidates} Photometric magnitudes for transition disk object candidates.}\\
\hline\hline
& RA & DEC & u$'$ & g$'$ & r$'$ & i$'$ & z$'$ & J & H &K$_{s}$ & [3.6] &[4.5] &[5.8] &[8.0] &[24] \\
ID& (J2000) & (J2000) & (mag) & (mag) & (mag) & (mag) &(mag)& (mag) & (mag) &(mag)  & (mag)  & (mag)  & (mag)  & (mag) & (mag)\\
\hline
\endfirsthead
\caption{continued.}\\
\hline\hline
& RA & DEC & u$'$ & g$'$ & r$'$ & i$'$ & z$'$ & J & H &K$_{s}$ & [3.6] &[4.5] &[5.8] &[8.0] &[24]\\
ID& (J2000) & (J2000) & (mag) & (mag) & (mag) & (mag) &(mag)& (mag) & (mag) &(mag) & (mag)  & (mag)  & (mag)  & (mag) & (mag)\\
\hline
\endhead
\hline
\endfoot
﻿﻿     1 &05 42 56.39 &$-$08 37 45.9   & ...              & ...              &17.76$\pm$ 0.01   &16.26 0.01   & ...            &13.74$\pm$ 0.02 &12.87$\pm$ 0.02 &12.57$\pm$ 0.02 &12.41$\pm$ 0.06 &12.17$\pm$ 0.06 &12.27$\pm$ 0.06 &12.36$\pm$ 0.09  &8.85$\pm$ 0.06\\
     2 &05 42 45.03 &$-$08 33 36.2   & ...              & ...              &20.31$\pm$ 0.01   &18.06 0.01   &16.79$\pm$ 0.01 &13.26$\pm$ 0.03 &12.02$\pm$ 0.04 &11.40$\pm$ 0.02 &10.90$\pm$ 0.06 &10.96$\pm$ 0.06 &10.46$\pm$ 0.06 &10.23$\pm$ 0.06  &6.40$\pm$ 0.06\\
     3 &05 42 30.62 &$-$08 32 55.2   & ...              &13.97$\pm$ 0.01   &12.92$\pm$ 0.01   &12.48 0.01   &12.52$\pm$ 0.01 &10.82$\pm$ 0.02 &10.13$\pm$ 0.02 & 9.76$\pm$ 0.02 & 9.20$\pm$ 0.06 & 8.92$\pm$ 0.06 & 8.70$\pm$ 0.06 & 8.47$\pm$ 0.06  &3.89$\pm$ 0.06\\
     4 &05 43 13.53 &$-$08 31 00.5   & ...              & ...              &20.88$\pm$ 0.01   &18.70 0.01   &17.25$\pm$ 0.01 &14.22$\pm$ 0.03 &12.54$\pm$ 0.03 &11.77$\pm$ 0.02 &11.21$\pm$ 0.06 &10.87$\pm$ 0.06 &10.78$\pm$ 0.06 &10.82$\pm$ 0.06  &5.90$\pm$ 0.06\\
     5 &05 42 39.33 &$-$08 26 28.2   & ...              &17.48$\pm$ 0.01   &15.59$\pm$ 0.01   &14.57 0.01   &14.05$\pm$ 0.01 &12.59$\pm$ 0.02 &11.73$\pm$ 0.02 &11.53$\pm$ 0.02 &11.39$\pm$ 0.06 &11.20$\pm$ 0.06 &10.78$\pm$ 0.06 & 9.89$\pm$ 0.06  &6.38$\pm$ 0.06\\
     6 &05 42 53.59 &$-$08 20 22.7   & ...              &19.02$\pm$ 0.01   &16.89$\pm$ 0.01   &15.72 0.01   &14.72$\pm$ 0.01 &12.80$\pm$ 0.02 &11.77$\pm$ 0.02 &11.35$\pm$ 0.02 &10.88$\pm$ 0.06 &10.39$\pm$ 0.06 &10.34$\pm$ 0.06 & 9.52$\pm$ 0.06  &5.68$\pm$ 0.06\\
     7 &05 43 04.56 &$-$08 16 39.1   & ...              & ...              &18.92$\pm$ 0.01   &17.12 0.01   &15.88$\pm$ 0.01 &13.63$\pm$ 0.02 &12.49$\pm$ 0.03 &11.94$\pm$ 0.02 &11.69$\pm$ 0.06 &11.04$\pm$ 0.06 &10.96$\pm$ 0.06 &10.07$\pm$ 0.06  &6.87$\pm$ 0.06\\
\end{longtable}
\end{landscape}
 }}

\end{document}